\documentclass{pasa}

\jid{PASA}
\doi{10.1017/pas.\the\year.xxx  Draft dated 2013/10/02}
\jyear{2013}

\usepackage{graphicx}
\usepackage[figuresright]{rotating}

\usepackage[shortcuts]{extdash}

\def\degr{\hbox{$^\circ$}}
\def\arcmin{\hbox{$^\prime$}}

\def\fdg{\hbox{$.\!\!^\circ$}}
\def\farcm{\hbox{$.\mkern-4mu^\prime$}}

\newcommand{\MAD}{{\sc mad}}
\newcommand{\med}{\mathrm{med}}
\newcommand{\sub}[1]{_\mathrm{#1}}
\newcommand{\hr}{^\mathrm{h}}

\newcommand{\D}[1]{{\rm d}#1}
\newcommand{\Dfrac}[2]{\frac{\D{#1}}{\D{#2}}}

\newcommand{\Tatm}{T\sub{atm}}
\newcommand{\Tsys}{T\sub{sys}}

\usepackage[usenames]{color}
\newcommand{\figlab}[1]{\fboxsep=2pt\fcolorbox{black}{white}{\scriptsize{#1}}}

\definecolor{linkcol}{rgb}{0.0,0.0,0.7}
\usepackage[colorlinks=true, allcolors=linkcol]{hyperref}

\title[1.4\,GHz Radio Continuum Map of $\delta < +25\degr$]
      {A New 1.4\,GHz Radio Continuum Map of the Sky South of Declination
       $+25\degr$}

\author[Calabretta, Staveley-Smith and Barnes]
  {Mark R. Calabretta\thanks{Email: mcalabre@atnf.csiro.au},\\
   \affil{CSIRO Astronomy and Space Science,
          PO Box 76, Epping, NSW 1710, Australia}
   Lister Staveley-Smith\\
   \affil{International Centre for Radio Astronomy Research,
          M468, University of Western Australia,
          35 Stirling Highway, Crawley, WA 6009, Australia}
   \affil{ARC Centre of Excellence for All-sky Astrophysics}
   \and David G. Barnes\\
   \affil{Monash e-Research Centre, Monash University, Clayton, Vic 3800}
   \affil{Clayton School of Information Technology, Monash University,
	  Clayton, Vic 3800}
  }

\begin{document}

%==================================================================== Abstract

% No more than 200 words.

\begin{abstract}
Archival data from the HI Parkes All-Sky Survey (HIPASS) and the HI Zone of
Avoidance (HIZOA) survey have been carefully reprocessed into a new 1.4\,GHz
continuum map of the sky south of $\delta~= +25\degr$.  The wide sky coverage,
high sensitivity of 40\,mK (limited by confusion), resolution of $14\farcm4$
(compared to $51\arcmin$ for the Haslam et al.\ 408\,MHz and $35\arcmin$ for
the Reich et al.\ 1.4\,GHz surveys), and low level of artefacts makes this map
ideal for numerous studies, including: merging into interferometer maps to
complete large-scale structures; decomposition of thermal and non-thermal
emission components from Galactic and extragalactic sources; and comparison of
emission regions with other frequencies.  The new map is available for
download.
\end{abstract}

%==================================================================== Keywords


\maketitle

%================================================================ Introduction

\section{INTRODUCTION}
\label{sec:intro}

Modern maps of radio continuum emission have tended towards increasing
sensitivity and angular resolution.  Increased sensitivity can be obtained
with larger collecting area, but increased angular resolution normally means
resorting to interferometers with widely-spaced elements.  Notable
high-resolution, large-area interferometer surveys at radio wavelengths
include SUMSS \cite{Bock}, FIRST \cite{Becker}, NVSS \cite{Condon}, and
AT20G \cite{Murphy}, which mapped the sky at frequencies from 843\,MHz to
20\,GHz.

However, interferometers are inherently insensitive to large-scale emission.
That is, emission on angular scales greater than $\lambda/B_s$, where $B_s$ is
the shortest projected baseline, is lost.  Whilst not an issue for discrete
sources, this a serious shortcoming for mapping large objects including:
cluster halos; the lobes of giant radio galaxies; nearby supernova remnants
and HII regions; and the diffuse thermal and non-thermal emission from the
Milky Way and other nearby galaxies.  Sensitive single-dish maps of the sky
therefore remain useful for filling in the ``missing information'', and can
often be directly merged with interferometer maps of similar frequency and
with overlapping coverage in the $u$-$v$ domain.

Single-dish, single-beam millimetre wavelength continuum studies by the COBE,
WMAP and now Planck space missions have been spectacularly successful in
mapping the Cosmic Microwave Background.  Complemented by lower and higher
frequency maps, to separate out thermal and non-thermal foregrounds, analysis
of such maps have fundamentally improved the measurement accuracy of important
cosmological parameters, e.g.\ Spergel et al.\ \shortcite{Spergel}.

Recently, there has been renewed interest in low-frequency maps of the sky.
This interest arises from the desire to: (a) constrain possible emission from
spinning dust and large dust grains; (b) provide a low-resolution complement
to all-sky maps from the Square Kilometre Array (SKA) pathfinders such as the
Murchison Widefield Array (MWA) and LOFAR; (c) provide missing polarisation
information for characterising foreground emission which contaminates EOR and
CMB B-mode studies.  The well-known map of Haslam et al.\ \shortcite{Haslam}
has stood out for many years as a low-frequency (408\,MHz) reference map.
However, it has fairly low resolution ($0\fdg85$), contains a number of
artefacts, and has no polarisation information.  It was a precursor to later
studies including those at 1.42\,GHz by Reich \shortcite{Reich1}, Reich \&
Reich \shortcite{Reich2}, Reich, Testori, \& Reich \shortcite{Reich3} and the
2.3\,GHz S-PASS survey by Carretti \shortcite{Carretti}.  Unfortunately, these
higher resolution studies using single-beam receivers are expensive in
telescope time -- S-PASS took around 2000\,h of observing time on the Parkes
telescope -- and are therefore not straightforward to conduct.

Previous surveys that overlap the HIPASS/ZOA coverage at frequences close to
1.4\,GHz and with resolution comparable to the $14\farcm4$ HPBW are: the
all-sky survey at 408\,MHz and $51\arcmin$ by Haslam et al.\
\shortcite{Haslam}; the 1.42\,GHz surveys of Reich \shortcite{Reich1} with
$+20\degr < \delta$, Reich \& Reich \shortcite{Reich2} with $-19\degr \leq
\delta \leq +24\degr$, and Reich, Testori, \& Reich \shortcite{Reich3} with
$\delta < -10\degr$ which between them cover the whole sky with HPBW
$35\arcmin$; the complementary polarisation surveys at 1.42\,GHz of Wolleben
et al.\ \shortcite{Wolleben} with $-29\degr < \delta$ and Testori, Reich, \&
Reich \shortcite{Testori} with $\delta < -10\degr$, each with HPBW
$36\arcmin$; and the 2.3\,GHz survey of Jonas, Baart, \& Nicolson
\shortcite{Jonas} covering most of the range $-83\degr < \delta < +32\degr$
with HPBW $20\arcmin$.  Wielebinski \shortcite{Wielebinski} provides an
historical overview of these surveys.

The 408\,MHz survey in particular has proved important for foreground
subtraction of the {\it WMAP} CMB surveys at 23, 33, 41, 61, and 94\,GHz
\cite{Hinshaw, Gold}, its $51\arcmin$ resolution being well-matched to the
23\,GHz data.  A resolution of $14\farcm4$ would be a better match to that of
the higher {\it WMAP} frequency bands, and also to those of the {\it Planck}
mission which aims to achieve a final map of the CMB anisotropies in the
vicinity of 5\,-- $10\arcmin$.

The HI Parkes All Sky Survey (HIPASS) and Zone of Avoidance (ZOA) surveys were
conducted using the 13-beam Parkes multibeam system starting on 1997 February
28, with HIPASS and its northern extension concluding on 2001 December 14, and
ZOA with its extension into the Galactic bulge continuing until 2005 August
08.  The raw data archive consists of 33,967 HIPASS and 15,305 ZOA scans, each
of 100 $\times$ 5\,s integrations, 13 beams and 2 linear polarisations for an
effective single-beam, single-polarisation equivalent total integration time
of 640\,Ms ($\sim$20 years).  The multibeam instrumentation, HIPASS survey
strategy, and data reduction software have been described by Barnes et al.\
\shortcite{Barnes}, and the ZOA survey strategy by Staveley-Smith et al.\
\shortcite{Staveley-Smith}.

Numerous papers based on this data set have been published, though to date
these have been concerned almost exclusively with analysis of the 21\,cm
neutral hydrogen line, e.g.\ Meyer et al.\ \shortcite{Meyer}; Zwaan et al.\
\shortcite{Zwaan}.  It has long been recognised that the survey data also
contains a wealth of information on the hydrogen recombination lines between
H166$\alpha$ and H168$\alpha$.  These are being studied in an ongoing series
of papers by Alves et al.\ \shortcite{Alves}.  It is also relatively simple to
measure continuum flux densities for compact sources and studies have shown
that it should be possible to measure spectral indices for the strongest of
them \cite{Melchiori}.  However, the task of producing large-scale continuum
maps is complicated by a number of factors, the nature of which, and the
solutions developed are the subject of this work.

In this paper, we introduce the surveys in Section~\ref{sec:surveys} and
extensively discuss all aspects of the the data reduction in
Section~\ref{sec:reduction}, including presentation of the final downloadable
maps and comparison with previous studies.  We briefly discuss the results in
Section~\ref{sec:discussion}.

%----------------------------------------------------------------- floating --

\begin{table}
  \caption[]{Summary of the survey parameters.  Beam parameters are given for
    the central beam and the inner and outer rings of six beams each.}
  \begin{center}
  \protect\begin{tabular}{lc}
    \hline\hline
    \noalign{\smallskip}
    \noalign{\smallskip}
    Sky coverage            & $\delta < 25\degr$ \\
    Beams                   & 13, hexagonal close-packed \\
    Beam FWHM               & $14\farcm0$, $14\farcm1$, $14\farcm5$ \\
    Beam ellipticity        & 0.0, 0.03, 0.06 \\
    Centre beam efficiency  & 63\% \\
    Polarisations           & 2 orthogonal linear \\
                            & (Stokes I only) \\
    Centre frequency        & 1394.5\,MHz \\
    Bandwidth               & 64\,MHz (in 1024 channels) \\
    HIPASS zone centres     & $\delta = -87\degr$, $-82\degr$ to $+22\degr$ \\
                            & in steps of $8\degr$ \\
    ZOA zoan centres        & $\ell = 200\degr$ to $48\degr$ \\
                            & in steps of $8\degr$ \\
    \parbox{30mm}{
    Scan length \\
    \\}                     & \parbox{28mm}{
                              $8\fdg5$ except \\
                              $4\fdg6$ in zone $-87\degr$, \\
                              $7\fdg5$ in zone $+22\degr$} \\
    Scan rate               & 1\arcmin\,s$^{-1}$ \\
    Integration time        & 5\,s \\
    Average system temp.    & 21\,K ($\approx$ 33\,Jy) at \\
                            & elevation $55\degr$ \\
    \noalign{\smallskip}
    \hline\hline
  \end{tabular}
  \end{center}
  \label{ta:obsparm}
\end{table}

%================================================================= The Surveys

\section{THE SURVEYS}
\label{sec:surveys}

For the purpose of this work the most important aspects of the survey
strategy, summarised in Table \ref{ta:obsparm}, are
\begin{itemize}
\item
  A central frequency of 1394.5\,MHz and bandwidth of 64\,MHz in 1024 spectral
  channels.

\item
  The scan rate in each survey was 1\arcmin\,s$^{-1}$ with 5\,s
  integrations.  The scan rate was used as part of the data validation as
  discussed in Section \ref{sec:validation}.

\item
  HIPASS scanned in declination in 15 zones centred on declinations
  $-87\degr$, and $-82\degr$ to $+22\degr$ in steps of $8\degr$.  The scan
  length was $8\fdg5$ except for the shorter $4\fdg6$ scans in the $-87\degr$
  zone, and $7\fdg5$ in the $+22\degr$ zone.  The overlap between zones is
  significant for this work and is discussed in Section \ref{sec:zonelevs}.

  \begin{sloppypar}
  As the Parkes telescope is based on a master equatorial drive system, scans
  in the $-87\degr$ zone stopped just short of the south celestial pole which
  consequently is surrounded by a small dead zone.
  \end{sloppypar}

\item
  The ZOA survey scanned in Galactic longitude in 27 zones centred on
  longitudes $200\degr$ to $48\degr$ in steps of $8\degr$.  The scan length
  was again $8\fdg5$.  Galactic latitudes between $\pm5\degr$ were covered.
  In zones $336\degr$ to $32\degr$ the coverage was extended to $\pm10\degr$,
  and then to $+15\degr$ to cover the Galactic bulge in zones $352\degr$ to
  $16\degr$.  As the survey was designed primarily for detecting HI line
  emission, scanning was done in longitude rather than latitude in order to
  minimise the variation in Galactic continuum emission at 21\,cm.  By a quirk
  of geometry, this was a fortuitous choice for this work as well, as
  discussed in Section \ref{sec:zonelevs}.

\item
  \begin{sloppypar}
  Additional to HIPASS and ZOA, deep surveys were conducted of the Centaurus
  and Sculptor clusters of galaxies -- HIDEEP \cite{Minchin}.  The Centaurus
  survey comprised 623 scans centred on declination $-30\degr$ (excluding
  Cen\,A), so straddling the HIPASS $-34\degr$ and $-26\degr$ zones, while the
  439 Sculptor scans lay entirely within the $-34\degr$ zone.  These
  additional observations were used selectively as described.
  \end{sloppypar}

\item
  In each survey, the 13-beam multibeam feed array was rotated by $15\degr$
  with respect to the scanning direction so that the beams were almost
  uniformly spaced on the sky, thus optimising coverage in a single scan.
  At the midpoint of a scan, the spacing between adjacent beams on the sky was
  then close to an integer multiple of $7\arcmin$, just below the Nyquist rate
  of $5\farcm7$ viz $\{0$, $\pm8\arcmin$, $\pm13\arcmin$, $\pm21\arcmin$,
  $\pm28\arcmin$, $\pm36\arcmin$, $\pm49\arcmin\}$.

\item
  \begin{sloppypar}
  The feed array was not rotated to account for parallactic effects once a
  scan had started, though parallactification was later found to have been
  enabled inadvertently for about 1\% of the HIPASS scans.
  \end{sloppypar}

  In the completed HIPASS survey, the parallactic rotation in the course of a
  scan was found to be $> 5\degr$ for 64\% of the scans, $> 10\degr$ for 29\%,
  and $> 15\degr$ for 9\%.  The corresponding figures for ZOA are 72\%, 40\%,
  and 22\%.  The effect this had on the mapping is discussed in Section
  \ref{sec:gridding}.

\item
  Adjacent HIPASS scans in the same zone are stepped by $7\arcmin$ orthogonal
  to the scan direction (i.e.\ in right ascension) so that {\em each} of the
  13 beams mapped the sky at slightly below the Nyquist rate, whereby the
  survey is approximately $\times 10$ oversampled spatially.  However, because
  the step was close to the spacing between beams in an individual scan, the
  beam tracks in adjacent scans tended to overlay one another rather than fill
  in the spaces between them.  The effect of this is discussed in Section
  \ref{sec:gridding}.

\item
  The scans in each HIPASS zone were organised into five interleaved sets
  (labelled $a$\,-- $d$) with adjacent scans in each set stepped by
  $35\arcmin$.  Each interleave maps the sky at approximately twice the
  Nyquist rate and was completed before starting on the next.  This ensured
  that scans for a particular part of the sky were observed at widely
  separated times, thus minimising the possible impact of Radio Frequency
  Interference (RFI), the Sun, Moon, and planets.

\item
  Adjacent ZOA scans are stepped by $1\farcm4$ in Galactic latitude in 25
  interleaved sets ($a$\,-- $y$), i.e.\ $\times 5$ denser than HIPASS.

\item
  During observations, the system temperature was calibrated against a
  high-quality noise diode switched in and out of the signal path.  The diode
  itself was calibrated periodically against a flux density calibrator,
  typically 1934-638 (14.9\,Jy at 1420\,MHz) or Hydra A (40.6\,Jy at
  1395\,MHz, after correction for dilution in the beam of the telescope).

\item
  The average system temperature at elevation $55\degr$ was $21\degr$K.
  This rises towards higher elevations because of spillover effects, and
  likewise towards lower elevations with the additional contribution from
  atmospheric opacity becoming significant.  Refer to Section
  \ref{sec:elevation}.
\end{itemize}

%========================================================== The data reduction

\section{THE DATA REDUCTION}
\label{sec:reduction}

There were two main difficulties in extracting a 1.4\,GHz continuum map
of the sky from the 21\,cm HIPASS data.  The first was in correcting for
the elevation dependence of $\Tsys$.  A correction was derived from the HIPASS
data itself in regions well away from Galactic emission as discussed in
Section \ref{sec:elevation}.

The second difficulty was that the zero-level of each $8\degr$ HIPASS
declination scan could only be determined by the narrow overlap between
adjacent zones, with an arbitrary zero-level for the sky as a whole.  This is
discussed in Section \ref{sec:zonelevs}.

Firstly, however, there is the relatively simple but important step of
removing invalid or questionable data from the HIPASS/ZOA data set.

%============================================================= Data validation

\subsection{Data Validation}
\label{sec:validation}

Whereas spectral line processing concerns the signal response in a few
channels of a spectrum with a baseline defined by many, continuum processing
is concerned with the many channels that define that baseline itself.  Thus
while spectral line processing is less sensitive to external variations in
$\Tsys$, continuum processing is more sensitive and involves a certain element
of {\em dead-reckoning}.  For this reason extra care was taken to remove
invalid data from the survey datasets.

An important test for validating HIPASS/ZOA data arises in connection with
the way the Parkes telescope control system measures position in scanning
observations.  Positions are recorded at 5\,s intervals with a separate
timestamp which generally does not match the midpoint of the integration.
Consequently, {\em livedata}\footnote{
\url{www.atnf.csiro.au/computing/software/}}, the software that reads and
processes the data, must buffer it and interpolate the positions to match the
integration timestamp.

This position interpolation is based on a uniform scan rate, and as this is
known beforehand to be $1\arcmin$\,s$^{-1}$, it provides a useful check on
the positional accuracy.

However, in validating the scan rate, complications arise for the outer beams
in a multibeam system due to grid convergence effects, particularly near the
south celestial pole.  As the most extreme case, with the central beam at the
pole and moving north, the beams on either side will have a non-zero motion in
RA, either increasing or decreasing, with the trailing beams scanning towards
the pole, i.e.\ in {\em decreasing} declination!  Furthermore, because the
feed array was not (usually) rotated to account for parallactic effects once a
scan had started, a parallactic component of the scan rate arises for the
outer beams, particularly for scans close to the zenith with a large change in
azimuth.

Once these effects were accounted for, there was generally excellent agreement
between the measured and nominal rates for all 13 beams within an allowed
range of $\pm 0\farcm05$\,s$^{-1}$ except for data that was genuinely
discrepant.  Of the latter a number of problems were readily identified:
\begin{itemize}
\item
  The first and last few integrations of a scan often reported discrepant scan
  rates.  Often these were flagged as bad by the system but not always.  For
  some scans it was apparent that the feed assembly had not quite finished
  rotating into position before the scan started.

  It is important for this work that these integrations be culled; the overlap
  between zones used for level-adjustment in Section \ref{sec:zonelevs} is
  normally limited to 11 integrations at most (23 for the $-87\degr$ zone), so
  two bad integrations could affect the results.

\item
  Occasional (and rare) mid-scan glitches, apparently related to movement of
  the feed assembly or possibly a glitch in the feed rotation encoder or
  control system.  Either way, the effect is to produce a questionable
  position which thus necessitated culling of the affected integrations.

\item
  A small number of scans were made with the wrong scan rate.
\end{itemize}
Detection of a bad scan rate resulted in all 13 beams $\times$ 2 polarisations
being culled for the particular integration.  The total number of integrations
amounted to the equivalent of about 39 full scans but with bad integrations
concentrated in the sensitive beginning and end integrations as discussed
above.

The scan rate test also uncovered a number of correctable problems, some of
which originated from causes that were understood, and others for which the
cause could be reliably inferred.  These data were corrected on-the-fly by
{\em livedata}.  This affected the equivalent of about 23 full scans.

Finally, in the process of validating and subsequently processing the data,
an additional 227 complete scans were uncovered that needed to be culled for a
variety of reasons, not necessarily relating to position or scan rate errors.
For example, observations made with the wrong centre frequency or bandwidth.
With the removal of these bad observations, 33,967 HIPASS and 15,305 ZOA scans
remained.

After the data had been completely processed, it became apparent that the
results were affected by sources that were strong enough to saturate the
receiver electronics.  This was particularly so for the classical radio
sources such as Sgr\,A, Tau\,A, Ori\,A, Cen\,A, etc.  In HIPASS/ZOA spectral
processing, integrations were culled by the reader if all beams and
polarisations were flagged because of saturation.  However, partially-flagged
integrations were allowed to pass through the bandpass calibration phase and
only culled when the data was gridded into spectral cubes.  Again, a more
rigorous treatment was required for continuum processing, with the bandpass
calibration phase modified to handle partially-flagged data properly.

%----------------------------------------------------------------- floating --

% /DATA/GRUS_1/CHIPASS/poly2fit/CHIPASS+Cen+Scl+ZOA_GAL/-
%   CHIPASS+Cen+Scl+CZOA_GAL_WGTMED1.continuum.fits
%
% kvis levels 0 to 54.5 (1.09*50), logarithmic 3-cycle, Greyscale1.
\begin{figure*}
  \centering
  \includegraphics[width=460pt]{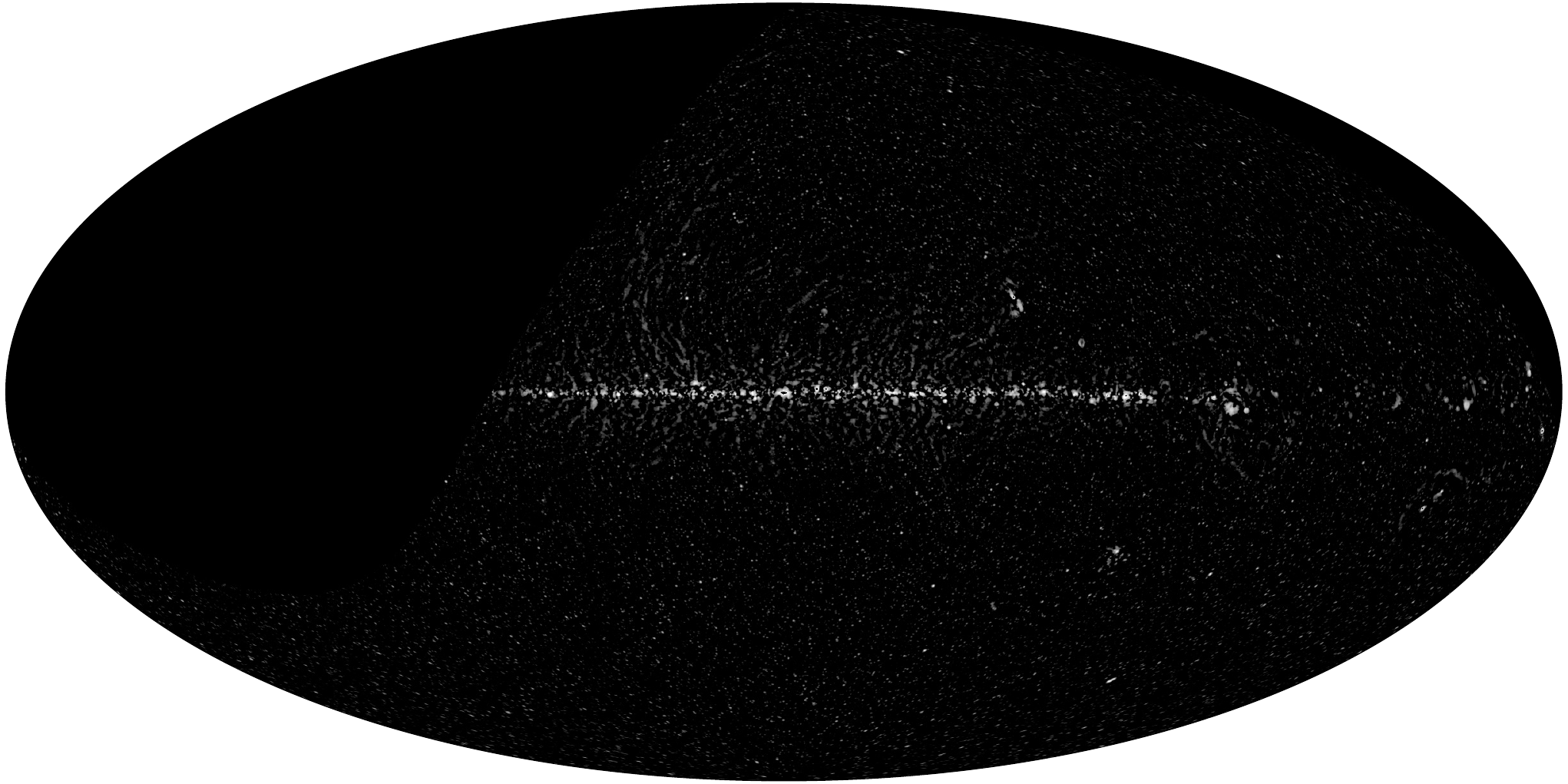}
  \caption[]{HIPASS ``point-source'' continuum map at 1.4\,GHz produced using
    the HIPASS/ZOA compact source algorithm which is sensitive only to regions
    of emission much less than $8\degr$ in extent.  The high-pass spatial
    filter also produces pronounced negatives on either side of the Galactic
    plane.  This map and those following are on a Hammer-Aitoff projection in
    Galactic coordinates centred on $(\ell,b) = (0\degr,0\degr)$ at which
    point the pixel spacing is $4\arcmin$ in either direction.  Longitude
    increases towards the left as usual for a celestial map.  A three-cycle
    logarithmic base-10 intensity scale is used for this and the following
    images, the greyscale range is 0.05--50\,Jy/beam (nominal calibration).}
  \label{fig:Compact}
\end{figure*}

%----------------------------------------------------------------- floating --

% /DATA/GRUS_1/CHIPASS/median/CHIPASS_GAL/-
%   CHIPASS_GAL_WGTMED.continuum.fits
%
% kvis levels 0 to 50, logarithmic 3-cycle, Greyscale1.
\begin{figure*}
  \centering
  \includegraphics[width=460pt]{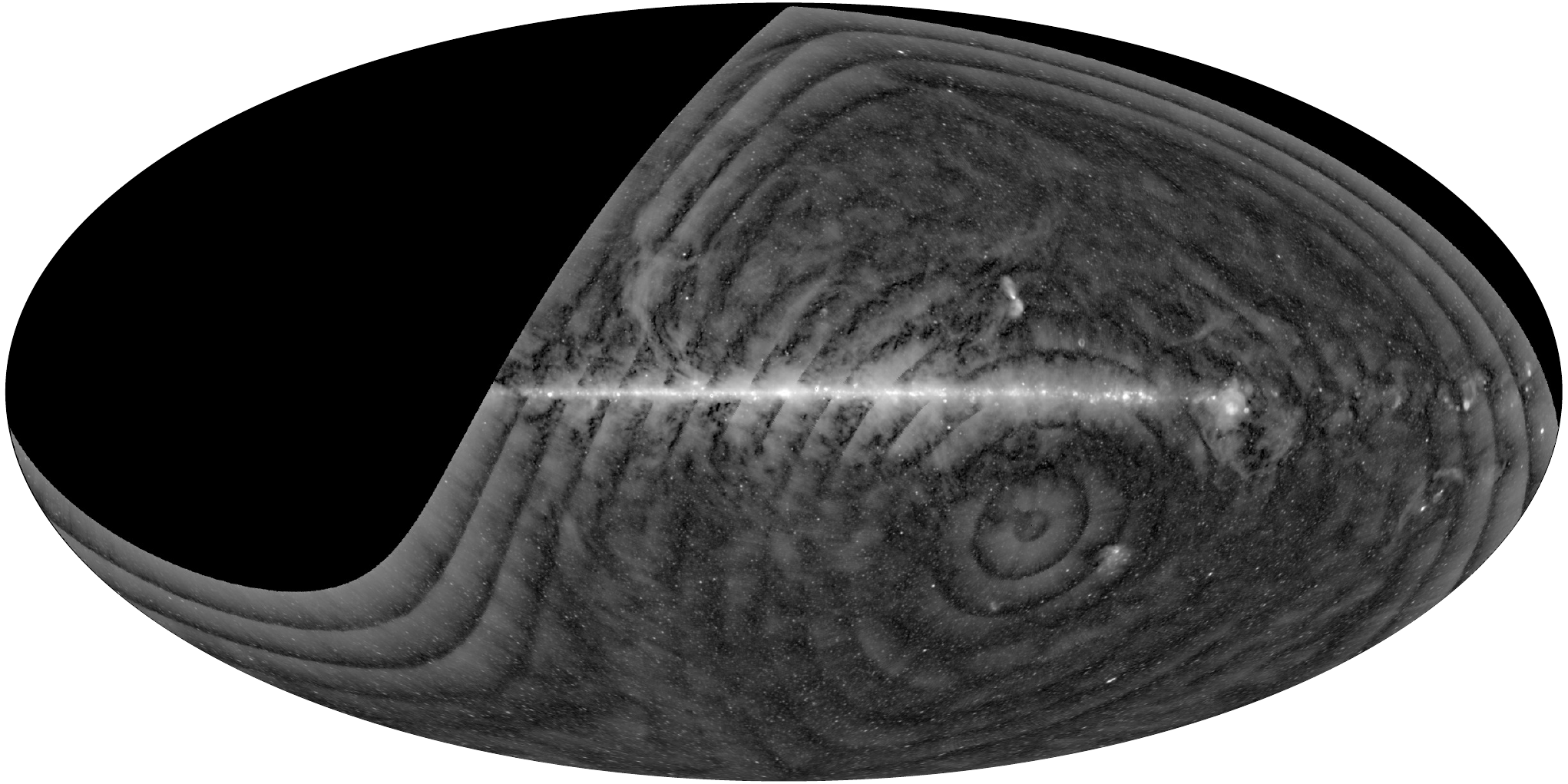}
  \caption[]{HIPASS 1.4\,GHz continuum data now processed with the
    running-median algorithm which is sensitive to extended emission.  The 15
    HIPASS declination zones are readily apparent.  This map illustrates the
    two effects requiring calibration: the elevation dependence of $\Tsys$
    discussed in Section \ref{sec:elevation} which produces the gradual
    brightening away from the zenith, and the zone level-adjustment discussed
    in Section \ref{sec:zonelevs} which delineates the 15 declination zones.
    Same logarithmic greyscale as Figure \ref{fig:Compact}.}
  \label{fig:Extended}
\end{figure*}

%======================================================== Bandpass calibration

\subsection{Bandpass calibration}
\label{sec:bandpass}

The bandpass calibration algorithm described by Barnes et al.\
\shortcite{Barnes} was modified for continuum processing in several key
respects.
\begin{itemize}
\item
  In representational terms, HIPASS/ZOA bandpass calibration was based on the
  following equation which is applied in turn to each spectral channel:
  \begin{equation}
    S'_i = S_i B({\Tsys}_i) / B(S_i) - {\Tsys}_i .
    \label{eq:oldbp}
  \end{equation}
  Here $S_i$ and $S'_i$ are the raw and calibrated values in the spectral
  channel for integration $i$, and $B()$, estimates the baseline response for
  a quantity that varies with $i$ (because of scanning), whether ${\Tsys}_i$
  or $S_i$.  In essence, $B()$ determines the baseline in source-free
  (reference) regions of the scan and interpolates into the integrations
  occupied by sources.  In HIPASS/ZOA it was implemented as a running-median
  filter.

  Equation (\ref{eq:oldbp}) is applied in turn to each of the 1024 channels,
  separately for each of the 13 beams $\times$ 2 polarisations.

  Besides receiver noise, ${\Tsys}_i$ contains the signal from cosmic
  continuum emission.  By subtracting it, the continuum is thereby discarded.
  Instead we use
  \begin{equation}
    S'_i = S_i / B(S_i / {\Tsys}_i) - B({\Tsys}_i)
    \label{eq:newbp}
  \end{equation}
  Here we have only subtracted the baselevel value of ${\Tsys}_i$ which should
  consist of receiver noise only, thus preserving the cosmic signal of
  interest.

  Note in Equation (\ref{eq:newbp}) that each spectrum is first divided by
  $\Tsys$ as a prelude to bandpass calibration in order to avoid statistical
  biasses that may be introduced by integrations affected by strong continuum
  emission.  That is, the scale factor is
  $1/B(S_i / {\Tsys}_i)$, not $B({\Tsys}_i) / B(S_i)$.

  Note also that the factor is not $B({\Tsys}_i / S_i)$ as can be understood
  simply by considering that integrations could conceivably have $S_i = 0$ in
  channels close to the edge of the band, whereas ${\Tsys}_i > 0$ always.

\item
  21\,cm continuum emission extends over many degrees, often exceeding the
  $8\degr$ scan length.  However, as discussed by Putman et al.\
  \shortcite{Putman}, the HIPASS/ZOA compact source algorithm is designed for
  extragalactic objects of typically $< 1\degr$ in extent and is blind to
  regions of emission much larger than this.

  Figure\footnote{High-resolution images in this paper are suitable for
  ``zooming'' in PDF viewers.} \ref{fig:Compact} shows the result of applying
  a compact source algorithm to continuum data.  The algorithm differed from
  the running median used by HIPASS/ZOA, instead utilising a robust polynomial
  fitting technique as described in McClure-Griffiths et al.\
  \shortcite{GASS}.  Most of the extended emission is lost, leaving behind
  only point sources and ridges.  Such maps are useful in their own right for
  producing point-source catalogues.

  For their work on the Magellanic stream and high-velocity clouds, which have
  scales extending well beyond $1\degr$ in size, Putman et al.
  \shortcite{Putman} used the {\sc minmed5} algorithm for function $B()$ in
  Equation (\ref{eq:oldbp}).  For each spectral channel, the scan is divided
  into five fixed segments $1\fdg6$ in length and the median of the $\sim 20$
  integrations computed for each.  The segment with the lowest response of the
  five is then used as the zero-level for the whole scan.

  This work uses a refinement of {\sc minmed5}.  Instead of five fixed
  segments, the running median in a box of chosen size is computed for the
  whole scan.  Ideally the box should be small, but not so small as to be
  affected by the intrinsic noise associated with extremal statistics.  In
  practice, a box of size 10 integrations, being half that of {\sc minmed5},
  was found to be an acceptable compromise.  This means that we should expect
  a minimum of 5 negative integrations per channel, rather than 10, out of the
  $~100$ integrations in a bandpass-calibrated scan.

  This {\em running-median} technique provides the bandpass calibration
  function $B()$ in Equation (\ref{eq:newbp}).

\item
  In the post-bandpass calibration phase, a robust polynomial fit of degree
  four (quartic) is done over frequency for each spectrum in a scan.  The
  robust polynomial fitting algorithm used by {\em livedata} has been
  described by McClure-Griffiths et al.\ \shortcite{GASS}.  In this instance
  an initial mask 41 channels wide was set to exclude Galactic HI emission.

  The continuum response is defined by the zeroth and first order coefficients
  of the best-fit polynomial, the latter providing a measure of the spectral
  index; the continuum flux density varies by $\sim 10\%$ across the 64\,MHz
  band at 1394.5\,MHz for a source with $\alpha = \pm 2$.
\end{itemize}

The map shown in Figure \ref{fig:Extended} is the result of applying the
running-median bandpass calibration technique.  It illustrates the two effects
requiring calibration.  Firstly the elevation dependence of $\Tsys$ which
manifests itself as an apparent increase in continuum emission towards lower
elevations, here the celestial equator and south celestial pole.  This is
discussed in Section \ref{sec:elevation}.  Secondly the discontinuity along
the edges of the $8\degr$ declination zones.  Zone level-adjustment, required
to eliminate this, is discussed in Section \ref{sec:zonelevs}.

%----------------------------------------------------------------- floating --

% Generated using program chipass_tracks.
\begin{figure*}
  \centering
  \includegraphics[width=460pt]{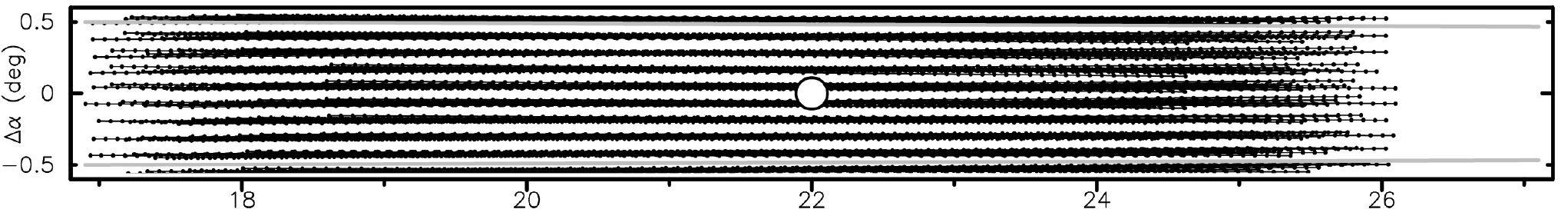}
  \includegraphics[width=460pt]{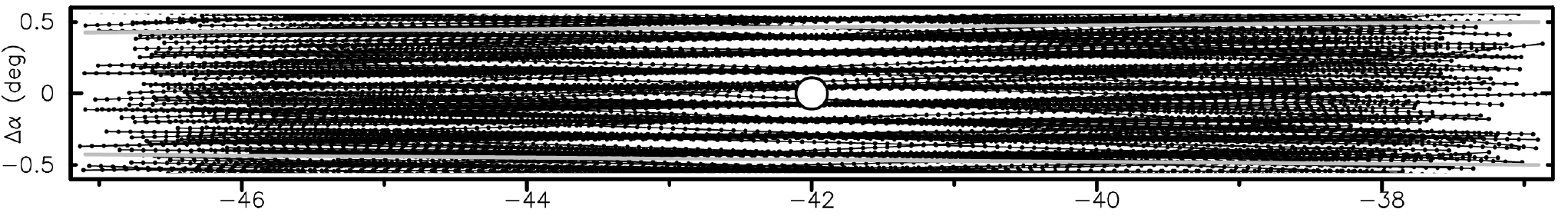}
  \includegraphics[width=460pt]{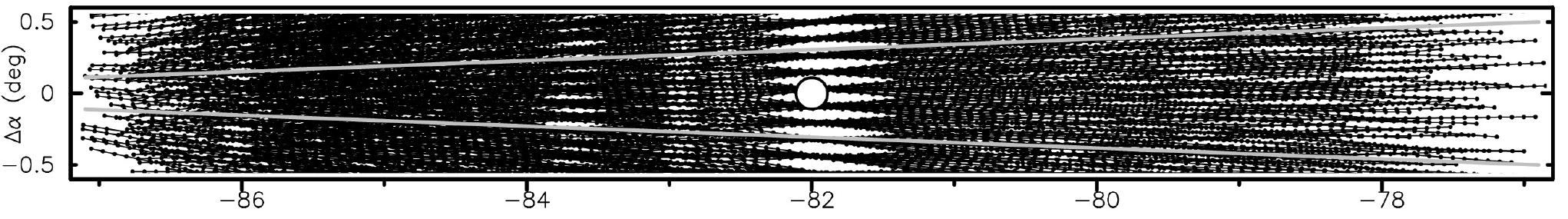}
  \caption[]{HIPASS beam tracks within $\pm 0\fdg5$ of arc of RA$ = 12\hr$
    in the $+22\degr$, $-42\degr$, and $-82\degr$ declination zones.  The
    effect of parallactic rotation of the outer beams is evident in this
    equi-scaled, equiareal Sanson-Flamsteed projection.  The ``windows'' near
    the centre of the scan arise because the rotation angle of the multibeam
    feed assembly was tuned for this point.  Superposed on the tracks, the
    $14\farcm4$ HPBW beamwidth is represented as the outer diameter of the
    central black circle whose inner diameter corresponds to the $6\arcmin$
    radius cut-off in the gridding kernel used in HIPASS spectral line
    processing.  A pair of RA meridians is superposed in grey to indicate grid
    convergence towards the poles, the central beam follows such a track.
    Dots on the tracks indicate sample points.}
  \label{fig:tracks}
\end{figure*}

%==================================================================== Gridding

\subsection{Gridding}
\label{sec:gridding}

While map production is normally the last step in data processing, this work
relied on various forms of gridding at all stages of the reduction.  Two
intermediate maps have already been shown in Figures \ref{fig:Compact} and
\ref{fig:Extended}.  Therefore it is appropriate to discuss gridding at this
point.

The large number of gridding techniques in common use -- {\em kriging}, {\em
minimum curvature}, {\em Delaunay triangulation}, {\em bi-directional line},
{\em thin plate spline}, {\em inverse distance weighted}, etc.\ -- suggests
that there is no single general method that suits all purposes.  Many of these
gridding algorithms are designed for sparse, well-determined measurements,
such as pertain in the geospatial fields, where the problem is to interpolate
from an irregular to a regular grid.  On the contrary, we are faced with a
large, over-determined data set, which however, is subject to measurement
error and may have severe outliers.   Use of a robust statistical gridding
technique is therefore indicated.

Barnes et al.\ \shortcite{Barnes} describe the robust gridding algorithm used
to produce the final HIPASS/ZOA spectral cubes as implemented by
{\em gridzilla}\footnote{Provided in the same package as {\em livedata}.}.
For this work, adaptations were required to handle very wide fields of view
correctly, to implement the FITS celestial and spectral world coordinate
systems \cite{WCS1, WCS2, WCS3}, and to deal with the much larger amount of
input data and output image size.

Another significant change in this work was in the gridding statistic used.
As the beam normalisation applied by Equation (5) of Barnes et al.\
\shortcite{Barnes} is not appropriate for extended emission, their Equation
(8) was replaced by a weighted median estimator which is equally robust
against RFI and other time-variable sources of emission such as the Sun, Moon,
and planets.

The weighted median is the {\em middle-weight value} - the sum of weights of
all measurements less than it being equal to that of all measurements greater.
Pro rata interpolation is used to bisect the sum of weights if required.  This
statistic is a generalisation of the median in the same way that the weighted
mean generalises the mean.  To see this, first consider the case where all
weights, $w_i$, are integral.  Computation of the weighted mean and median is
then equivalent to replacing each measurement $x_i$ with weight $w_i$ by $w_i$
measurements of value $x_i$ and computing the unweighted mean or median in the
usual way.  For non-integral weights, each weight may be replaced by an
integral approximation $\lfloor w_i 10^k \rfloor$ and the limit taken as
$k \rightarrow \infty$.

As previously described, HIPASS scanned in declination alone, in general there
are no orthogonal scans with which to form a ``basket-weave'' \cite{Sieber,
Emerson}.  Normally one would expect to see noticable striping in maps
produced from such data.  However, two factors mitigate against this in
HIPASS: the $\times 10$ oversamping above the Nyquist rate, and parallactic
rotation of the outer beams.  The latter means that the outer beams do not
actually scan in declination but instead follow a path that either tends to
converge on the central beam or diverge from it.  Consequently, when the
tracks from all scans that contribute to a particular patch of the sky are
mapped, the result tends to be a jumble of obliquely intersecting lines rather
than a parallel array, as shown in Figure \ref{fig:tracks}.

As seen in Figure \ref{fig:tracks}, parallactic rotation in the $+22\degr$
zone is minimal and increases progressively towards the southern zones.  The
$+22\degr$ zone is just above the elevation limit of the Parkes dish, in fact
the last degree or so is below the limit and the scans are truncated.
Consequently, these northerly scans are performed mainly in elevation on the
local meridian, the change in azimuth being limited to that required to follow
diurnal rotation.  Since the parallactic angle generally changes more quickly
with azimuth than elevation, parallactic rotation of the outer beams is
thereby minimised.

Because the step size of $7\arcmin$ between scans matches the basic separation
between the beams in an individual scan, the beam tracks of successive scans
tend to overlay one another.  This creates a raster pattern that is more
pronounced closer to the equator and near the midpoint of the scans, but tends
to be obliterated by parallactic rotation in the more southerly zones as is
shown in Figure \ref{fig:tracks}.  The raster was of little consequence for
HIPASS spectral line processing which was concerned with compact sources, and
beam normalisation probably also helped to obscure it.  In fact, it might be
considered advantageous to have all 13 beams contribute equally to each point
in the map.  However, the raster does manifest itself, albeit at a low level,
in various parts of the continuum map because of the extended scale of the
emission.

%========================================================== Iterative gridding

\subsubsection{Iterative gridding}
\label{sec:iterative}

Weighted median gridding is a non-linear process.  It has the desirable
property of robustness against outliers such as might arise from RFI, the Sun,
Moon, or satellites.  With the parameters typically used, it also tends to
produce a gridded beam that is very close to Gaussian with FWHM that is
remarkably insensitive to the cut-off radius, i.e.\ the radius of the circular
``catchment area'' surrounding each pixel in the map.  For example, in
simulations using a circular Gaussian of FWHM $14\farcm40$ and a cut-off
radius of $6\arcmin$, the gridded FWHM was measured at only $14\farcm45$,
increasing to only $14\farcm90$ for a cut-off of $12\arcmin$.

However, as explained in Barnes et al.\ \shortcite{Barnes}, one adverse effect
of median gridding is that the flux density scale differs depending on the
source size.  Without beam normalisation, compact sources appear in the map
with reduced peak height, while extended sources appear at their correct peak
height.  While this is also true of mean gridding, in that case the volume is
preserved via broadening of the gridded beam, whereas with median gridding it
is not because the gridded beam is not significantly broadened.  The magnitude
of the effect on compact source peak height depends on the gridding
parameters, particularly the cut-off radius.

Unless otherwise stated, the simulations referred to below were done using a
top-hat kernel, natural (unity) beam weighting, and without beam
normalisation.  For HIPASS data, with cut-off~=~$6\arcmin$, point sources
appear with peak height of about 86\% their true value.  With a $12\arcmin$
cut-off radius this quickly degrades to 61\%.  While compact sources can be
corrected by beam normalisation (as used in HIPASS), this inflates, and could
introduce other adverse effects for extended sources.  On the other hand, if a
map consists only of extended sources then there is no problem.  However, the
all-sky continuum map is composed of extended and compact emission of equal
import.

Iteration provides a solution to this problem for maps that contain a mix of
compact and extended sources.  Initially the zeroth-iteration map (henceforth
{\em 0-iter}) is produced as normal.  In the first iteration, the 0-iter map
is subtracted from the raw data which is then gridded to produce a residual
map.  Compact and slightly extended sources, which were underestimated in the
0-iter map, will be recovered, at least partially, in the residual map which
is then added to the 0-iter map to produce the first-iteration map {(\em
1-iter)}.  Further iterations may be performed in like vein.

As the raw data serves as the standard against which the map in each iteration
is compared, {\em gridzilla} never modifies it after reading it into memory.
There is also too much of it to make a full working copy.  Instead, each datum
has the map value from the previous iteration subtracted only at the point
that it is used, and as most data contributes to several pixels this happens
multiple times for each datum.  Iterative gridding thus requires a fast and
accurate interpolation method for a practical implementation.  {\em gridzilla}
uses barycentric bivariate parabolic interpolation \cite{Rodriguez} on the
nine nearest pixels\footnote{The generalisation to 2D is actually described in
the {\sc traceo} manual (\url{www.siplab.fct.ualg.pt}).}.  In its general form
  \begin{eqnarray}
    f(x,y) & = & \sum_{i=1}^{3}\sum_{j=1}^{3} f(x_i,y_j)
                 \left[ \frac{p_i(x) p_j(y)}{p_i(x_i) p_j(y_j)} \right],
                 \nonumber \\
    p_i(x) & = & \prod_{k=1,k \neq i}^{3} (x - x_k),
    \label{eq:interp}
  \end{eqnarray}
where $f(x,y)$ is the map value at $(x,y)$ and $(x_i,y_j)$ are the surrounding
pixel coordinates.  In effect, this fits a quartic polynomial with highest
power $x^2y^2$ to the nine pixel values and evaluates it at $(x,y)$.  Although
the polynomial is a quartic, the interpolation is legitimately described as
{\em parabolic} as each cross-section in $x$ and each in $y$ is a quadratic.
In practice, the term in brackets can be simplified somewhat because $x_1$,
$x_2$, and $x_3$ are equispaced, and likewise $y_1$, $y_2$, and $y_3$.
Barycentric linear interpolation on the four surrounding pixels is used on the
map edges or if any of the nine pixels is blanked.

Iteration may also be used productively for mean gridding where it acts to
reduce the gridded beam size.  Effectively, by driving the residual map
towards zero, it makes the gridded map look more like the data in the sense
that if the raw data were plotted on the map they would sit close to the
surface defined by it.

Iteration in {\em gridzilla} is performed after the data has been indexed and
read in and so adds only a relatively small overhead in processing time.  It
can also be sped up by applying a gain factor to the residual map before
adding it.  This factor compensates for the fact that point sources will
appear in the residual map still with underestimated peak height.  If the
factor used is too high then when the corrected map is subtracted from the raw
data it might drive some of it negative.  However, negatives appearing in the
residual map on the next iteration would then tend to correct this.

However, it is important to chose the gain factor appropriately to minimize
the number of iterations, preferably to just one, because the baselevel noise
increases steadily with each successive iteration.  This can be understood by
considering a map produced from data that consists solely of noise.  Because
of the averaging associated with statistical gridding and the fact that the
noise in the data is spatially uncorrelated, the rms noise in the 0-iter map
will be much less than in the data (about 10\% for a $6\arcmin$ cut-off).
Therefore, subtracting the 0-iter map from the data will change it only
weakly, whence the residual map will be highly correlated with the 0-iter map,
i.e.\ the the noise pattern will be similar.  Thus when the residual is added
to the 0-iter map the noise tends to be significantly worse than the sum in
quadrature expected for uncorrelated noise.  In simulations performed with a
$6\arcmin$ cut-off, the rms noise in the 1-iter map was 50\% higher than the
0-iter map for both mean and median.  For the next two iterations the rms
increased by roughly 20\% per iteration.

However, because the residual map is only added to correct the profile of
sources in the map, it only makes sense to add those parts of the residual map
that do in fact achieve this.  The rest is set to zero via a censoring
algorithm.  The residual values for a pixel and its eight neighbours are
summed, giving half-weight to those in the corners.  If this sum is below
$\times$1.5 the map rms the value is zeroed, i.e.\ the 9 pixels must be a
little above twice the rms on average.  This criterion helps to exclude
isolated noise spikes, and to retain weak residuals in the vicinity of strong
ones, i.e.\ on the edge of sources.  While this censoring ameliorates the
adverse noise behaviour of iterative gridding on the background regions, it
does tend to complicate the noise characteristics of the map as a whole.

Considering the unfavourable noise behaviour of iterative gridding, it is
indeed fortunate that only a single iteration is required to achieve
high-precision results for typical gridding parameters.  If point sources
appear in the 0-iter map at ratio $g$ of their true peak height, then they
would be expected to appear in the 1-iter map at ratio $g + g(1-g)f$, where
$f$ is the loop gain factor.  A factor $f = 1.16$ is thereby indicated for
$g = 86\%$ (as above).  However, this factor only applies to point sources --
slightly extended sources would be overestimated.  Sources that appear in the
0-iter map at ratio $h$ ($> g$) will appear in the 1-iter map at
$r = h + h(1-h)f$.  The worst case occurs for $h = (1+f)/2f$.  With $f = 1.16$
here, $h = 0.93$, whence $r = 1.006$, i.e.\ a precision of 0.6\%.

In simulations using HIPASS data with a $6\arcmin$ cut-off, the integrated
flux density and peak height of Gaussian test sources of various sizes were
indeed recovered to within 1\% with a single iteration using a gain of 1.2 for
both the median and mean.  In fact, acceptable results, to within about 3\%,
were obtained with gain factors in the range 1.1 to 1.3.  The implied HPBW,
i.e.\ $\sqrt(F^2 - T^2)$ where $F$ is the measured FWHM and $T = 14\farcm4$ is
the FWHM of the test source, was $14\farcm3$ for median gridding (indicating a
slight tendency to super-resolve), and at $14\farcm6$ was only slightly
greater than it for mean gridding.

Minor edge effects are associated with iterative gridding because spectra that
contribute to edge pixels, though lying outside the map, cannot be reliably
corrected.  The width of the edge strip affected is slightly greater than the
cut-off radius.

%=============================================== Robust measures of dispersion

\subsection{Robust measures of dispersion}
\label{sec:Sn}

As previously stated, the HIPASS/ZOA data set contains severe outliers arising
from RFI, the Sun, Moon, and satellites.  Rather than attempting to censor the
data, robust statistical methods are used to ameliorate their effects.
Principal amongst these methods is the use of the weighted median in place of
the mean when estimating an average value.

While the Gaussian asymptotic efficiency\footnote{That is, the relative
efficiency of the statistic for Gaussian distributions as the sample size
grows.} of the median is only 64\% with respect to the mean ({\em increasing}
marginally for sample sizes less than 9, Snedecor \& Cochran \shortcite{SC}),
the efficiency of the median absolute deviation from the median (\MAD),
\begin{equation}
  \mathrm{MAD} = 1.4826 \med_i |x_i - \med_j x_j|,
  \label{eq:MAD}
\end{equation}
a commonly-used robust measure of dispersion, is even lower at only 37\% that
of the root mean square deviation.  This is a high price to pay for robustness
in certain demanding situations encountered in this work.  The \MAD\ also
finds the symmetric interval that contains 50\% of the data, which does not
seem to be a good approach if outliers are either strongly positive or
negative as is usually the case here.  Thus, the $S_n$ estimator developed by
Rousseeuw \& Croux \shortcite{RC} was routinely used in its place,
\begin{equation}
  S_n = 1.1926 \med_i(\med_j(|x_i - x_j|)).
  \label{eq:Sn}
\end{equation}
As in Equation (\ref{eq:MAD}), the scale factor is chosen so that $S_n$
matches the $\sigma$ parameter of Gaussian distributions, i.e.\ so that it may
substitute directly for the rms deviation.  $S_n$ has a Gaussian asymptotic
efficiency of 58\% with respect to the rms and is also more efficient than
the \MAD\ for small sample sizes, its measures of robustness are comparable
with the \MAD, and it works well with asymmetric distributions because it
deals only with differences between pairs of measurements, not differences
from a central location.

In this work, $S_n$ is always used in place of the rms whenever a measure of
dispersion is required.  $S_n$ has a higher computational cost than the
\MAD\ but that is irrelevant here.  Croux \& Rousseeuw \shortcite{CR} describe
an $O(n\log n)$ algorithm and provide an implementation in Fortran.  This
source code was translated\footnote{The C source code is available in the
{\em livedata} package.} to C for use by {\em livedata} and {\em gridzilla}.

%================================================ Continuum baseline signature

\subsection{Continuum baseline signature}
\label{sec:CBS}

A portion of the signal from a cosmic radio source may undergo multiple
reflections from the structure of the Parkes radio telescope before being
detected.  Principal among these is the reflection from the base of the prime
focus cabin which returns to the receiver after secondary reflection from the
dish.  Constructive or destructive interference with the primary signal may
occur depending on the frequency and this gives rise to a ripple pattern of
small relative amplitude in the spectral response of strong continuum sources.
The period of this ripple corresponds to twice the distance from the base of
the feed cabin to the dish which is a little less than the focal length of
27.4\,m.  At Parkes the ripple has a period of 5.7\,MHz, thus with 11 full
cycles in the 64\,MHz bandwidth.  In reality, the ripple pattern is not a pure
sinusoid, being complicated by reflections from the feed legs as well.

As discussed by Barnes et al.\ \shortcite{Barnes}, Solar emission is the most
common cause of ripple within individual scans taken during daylight hours.
However, as it is a moving celestial source, the Sun differently affects
spectra taken at a particular point in the sky at different times and its
influence does not survive robust gridding (Section \ref{sec:gridding}).  This
is not the case for strong, fixed cosmic continuum sources such as Sgr\,A and
a host of other Galactic and extra-Galactic sources whose effect is constant
for each scan.  We distinguish between ripple caused by sources close to the
optical axis of the telescope, which has a fixed pattern (Section
\ref{sec:onAxis}), and that caused by sources away from the optical axis,
which doesn't (Section \ref{sec:offAxis}).

In addition to the baseline ripple, Barnes et al.\ \shortcite{Barnes} also
reported that the multibeam receivers exhibit a frequency-dependent response
which results in an accelerating rise in the continuum response towards low
frequencies (Section \ref{sec:curvature}).  Like the baseline ripple, this
effect is also proportional to the strength of the continuum source.

While these baseline effects pose significant problems for spectral line
observations, they are also a nuisance factor for continuum work.  Barnes et
al.\ \shortcite{Barnes} described measures taken to correct the gridded maps,
but also speculated that future reprocessing might correct the individual
spectra before gridding.  That approach is to be preferred as the ripple and
curvature response, hereafter {\em continuum baseline signature} or CBS,
varies with beam and polarisation and so are best removed before the spectra
are combined in the gridded map.

The work reported in this section is also of particular interest to two
current projects involving the HIPASS/ZOA data set, that of extracting
hydrogen recombination line (RRL) maps \cite{Alves}, and of reprocessing the
HIPASS/ZOA survey itself using improved techniques \cite{Koribalski}.

%============================================================== On-axis ripple

\subsubsection{On-axis ripple}
\label{sec:onAxis}

The {\em scaled template method} of correcting the CBS described by Barnes et
al.\ \shortcite{Barnes} relied on the first-order constancy of the shape of
the CBS.  For each data cube, the first step was to determine the normalised
CBS empirically from the spectra of strong continuum sources.  Then for each
spectrum in the cube, the appropriate scale factor was measured directly from
the data and the scaled template CBS subtracted.

Essentially the same method is used here, the difference being that the
normalised CBS is determined separately for each beam and polarisation, this
being done once and for all using a large subset of the HIPASS survey data.
The appropriately scaled template CBS is then subtracted from each spectrum by
{\em livedata} during post-bandpass processing.

To determine the CBS templates, the entire HIPASS/ZOA data set was reprocessed
using robust polynomial bandpass calibration, with the continuum preserved as
per Section \ref{sec:bandpass}, but without Doppler shifting nor spectral
baseline removal.  Preliminary investigation found that the CBS is produced
only by point sources, extended continuum emission makes no contribution and
so interferes with the normalisation of the spectra by the continuum level.
Nor could any significant dependence on elevation nor feed rotation angle be
discerned.  Consequently, in order to exclude strong extended emission in the
Galactic plane, only a subset of the HIPASS data with $|b| > 15\degr$ was
used to derive the CBS templates.  For each beam and polarisation, using only
spectra with a continuum level of at least 0.25\,Jy/beam, the spectra were
normalised by the measured continuum level, and the template then formed as
the weighted median of these normalised spectra, with the continuum flux
density used as the weight.

Galactic HI emission was mostly removed from the CBS templates by taking
advantage of the annual Doppler shift which pushes the line around
sufficiently so that each of the relevant channels is unaffected for at least
part of the year.  The number of spectra used to generate the templates varied
between 74,062 and 120,000 spectra depending on beam (mainly) and
polarisation.  This number greatly exceeds that used in the original HIPASS
processing with a consequent reduction in the noise to the extent that small
birdies became visible.  These were removed via a 9-point running median
smooth as was done by Barnes et al.\ \shortcite{Barnes}.

%----------------------------------------------------------------- floating --

% Generated from CBS_HIPASS-T8.fits using a version of luthplot.
\begin{figure*}[t]
  \centering
  \includegraphics[width=230pt]{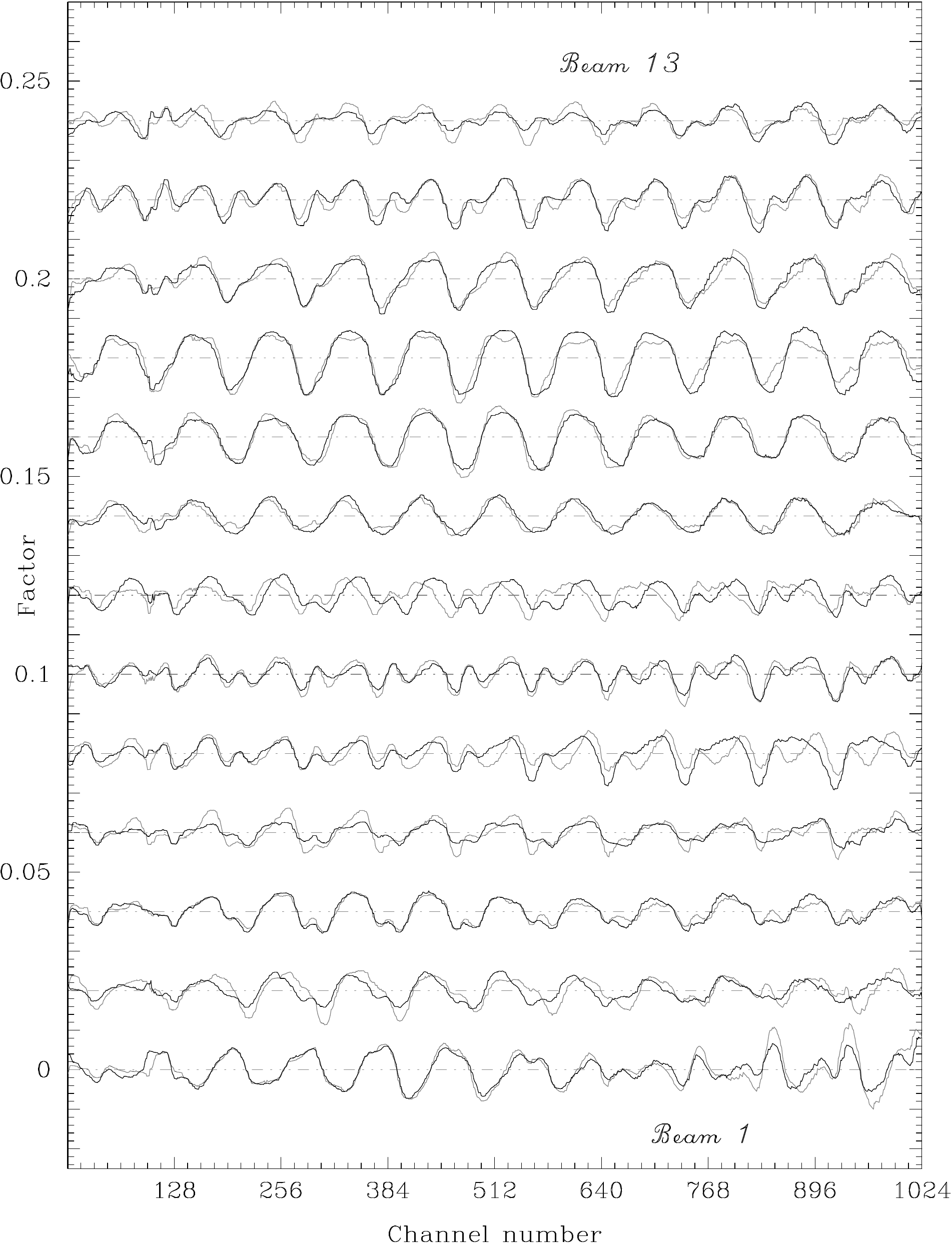}
  \hspace{-10pt}
  \includegraphics[width=230pt]{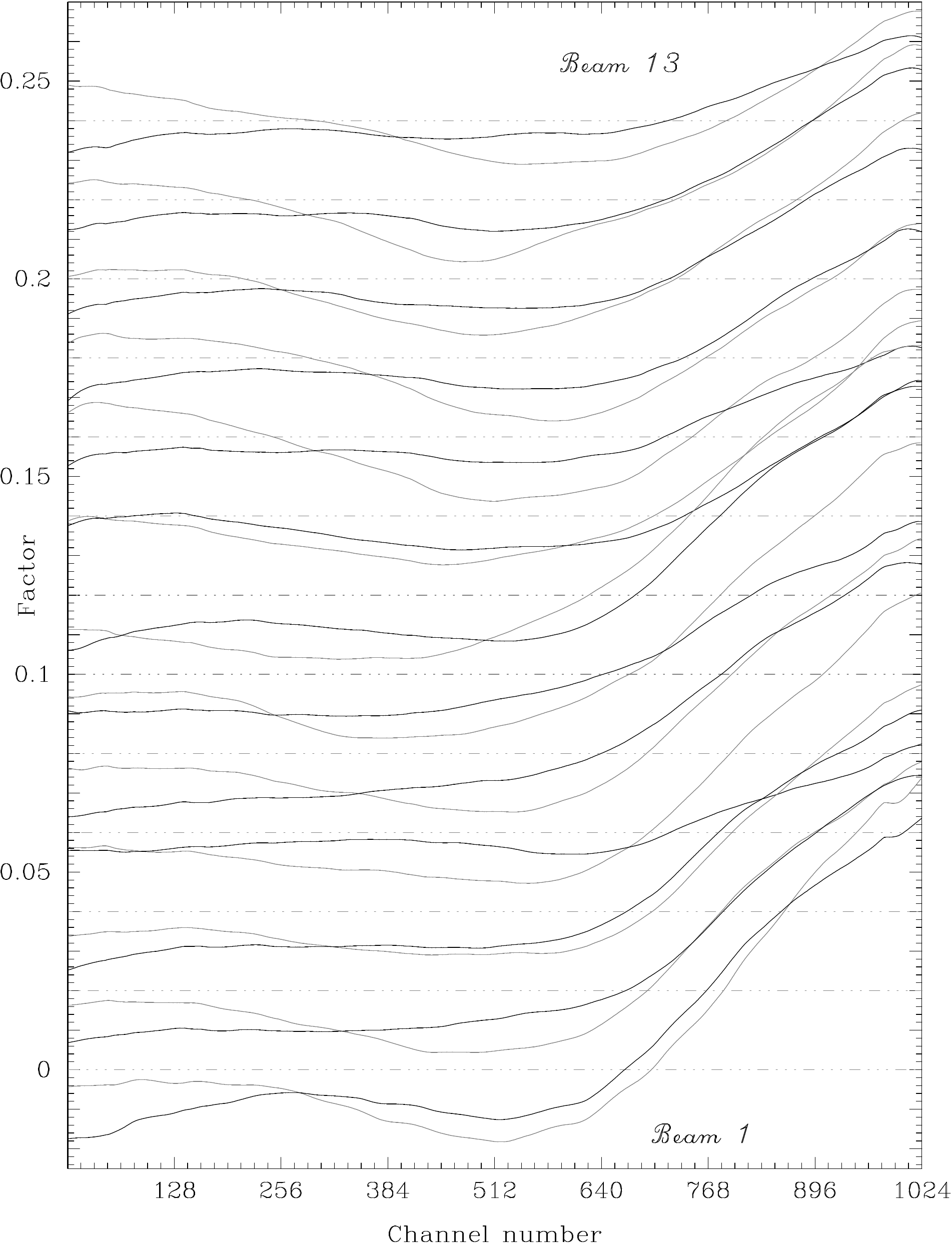}
  \caption[]{Continuum baseline signature (CBS) templates for each beam and
    polarisation decomposed into the ripple (left) and curvature components
    (right).  Channel 1024 is the low frequency end.  Beam 1 is at the bottom
    and successive beams are offset by 0.02 as indicated by the dashed
    horizontal lines.  The two polarisations are shown for each beam as black
    and grey traces.  For the ripple component there is such a close
    correspondence between them that the two traces are often
    indistinguishable.}
  \label{fig:CBS}
\end{figure*}

%========================================================== Baseline curvature

\subsubsection{Baseline curvature}
\label{sec:curvature}

Having derived the CBS templates, the task remains of determining the
appropriate scaling factor for removing the CBS from each individual spectrum.

Preliminary investigation revealed that simple scaling of the template CBS by
the continuum level, measured as the median value of the spectrum (excluding
Galactic HI), was adequate for weak sources for which the correction is barely
significant anyway.  However, a small residual of the ripple component often
remained for very strong sources.  Strong, extended continuum emission is a
complicating factor as it increases the continuum level without augmenting the
CBS.

A different, more computationally intensive strategy was therefore adopted for
the 10\% of spectra with a continuum level of 1\,Jy/beam or stronger.  This
method optimised removal of the ripple component at the expense of possibly
leaving a small residual of the curvature component, which, however, could
easily be removed later with a polynomial baseline fit.  The first step
therefore was to separate the ripple from the baseline curvature component.

The ripple has a basic frequency of 5.7\,MHz which corresponds to 91 channels
in the HIPASS/ZOA spectra.  Applying a running mean of this width to the CBS
templates therefore removes the ripple leaving behind only the curvature
component.  The ripple component may then be recovered by subtracting the
curvature component from the CBS.  This decomposition is shown in Figure
\ref{fig:CBS}.

For stronger sources, the scale factor for the CBS template was determined
from the ripple component alone.  A nine-point median smooth was applied to
the spectrum being corrected to match what was done in deriving the CBS
templates.  This eliminates H-recombination and narrow RFI lines and reduces
the noise a little.  Smoothing with a running mean over 91 channels then
removes the ripple, leaving only the baselevel which was then subtracted from
the spectrum to isolate the ripple component.  This was then divided by the
template ripple component shown in Figure \ref{fig:CBS}, and the scale factor
determined as the weighted median, with the weight taken as the ripple
template value.  In fact, RFI that occupies up to one-quarter of the band can
significantly affect the computation of the median across the whole band so
this potential source of contamination was minimised by taking the median of
the medians of each three-quarters of the band.

As expected, spectral cubes produced from data processed with this method of
CBS removal compare favourably with the older scaled template method.
Because separate templates are used for each beam and polarisation, the ripple
is more effectively removed.  The baseline noise is also significantly reduced
because the CBS templates are constructed from a much larger number of spectra
and consequently are virtually noise-free.  Also, because the CBS correction
is applied {\em before} gridding, rather than after, there is less scatter in
the data used to compute each pixel value in the gridded cube.

%----------------------------------------------------------------- floating --

% Test file: 1998-10-20_0407_164206-58_228d.hpf (used in H096).
% Extended/polyfit, 4th-deg polynomial, no Doppler shift.
% Beam 13, pol B.  Range in MultibeamVis: -0.5 to +0.5.
% Doctored versions of pksbandpass.cc were used to produce the intermediate
% steps.  The whole lot then composed using gimp.
% Refer to emails dated 2011/09/13.
\begin{figure}
  \centering
  \includegraphics[width=\columnwidth]{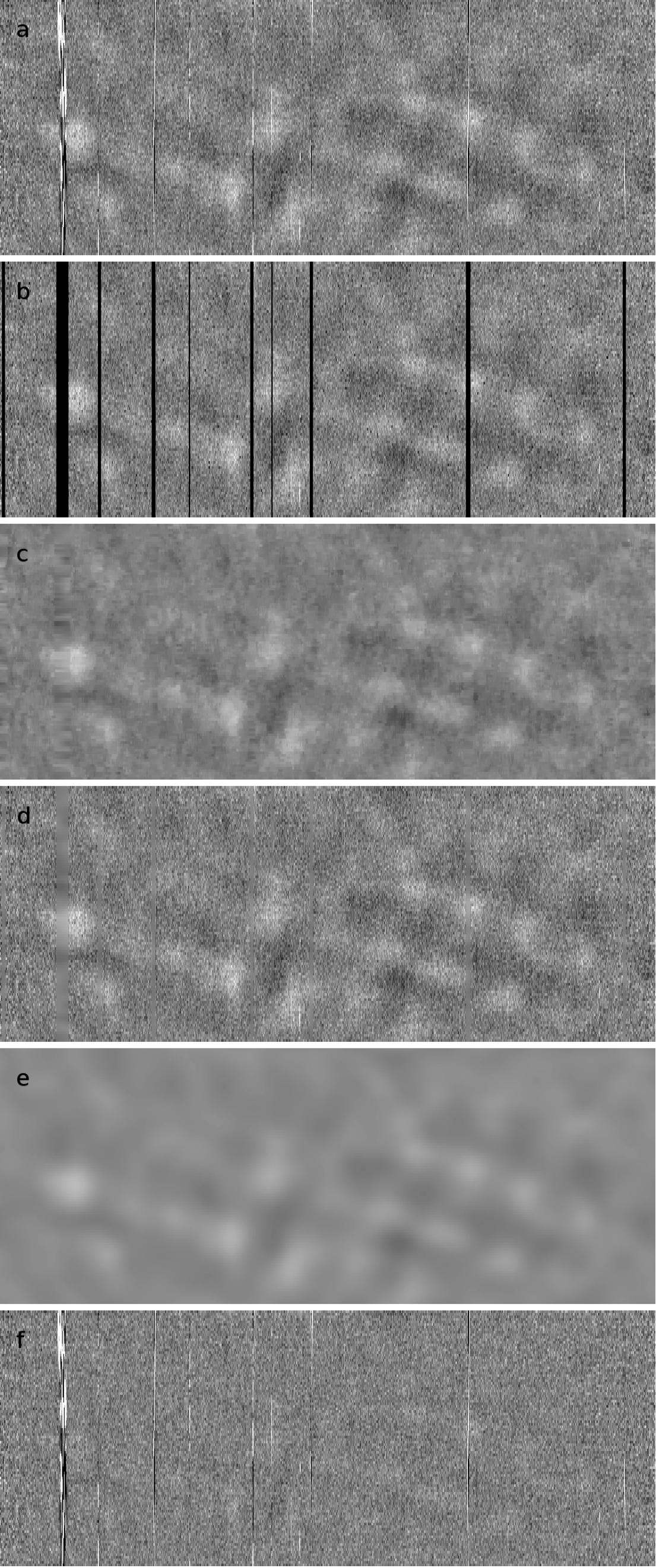}
  \caption[]{Steps in off-axis ripple filtering.  The spectral axis has
    frequency decreasing to the right from 1426.5 to 1362.5\,MHz, the vertical
    axis is scan position, 500\,s or equivalently 500' in extent.
    (a) the uncorrected scan;
    (b) masking of channels containing Galactic HI and RFI, as well as
        isolated discrepant pixels;
    (c) after linear interpolation and 9-point median smoothing in each
        direction;
    (d) masked pixels in (b) replaced with those of (c) after low-pass Fourier
        filtering;
    (e) ripple model obtained by low-pass Fourier filtering of (d); finally
    (f) the corrected scan obtained by subtracting (e) from (a).}
  \label{fig:osrf}
\end{figure}

%============================================================= Off-axis ripple

\subsubsection{Off-axis ripple}
\label{sec:offAxis}

As we have seen, strong on-axis point-source continuum emission creates highly
reproducible baseline distortions that can be characterised and removed.  Not
so for strong off-axis continuum sources which often manifest themselves via
irregular, quasi-periodic baseline disturbances which evolve with position as
the telescope scans.  Principal amongst these is the Sun which may affect
daytime observations.  Although it has a variable affect for each specific
point on the sky and so does not survive robust gridding, nevertheless any
increase in the scatter in the data acts to reduce the accuracy of the final
result.  On the other hand, strong fixed cosmic continuum sources, of which
there are many, especially in the Galactic plane, have a static effect which
can survive robust gridding.  Thus for spectral line work it is always
advantageous to remove such baseline distortions as far as is practicable,
although with no benefit at all for continuum work.

When the scan for a particular beam and polarisation is represented as an
image of channel versus position (i.e.\ time for a scanning observation),
off-axis ripple manifests itself as an irregular, quasi-periodic pattern,
usually with harmonics oriented at an angle to the horizontal or vertical as
seen in Figure \ref{fig:osrf}a.  This suggests the use of a low-pass 2D
Fourier filter to model the ripple pattern.

The principal difficulty is that such filtering is not robust against strong
RFI nor line emission which must therefore be excised.  Figure \ref{fig:osrf}
shows schematically the steps involved.  Because the ripple pattern is
broad-scale, whereas the RFI or line emission usually occupies a narrow range
of channels or integrations, the strategy is to interpolate across it.

Bandpass calibration effectively ensures that the mean value in each spectral
channel is close to zero.  However, line emission and RFI still clearly
manifest themselves via the dispersion, as seen in Figure \ref{fig:osrf}a.  As
a prelude, isolated narrow-, and broad-band RFI transients are masked on a
channel-by-channel basis by censoring pixels outside $3 \times S_n$ for the
channel, where $S_n$ is the robust measure of dispersion discussed in
Section \ref{sec:Sn}.  This discriminant corresponds to $3 \sigma$ for a
normal distribution.  As it is not sufficient simply to zero these pixels,
their values are replaced by linear interpolation across neighbouring values.

Discrepant channels are then identified by a discriminant based on the
dispersion of the dispersion, i.e.\ channels that have a value of $S_n$ much
higher than normal, such that they correspond to $4 \sigma$ outliers.  The
channel mask is then broadened by flagging immediately neighbouring channels
if they have $S_n$ outside $1 \sigma$.  Linear interpolation across the
channels provides values for the masked channels, followed by 9-point median
smoothing to eliminate any small patches of narrow RFI that may have survived
to this point, as in Figure \ref{fig:osrf}c.  This first-pass interpolation is
then subjected to low-pass Fourier filtering using a Gaussian filter function
with twice the FWHM of the required final filter function.  The result is used
to provide values for the masked pixels in the original unsmoothed image, and
this then subjected to low-pass Fourier filtering to obtain the ripple model
as in Figure \ref{fig:osrf}e.

A Gaussian with FWHM of 16 harmonics in the spectral direction and 8 in the
scanning direction proved to be an acceptable compromise for removing the
ripple while retaining compact spectral line sources, though inevitably
creating shallow moats around them.

%----------------------------------------------------------------- floating --

\begin{table}
  \caption[]{Regions used to determine $\Tsys(\eta,\zeta)$, and (below) as
    represented on a plate carr\'{e}e projection with $24\hr \ge \alpha \ge
    0\hr$, $-90\degr \le \delta \le +26\degr$ (i.e.\ with $\alpha = 0\hr$ on
    the right-hand edge).}
  \begin{center}
  \protect\begin{tabular}{rlll}
    \hline\hline
    \noalign{\smallskip}
    Zone & & RA ranges \\
    \hline
    \noalign{\smallskip}
    $+22\degr$ &            & 0820--1300 \\
    $+14\degr$ &            & 0840--1230 \\
    $+06\degr$ &            & 0800--1200 \\
    $-02\degr$ & 0000--0300 & 0815--1145 & 2200--2400 \\
    $-10\degr$ & 0000--0300 & 0810--1130 & 1300--1500 \\
    $-18\degr$ & 0000--0515 & 0840--1130 & 1230--1430 \\
    $-26\degr$ & 0000--0630 & 1030--1150 & 2000--2230 \\
    $-34\degr$ & 0000--0650 & 1010--1225 & 2130--2400 \\
    $-42\degr$ & 0000--0630 & 1120--1310 & 2130--2400 \\
    $-50\degr$ & 0000--0630 &            & 2100--2400 \\
    $-58\degr$ & 0000--0720 &            & 2100--2400 \\
    $-66\degr$ & 0000--0430 & 0600--0800 & 1845--2400 \\
    $-74\degr$ & 0130--0440 & 0600--1000 & 2000--2400 \\
    $-82\degr$ & 0000--1640 &            & 2020--2400 \\
    $-87\degr$ & 0000--2400 \\
    \noalign{\smallskip}
    \multicolumn{4}{c}{\includegraphics[width=205pt]{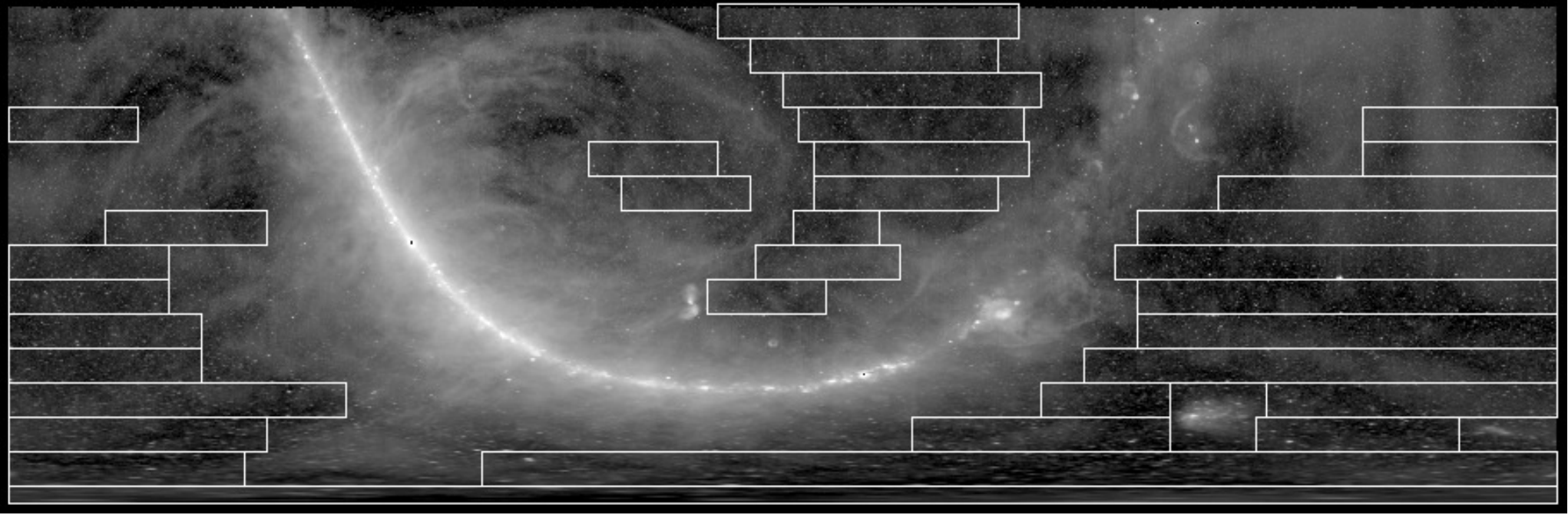}} \\
    \noalign{\smallskip}
    \hline\hline
  \end{tabular}
  \end{center}
  \label{ta:regions}
\end{table}

%=================================================== Tsys elevation dependence

\subsection{$\Tsys$ elevation dependence}
\label{sec:elevation}

$\Tsys$ increases systematically towards lower elevations due to a combination
of increasing atmospheric opacity and increasing spillover as distant
sidelobes intersect with the ground.  As is apparent in Figure
\ref{fig:Extended}, this effect is more pronounced in the northern declination
zones because $\Tsys$ changes more rapidly at lower elevations, and because
scans in these zones are predominantly in elevation.

As elevation, $\eta$, increases from its lower limit, $\Tsys$ drops to a
minimum at about $\eta = 58\degr$ and thereafter increases in a somewhat
irregular way towards the zenith because of spillover effects.  This is not
apparent in Figure \ref{fig:Extended} because the declination zones that can
be observed at elevations above $58\degr$ can also be observed at lower
elevations where the scans may be predominantly in azimuth.  Moreover,
HIPASS/ZOA observations tended to be made at elevations away from the zenith
in order to limit the change in parallactic angle.  Consequently the effect is
washed out.  Nevertheless, it must contribute noise to the map.

Figure \ref{fig:Extended} shows that the elevation dependence of $\Tsys$ has
an obvious signature which implies that it should be possible to deduce a
correction from the HIPASS data itself.  The Galactic plane and various other
strong, large-scale emission features that might contaminate the results were
excluded from the analysis; Table \ref{ta:regions} indicates the RA ranges
used.  Unfortunately, the north-polar spur and other large features precluded
using long stretches of the northern zones.  While low level emission is also
evident even in the selected regions, we rely on the essentially random
distribution of observations to cancel it out.  That is, if we consider all of
the observations made at a particular elevation, there are likely to be as
many cosmic emission features that contribute a positive gradient as a
negative gradient.  The effect then is to add statistical noise.  In any case,
this approach is likely to produce more accurate results than traditional
{\em sky-dips} which are likely to traverse the same emission features, albeit
unknowingly.

Because of the zone-level problem discussed in Section \ref{sec:zonelevs} it
is only possible to determine $\Tsys(\eta)$ in a piece-wise fashion within
each HIPASS zone, each with an arbitrary zero point.  The approach taken here
side-steps these zonal discontinuities by measuring the derivative,
$\D \Tsys/\D \eta$, and integrating.  The constant of integration is chosen
arbitrarily to set the minimum of $\Tsys(\eta)$ to zero which accords with our
lack of knowledge of the zero level of the continuum emission over the sky as
a whole.

The analysis was based on gridding using the observation that, in determining
the gradient,
\begin{equation}
  \frac{T_n - T_1}{{\eta}_n - {\eta}_1} =
    {\sum_{i=2}^n \Delta {\eta}_i (\frac{\Delta T_i}{\Delta {\eta}_i})} /
    {\sum_{i=2}^n \Delta {\eta}_i} ,
  \label{eq:deriv}
\end{equation}
where $\Delta {\eta}_i = {\eta}_i - {\eta}_{i-1}$ and likewise for
$\Delta T_i$.  That is, the gradient determined from the end-points of a scan
is equal to the weighted mean of the gradients determined from each pair of
integrations in the scan, provided that the weight is taken to be
$\Delta {\eta}_i$.  Thus, even though a measurement of the gradient obtained
from a small change in elevation might have a large uncertainty, when these
measurements are averaged with the correct weight, the uncertainty is no
larger than that relating to the longest span.  This result carries over to
combining different scans that have an unequal range of elevations.  This
approach allows all of the data to be used and also allows a more localised
measurement of the gradient, which is a better approximation to the
derivative.  Of course, when the data is gridded, it is these localised
measurements that are averaged, and in practice the weighted median replaced
the weighted mean for robustness.

%----------------------------------------------------------------- floating --

% /DATA/GRUS_1/CHIPASS/median/CHIPASS_ZPAEl/-
%   CHIPASS_AllAB_WGTMED.elgrad.fits
%
% kvis levels -0.2 to +0.1, linear, Greyscale1.
% Labelling added using gimp.
\begin{figure}
  \centering
  \includegraphics[width=220pt]{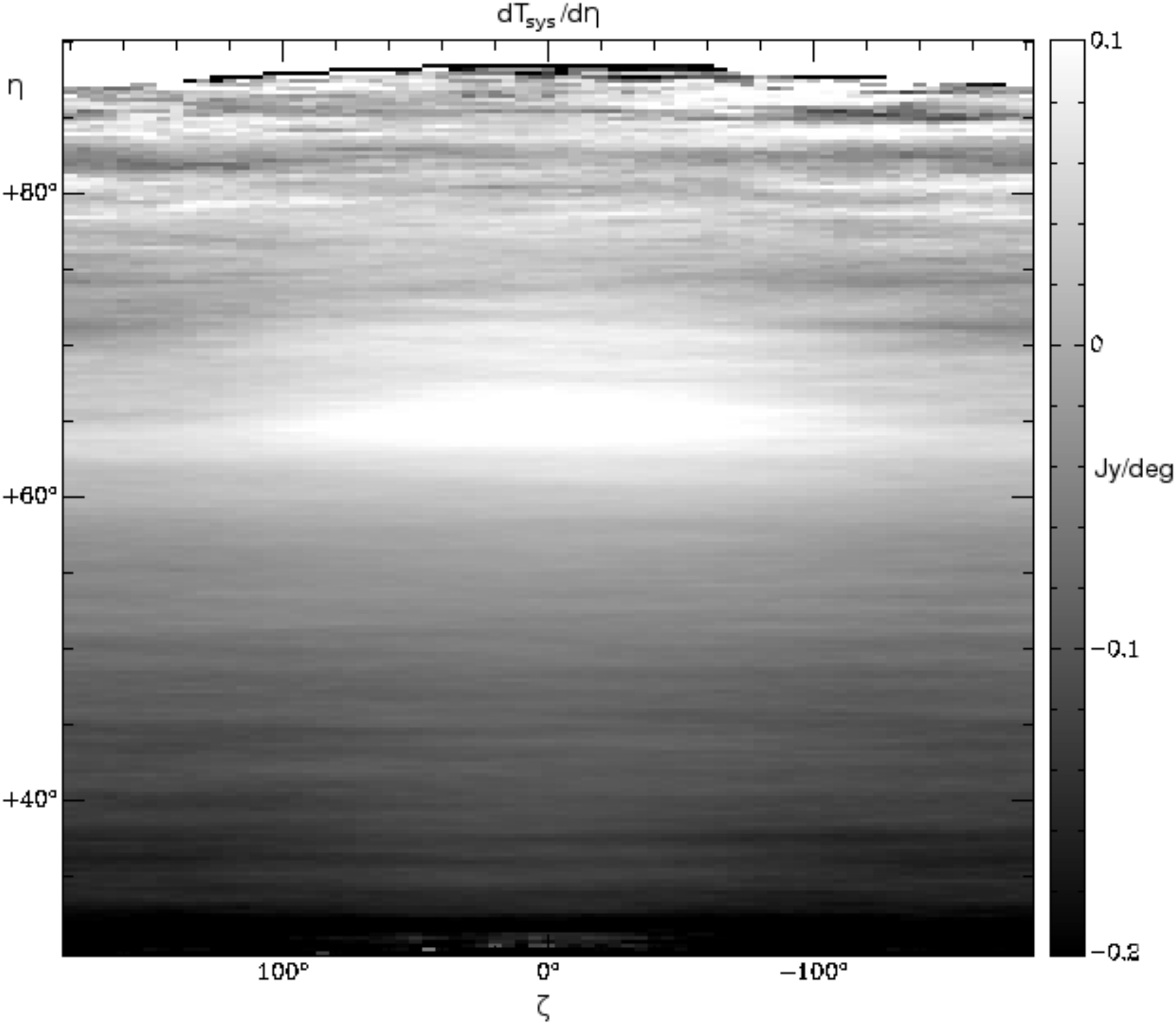}
  \caption[]{$\D\Tsys/\D\eta$ as a function of $\zeta$ and $\eta$ for all
    beams taken together, obtained by weighted median gridding of
    $\Delta T_i/\Delta {\eta}_i$.}
  \label{fig:ZPAEl}
\end{figure}

%----------------------------------------------------------------- floating --

The procedure then was to search for a dependence of $\Tsys$ as a function of
$\eta$ and any other likely parameters, where $\eta$ is the elevation of each
separate beam.  There was no dependence on azimuth, nor any discernible
dependence on polarisation.  However, there was a small but significant
dependence on the zenithal position angle, $\zeta$, which is defined as the
angle between the zenith, the central beam (assumed to be positioned on the
optical axis), and the beam in question.  For the central beam, $\zeta$ is
taken to be the rotation angle, $\rho$, of the feed assembly.  Thus
\begin{equation}
  \begin{array}{r@{}r@{\ }r@{\ }r@{\ }r@{\ }r@{\ }r}
  \zeta - \rho = &  (  0\degr, \nonumber \\
                 &   -60\degr, &   0\degr, &   60\degr, &
                     120\degr, & 180\degr, & -120\degr, \nonumber \\
                 &   -90\degr, & -30\degr, &   30\degr, &
                      90\degr, & 150\degr, & -150\degr)
  \end{array}
  \label{eq:zetarho}
\end{equation}
for beam numbers 1--13.  There was also a dependence on beam, but only between
the central beam (beam 1), beams in the inner ring as a set (beams 2--7), and
those in the outer ring (8--13).  Plausibly, the dependence on $\zeta$ and
beam may result from the asymmetry of the inner and outer beams as illustrated
by Kalberla et al.\ \shortcite{Kalberla}.

%----------------------------------------------------------------- floating --

% /DATA/GRUS_1/CHIPASS/median8/CHIPASS_ZPAEl/TsysEl.ps
% Generated using program TsysEl.
\begin{figure}
  \centering
  \includegraphics[width=220pt]{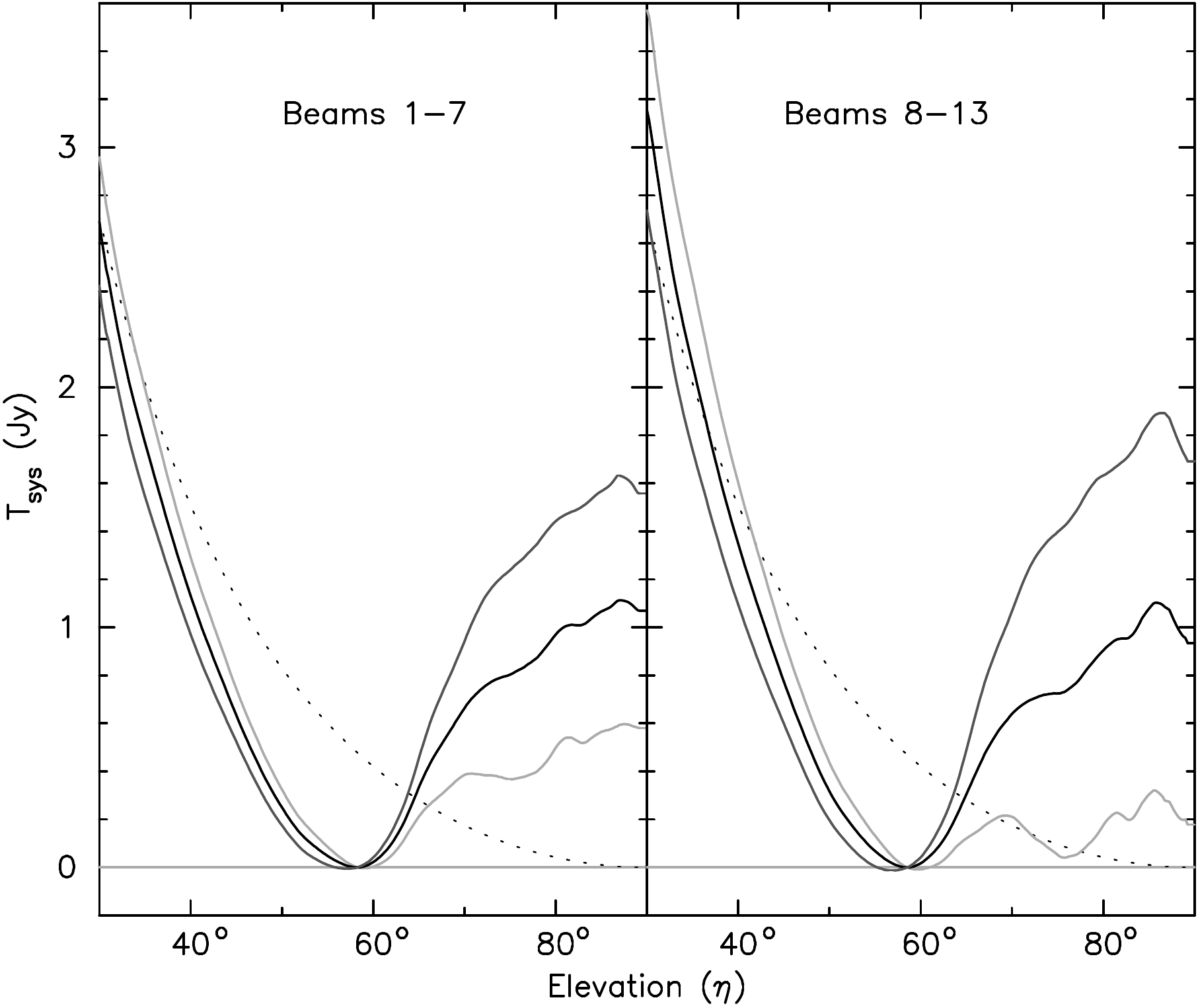}
  \caption[]{$\Tsys$ as a function of elevation, $\eta$, for beams 1--7 (left)
    and beams 8--13 (right).  The black curve at left applies for beam 1,
    independent of $\zeta$.  Envelope curves are shown for beams in the inner
    and outer rings with $\zeta = 0\degr$ (dark grey), $\pm 90\degr$ (black),
    and $180\degr$ (light grey).  The dotted curve in each is the nominal
    contribution from atmospheric opacity, $\Tatm (1 - \exp(-\tau/\sin\eta)$,
    normalised at the zenith, using canonical values of $\Tatm$ = 275\,K and
    $\tau = 0.01$.}
  \label{fig:TsysEl}
\end{figure}

%----------------------------------------------------------------- floating --

% /DATA/GRUS_1/CHIPASS/median8/CHIPASS_GAL/-
%   CHIPASS_GAL_WGTMED1.continuum.fits
%
% kvis levels 0 to 50, logarithmic 3-cycle, Greyscale1.
\begin{figure*}
  \centering
  \includegraphics[width=460pt]{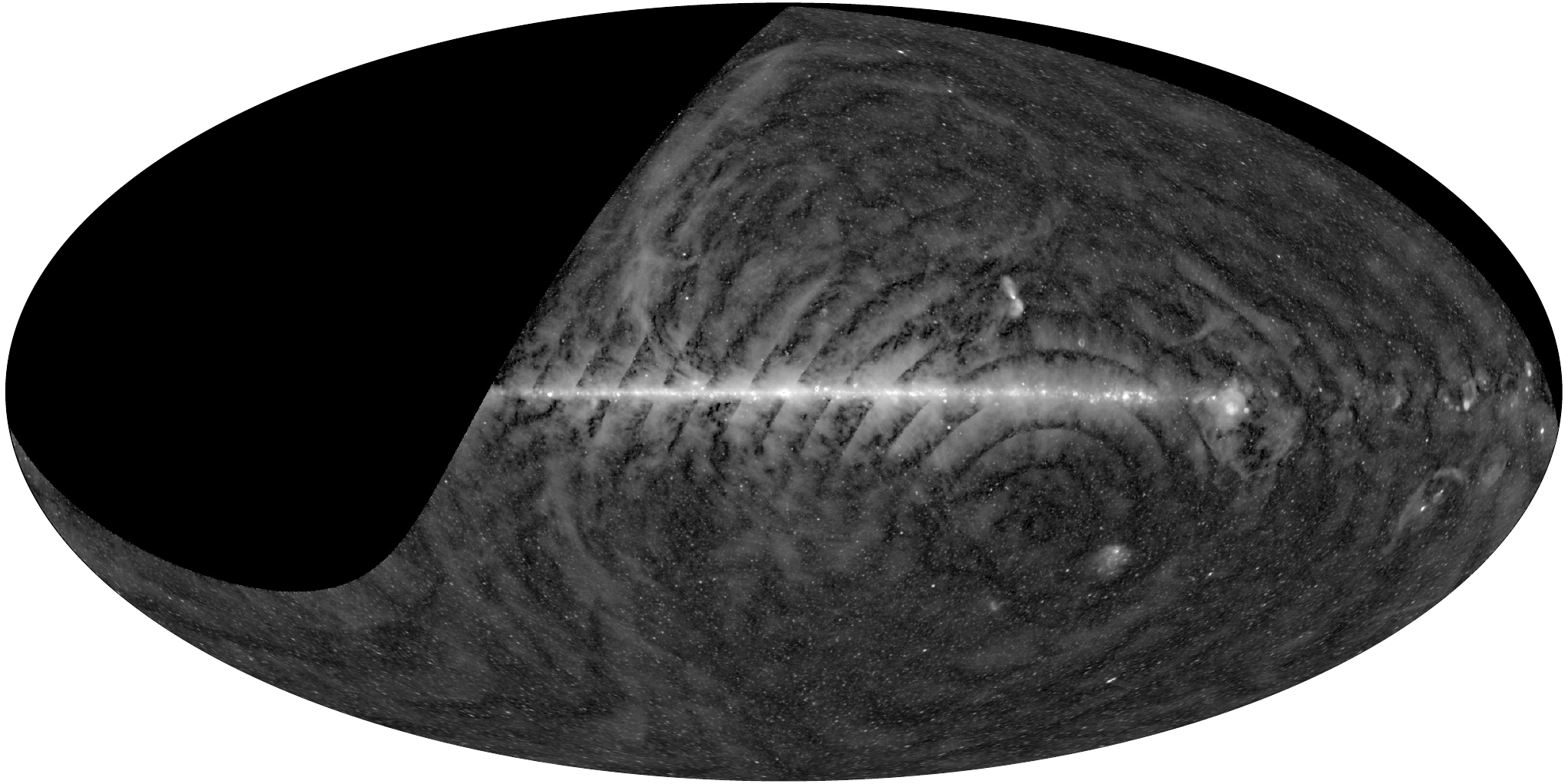}
  \caption[]{For comparison with Figure \ref{fig:Extended}, HIPASS 1.4\,GHz
    continuum data now corrected for the elevation dependence of $\Tsys$.  It
    still awaits zone level correction.  Logarithmic greyscale as for Figures
    \ref{fig:Compact} and \ref{fig:Extended}.}
  \label{fig:ElCorr}
\end{figure*}

%----------------------------------------------------------------- floating --

The general form of the dependence is
\begin{equation}
  \D{\Tsys} = \frac{\partial \Tsys}{\partial\eta}  \D{\eta}
            + \frac{\partial \Tsys}{\partial\zeta} \D{\zeta}.
  \label{eq:dT}
\end{equation}
However, as 99\% of HIPASS/ZOA observations were made without
parallactification, we have $\D{\zeta} = 0$ and this proved to be an important
simplification.  The dependence found was of the form
\begin{equation}
  \Dfrac{\Tsys}{\eta} = a(\eta) + b(\eta) \cos\zeta,
  \label{eq:dTdEta}
\end{equation}
where the functions $a(\eta)$ and $b(\eta)$ are beam-dependent.

As described, the analysis is based on weighted median gridding of
$\Delta T_i/\Delta {\eta}_i$ as a function of $\eta$ and $\zeta$ by virtue of
Equation (\ref{eq:deriv}).  Because there are no parameters in Equation
(\ref{eq:dTdEta}) that need to be determined as a function of $\zeta$, it was
possible to make the $\zeta$ grid much coarser than that of $\eta$ in order to
reduce the noise.  Because $\eta$ and $\zeta$ are true angles (not a spherical
coordinate pair), the most appropriate choice of projection was a plate
carr\'{e}e\footnote{Often incorrectly called ``Cartesian''.} with the
distance\footnote{{\em gridzilla} normally computes the distance as a
spherical arc length.} computed as
\begin{equation}
  d = \sqrt{(\Delta\zeta/\epsilon)^2 + (\Delta\eta)^2},
  \label{eq:distance}
\end{equation}
where $\epsilon$ is the elongation of the gridding kernel in $\zeta$, and
$(\Delta\zeta,\Delta\eta)$ is the difference in $(\zeta,\eta)$ between a datum
and a grid point.  This reduction used a top-hat kernel of diameter $1\degr$
with an elongation of $\times 120$.

Noting that the convolution of a cosine function with a top-hat function
produces a cosine of the same period but diminished amplitude, we would expect
that such heavy smoothing of $\cos\zeta$ should lead $b(\eta)$ to be
underestimated by 17\%.  However, smoothing was necessary to obtain reasonable
statistics for each beam, especially when studied in isolation from the
others, and particularly at higher elevations where the data is relatively
sparse.  This effect, and possibly others, were corrected by iterating the
solution.  Its magnitude was of the expected size.

At $5\degr \times 15\arcmin$, the $(\zeta,\eta)$ grid depicted in Figure
\ref{fig:ZPAEl} is considerably finer than the dimensions of the gridding
kernel.  As rotation of the feed assembly was limited to $\rho = \pm 60\degr$,
$\zeta$ for each beam spans up to $120\degr$ or a third of the full range
according to Equation (\ref{eq:zetarho}).  The portion of Figure
\ref{fig:ZPAEl} occupied by each beam was found to reproduce the behaviour
established by gridding all beams together as in Figure \ref{fig:ZPAEl}.

Individually, beams 2--7 in the inner ring were found to be indistinguishable,
within the noise, from all of them taken collectively.  Likewise for the outer
ring.  Also, $a(\eta)$ for beam 1 by itself did not differ significantly from
the inner ring, though we set $b(\eta) = 0$ for it by consideration of
symmetry.  There were small but significant differences in $a(\eta)$ for the
inner and outer beams, as became clear when one was plotted against the other.
However, $b(\eta)$ for the inner and outer rings was found to be sufficiently
similar within the noise, that they were taken to differ only by a scale
factor.  Consequently, there were three functions of $\eta$ and two scale
factors to be determined by least squares fits to Equation (\ref{eq:dTdEta}).
These were obtained from maps similar to Figure \ref{fig:ZPAEl} but produced
for beams 2--7 only, and for beams 8--13 only.

After integration, $\Tsys$ is shown in Figure \ref{fig:TsysEl} as a function
of $\eta$ for values of $\zeta$ that define the envelope.  For $\zeta = \pm
90\degr$, $\Tsys$ is set to zero at its minimum, whereas for other values of
$\zeta$ it may dip slightly below zero in the vicinity.  Clearly the effect of
the $b(\eta) \cos\zeta$ term in Equation (\ref{eq:dTdEta}) is significant,
especially for $\eta > 58\degr$.  The difference between the inner and outer
beams is subtle yet important for accurate calibration when it is considered
that this correction is added directly to the data.  Accuracy is much reduced
above $\eta = 80\degr$ because there is little HIPASS/ZOA data in this regime
but, by the same token, what little data there is to correct has a small
impact on the final map.  These curves are qualitatively similar to the curve
(their Figure 4) measured by Griffith \& Wright \shortcite{PMN} for the PMN
survey which used the NRAO 7-beam system at 4850\,MHz.

An option was added to {\em livedata} to apply the $\Tsys$ elevation
correction on the fly when processing the data.

%======================================================= Zone level adjustment

\subsection{Zone level adjustment}
\label{sec:zonelevs}

\begin{figure*}
  \includegraphics[width=\textwidth]{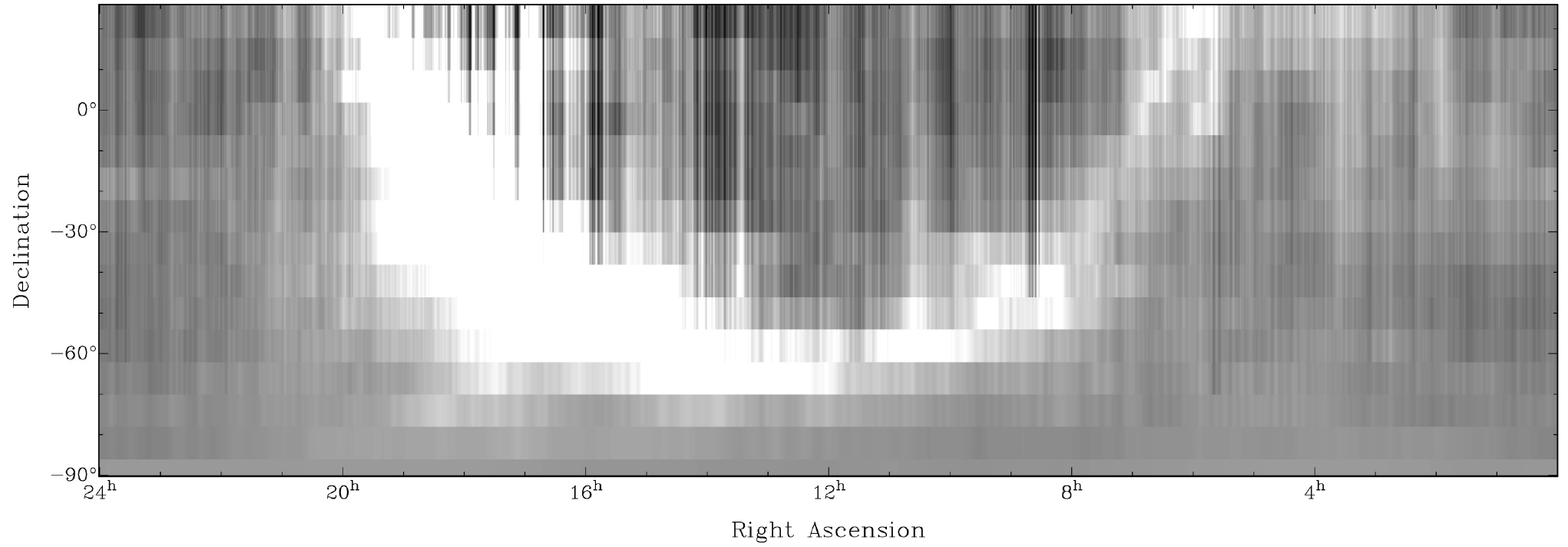}
  \caption[]{HIPASS zone levels derived from the overlaps, before correction
    provided by ZOA.  Particularly strong streamers eminate from the inner
    Galactic region (left).  In general, streamers can be negative as well as
    positive.  Plate carr\'{e}e equatorial projection, $5401 \times 15$
    pixels.  Linear greyscale from -1 to +1\,Jy/beam, emphasising streamers.}
  \label{fig:chipassLevs}
\end{figure*}

Figure \ref{fig:ElCorr} illustrates the final step to be overcome in producing
the full continuum map, that of repairing the discontinuities at the
declination zone boundaries, a process referred to here as {\em zipping}.  The
main handle that we have on this is the overlap between these zones.  In
principle it is clear that an offset can be computed by gridding the
overlapping strips separately and differencing them.  In practice, however,
the overlaps are narrow and it is by no means clear that the results will be
accurate enough, especially considering that errors will propagate from each
overlap to every successive zone when the levels are computed.  These form
what are referred to below as {\em streamers} such as manifest themselves as
the vertical artefacts in Figure \ref{fig:chipassLevs}.

The main source of streamers are strong compact continuum sources in the
Galactic plane and this is where the ZOA survey comes to the fore.  Because
scanning in the ZOA survey was done in Galactic longitude, which for the range
under study is always at an angle to the HIPASS scans in declination, the ZOA
scans provide a bridge across the HIPASS declination zones in the region where
the worst streamers are produced.

However, the ZOA zones (hereafter {\em zoans} to distinguish them from HIPASS
{\em zones}), themselves require levels, so we are faced with a bootstrapping
process.  In summary, the HIPASS levels are first deduced as accurately as
possible from the zone overlaps.  These levels are then used to derive the ZOA
zoan levels which, once applied to the ZOA maps, are used to correct the
HIPASS zone levels, particularly to remove streamers.

Thus firstly we must consider the HIPASS zone overlaps.  Numerically, 81,000
levels had to be determined, one every 16\,s of RA in 15 zones, though these
values are not independent.  Allowing one per $14\farcm4$ HPBW, the number of
independent values would be about 14,800.  However, considering that zone
levels arise from emission that is significantly extended with respect to the
$8\degr$ scan length, the true number is considerably less than this by
perhaps a factor of eight or more, leading to a number less than 2000.

%----------------------------------------------------------------- floating --

% /DATA/GRUS_1/CHIPASS/median8/CHIPASS_Overlaps/pass-6/-
%   chipass_levels.fits
%
% kvis levels -1 to 1, linear, Greyscale1

%============================================ HIPASS declination zone overlaps

\subsubsection{HIPASS declination zone overlaps}
\label{sec:zoneOverlaps}

Except for the two circumpolar zones, the HIPASS declination zones were
stepped by $8\degr$ ($= 480\arcmin$), i.e.\ $(-87\degr, -82\degr, -74\degr,
-66\degr, \ldots, +14\degr, +22\degr)$.  Typically the scans are $100 \times
5$\,s integrations at $1\arcmin$s$^{-1}$, ($500\arcmin$) so na\"{i}vely we
might expect a $10\arcmin$ overlap at each end.  However, as seen in Figure
\ref{fig:tracks}, the inner beams contribute another $29\arcmin$, and the
outer beams another $51\arcmin$ at each end of the scan.  Consequently, the
usable overlap between zones is more like $45\arcmin$.

However, at $1\fdg5$, the overlap between the $-87\degr$ and $-82\degr$ zones
is twice as wide, the data is also more uniformly distributed in this region
of the sky because of grid convergence effects (see Figure \ref{fig:tracks}),
and there happens not to be any significant sources of emission to upset the
measurements.  Consequently, the offsets between these two zones are better
determined than elsewhere and are considered not to require correction.

If the $-87\degr$ zone had extended to the south celestial pole (SCP), or
preferably slightly beyond it, then the southern edge of the zone would
effectively have overlapped itself, and it would have been possible to set
offsets for the zone by reference to the common point, the SCP, in each scan.
However, as the Parkes telescope is based on a master equatorial drive system,
the dish cannot scan through the SCP and necessarily stopped a short distance
from it.  Consequently, offsets for the $-87\degr$ zone were set by averaging
integrations in the three map rows $8\arcmin$, $12\arcmin$, and $16\arcmin$
north of the SCP.  Small offsets of up to 57\,mJy/beam were required to
achieve uniformity in this narrow strip.

In the first instance, the procedure was to map the overlapping strips
separately using weighted median gridding onto a plate carr\'{e}e projection
of size $5401 \times 11$ pixels (or $\times 23$ for the southern zones) spaced
by 16\,s in RA ($4\arcmin \cos\delta$) and $4\arcmin$ in declination.
Iterative gridding was not used because adverse edge effects would have
affected too high a proportion of each map.  Instead the whole process was
iterated as described later.

Overlapping strips were subtracted to produce a difference map, e.g.\ the
northern strip of zone $-34\degr$ minus the southern strip of zone $-26\degr$,
which ideally should have constant brightness in declination.  In practice,
such constancy usually was observed to a remarkable degree, but in certain
locations, particularly where the strip crossed the Galactic plane, there
might be significant departures.  Often this could be traced to a strong
compact source that did not cancel exactly between the two strips; although it
may have been a small fraction of the source strength, the residual would
still be unacceptably high in absolute terms.  This resulted in the streamers
that will be discussed later.  The zone offsets were computed by averaging the
11 (or 23) pixels across the width of the strip, potentially with censoring of
such outliers.

With as few as 11 measurements, it was important to use the data as
efficiently as possible.  Offsets were computed for each RA as the median of
the difference between the zone overlaps, southern minus northern.  To improve
statistics, $\lfloor 5 \sec\delta\sub{z} \rfloor$ image columns spanning
approximately $20\arcmin$ were aggregated, where $\delta\sub{z}$ is the
mid-zone declination.  However, as these samples are not independent, the
improvement is somewhat less than it would otherwise be.

The offsets were then subtracted from the difference map leaving a residual
map in which outliers could clearly be identified.  These tended to cluster
near the southern or northern edges, or in localised areas of strong emission
which failed to cancel exactly between the two zones.  $S_n$ computed for the
residual map as a whole was used to censor these outliers.  Starting from the
SCP, the values obtained for $S_n$ in each zone were: 15, 24, 27, 29, 40, 36,
35, 35, 37, 38, 41, 41, 41, 45, and 48\,mJy/beam.  Only 0.27\% of
normally-distributed data are expected to lie outside the $3 \sigma$
confidence interval.  However, averaging over all zones, 3.7\% were found to
lie outside $3 \times S_n$, indicating the degree of departure from normality.
Because one discrepant edge pixel could reasonably be expected in a scan of
length 11 pixels, the offset was recomputed if up to 10\% of the data was
outside $3 \times S_n$.  Of the 81,000 offsets being determined, 60\% were
computed from data that had no outliers, and 29\% were recomputed with a small
number of outliers censored.  The remaining 11\% were not recomputed on the
grounds that the value of $S_n$ for the whole scan probably was not applicable
to them and there was no way to improve on the median value.

The process of determining the offsets was iterated, with the offsets
determined on one pass used for zipping the overlap strips on the next.
The offsets provide everything required here and were used in preference to
the levels to avoid complications that might arise from the accumulation of
error in computing the latter.  Iteration results in a small relative change
because the offsets computed from the gridded maps are necessarily a smoothed
version of the offsets that must be applied to the ungridded data.  In
practice, however, it mainly only served to demonstrate self-consistency.  The
offsets obtained are referred to as the {\em core set} to which reference is
always made in subsequent processing.

%----------------------------------------------------------------- floating --

% /DATA/GRUS_1/CHIPASS/median8/-
%   CZOAZIP1: CHIPASS_ZOA/pass-1/CHIPASS_ZOA_WGTMED.continuum.fits
%   CZOAZIP2: CZOA_CAR/CZOA_CAR_WGTMED.continuum.fits
%   CZOAZIP3: generated by kvis
%   CZOAZIP4: CZOA_CARZ/pass-1/CZOA_CARZ_WGTMED.continuum.fits
%
% kvis levels 0 to 50, logarithmic 3-cycle, Greyscale1.
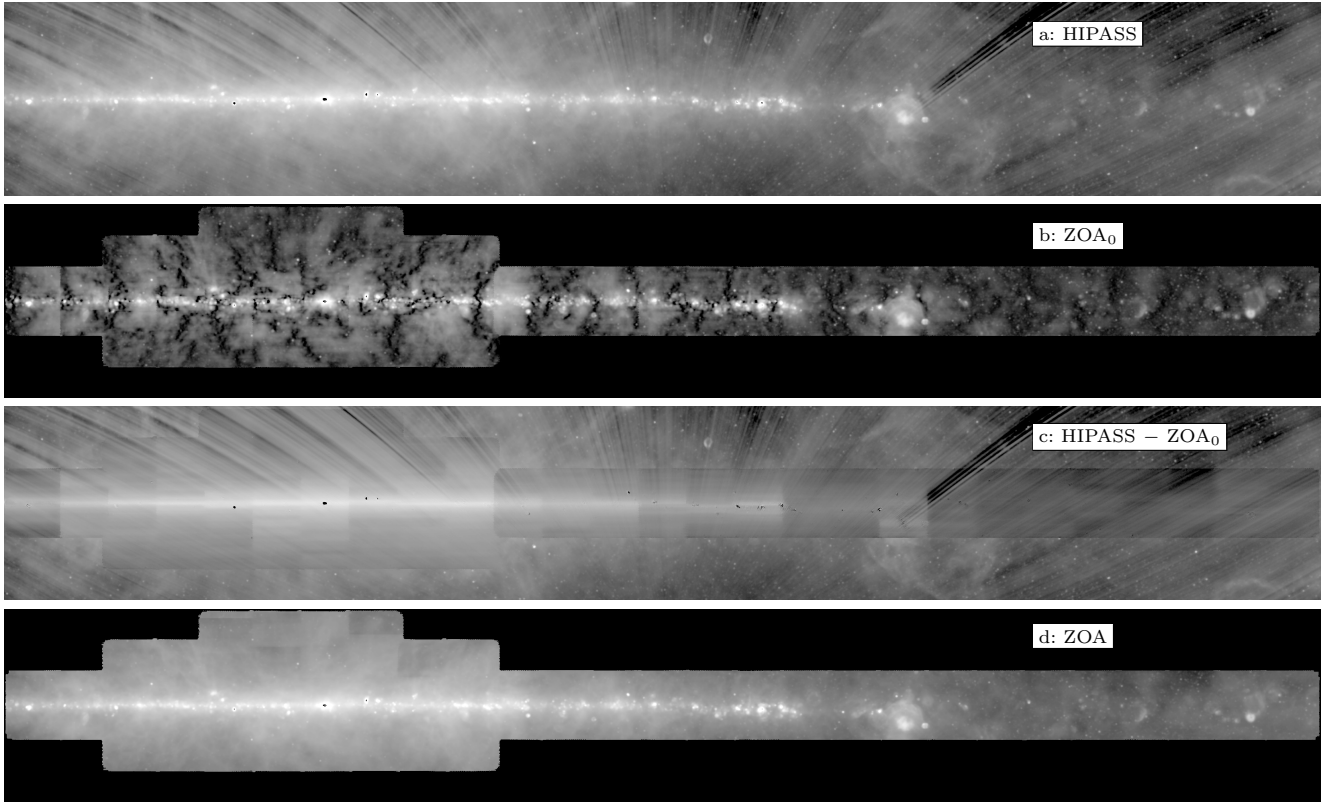
\begin{figure*}
  \centering
  \begin{tabular}{l}
    \noalign{\includegraphics[width=\textwidth]{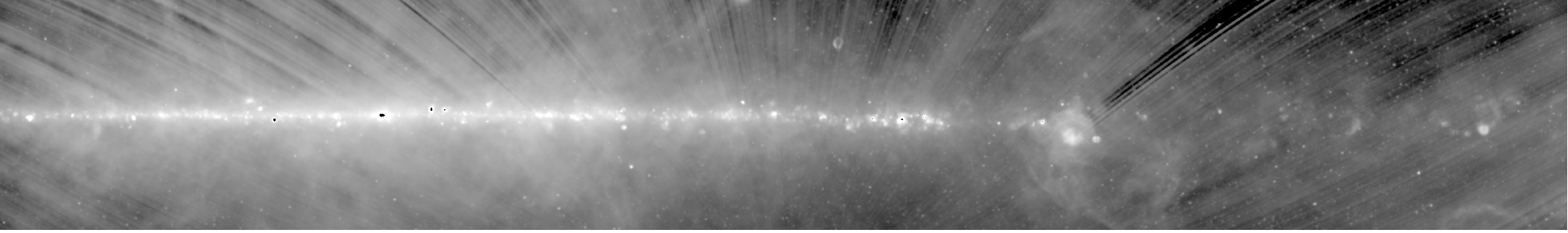}}
    \noalign{\vspace{1mm}}
    \noalign{\includegraphics[width=\textwidth]{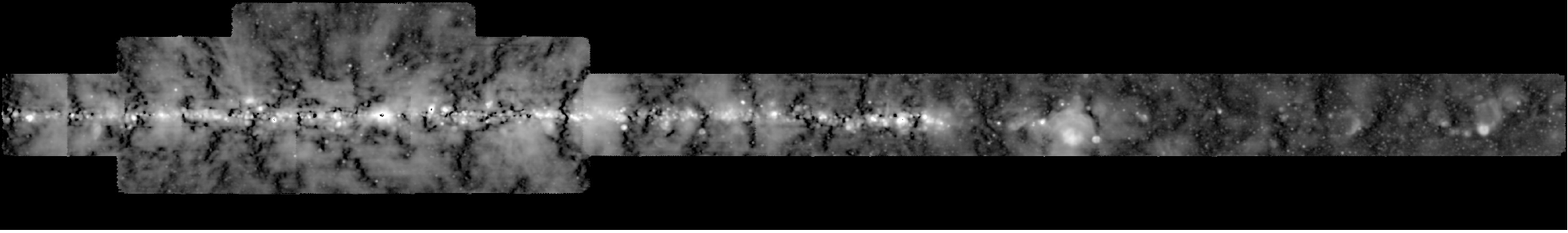}}
    \noalign{\vspace{1mm}}
    \noalign{\includegraphics[width=\textwidth]{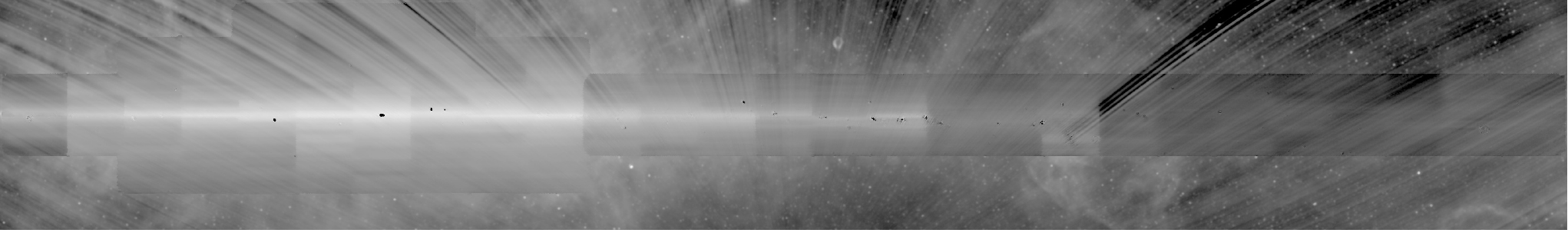}}
    \noalign{\vspace{1mm}}
    \noalign{\includegraphics[width=\textwidth]{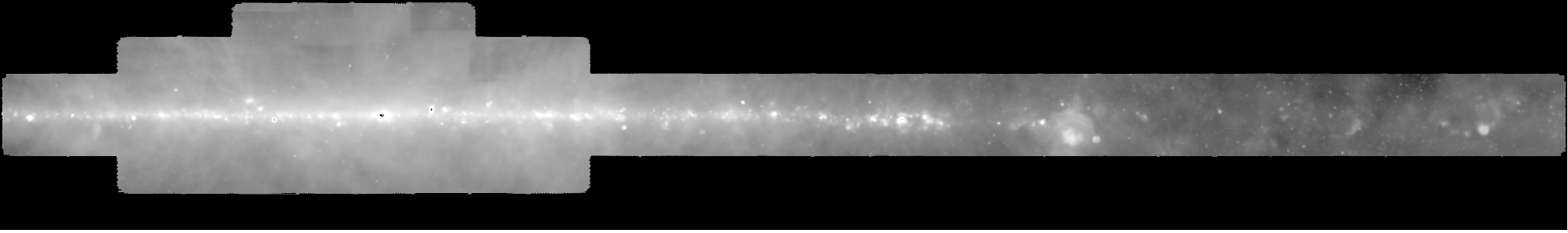}}
  \end{tabular}
  \vskip -280pt
  \begin{picture}(460,280)
    \put(370,290){\figlab{a: HIPASS}}
    \put(370,214){\figlab{b: ZOA$_0$}}
    \put(370,138){\figlab{c: HIPASS $-$ ZOA$_0$}}
    \put(370, 62){\figlab{d: ZOA}}
  \end{picture}
  \caption[]{Graphical depiction of the first iteration of ZOA zoan level
    determination based on HIPASS.  In each case the data are gridded onto a
    plate carr\'{e}e projection in Galactic coordinates, $+53\degr \ge \ell
    \ge -165\degr$, $|b| \le 16\degr$.  The inputs are (a) HIPASS zipped using
    levels deduced from the declination zone overlaps, and (b) ZOA without
    levels, analogous to Figure \ref{fig:ElCorr}.  The difference between
    these, (c), is used to derive the levels for ZOA which when applied yield
    (d).  In practice this is done separately for each ZOA zoan, with the
    levels derived by averaging in Galactic longitude.  Logarithmic greyscale
    as for previous figures.}
  \label{fig:CZOAZIP}
\end{figure*}

%================================================= Computing the HIPASS levels

\subsubsection{Computing the HIPASS levels}
\label{sec:HIPASS_levels}

The level for each of 5400 RAs in each of the 15 declination zones was then
computed by summing the core offsets from zone $-87\degr$.

Negative levels of up to $-133$\,mJy/beam then resulted in zone $-82\degr$
over an extended range of right ascension.  These could not be attributed to
noise alone, especially when, as previously explained, the offsets between
these two zones are better determined than elsewhere.  Also, there is no
reason to suppose that the minimum in the continuum emission occurs at the SCP
or anywhere near it.  Consequently, the offsets in zone $-87\degr$ were
adjusted upwards uniformly by 133\,mJy/beam which has the effect of increasing
all levels uniformly.

The levels were then smoothed lightly using a Gaussian of FWHM
$4 \sec\delta\sub{z}$ image columns, or approximately $16\arcmin$.
Considering that zone levels arise from emission that is significantly
extended with respect to the $8\fdg5$ scan length, much heavier smoothing in
RA would be justified on the basis that there are no long, narrow sources of
continuum emission that are oriented north-south on the sky.  However, that
was unnecessary and probably would have been counterproductive at this stage.

Determination of the zone levels by integrating the offsets amounts to a
simple one-dimensional random walk, the variance of the level at any point
being equal to the sum of the variances of the steps preceding it.  Noting
that the values of $S_n$ computed for each zone above were for the dispersion
of the data used to compute the offsets, the equivalent of the standard error
in the mean should be obtained by dividing by $\sqrt{n}$.  However, we have
used the median which is less efficient than the mean, but on the other hand,
aggregated image columns spanning $20\arcmin$, though they did not provide
independent samples.  Taking the overlap width divided by the HPBW, $n = 3$
(or 6), as a lower limit for $n$, and omitting the $-87\degr$ zone which
defines an arbitrary zero level, the $1 \sigma$ uncertainty in the levels in
each zone should be less than 0, 10, 18, 25, 34, 40, 45, 49, 53, 58, 62, 67,
71, 75, and 80\,mJy/beam.  These provide target figures for the potential
accuracy of the zipping operation.  However, as indicated previously, $S_n$ is
abnormally high for about 11\% of the measured offsets, mainly for those scans
that traversed the Galactic plane.  Thus the uncertainty in the levels
computed from these offsets can be much higher and this is what gives rise to
{\em streamers}.

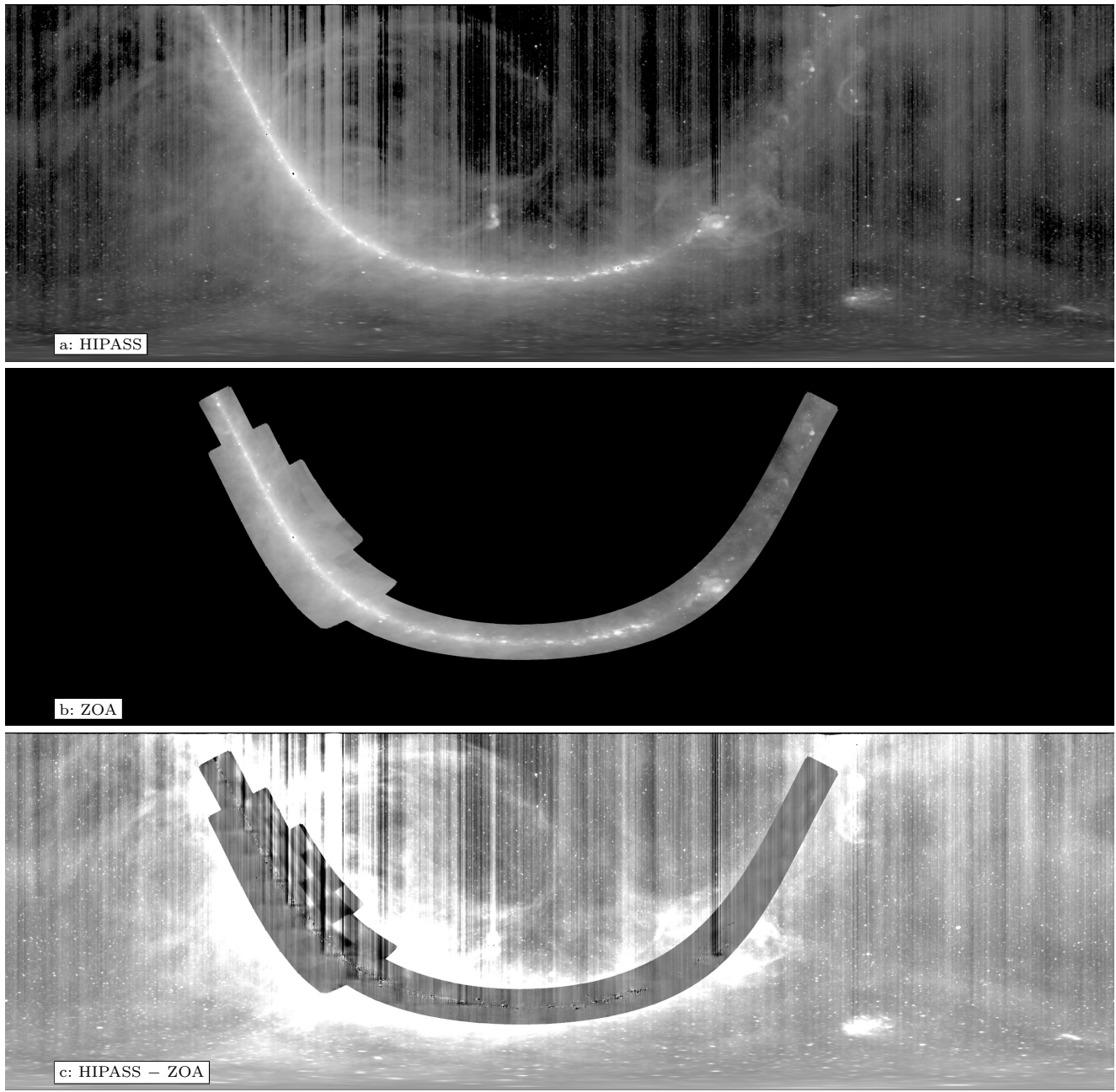
\begin{figure*}[ht!]
  \centering
  \begin{tabular}{l}
    \noalign{\includegraphics[width=\textwidth]{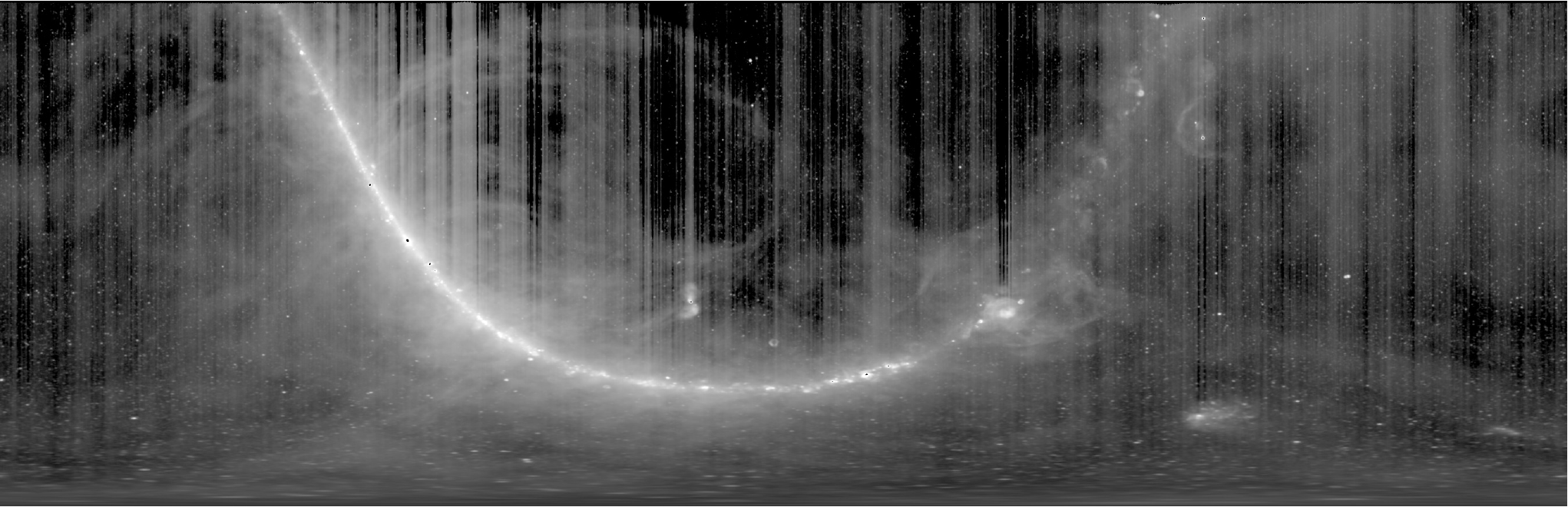}}
    \noalign{\vspace{1mm}}
    \noalign{\includegraphics[width=\textwidth]{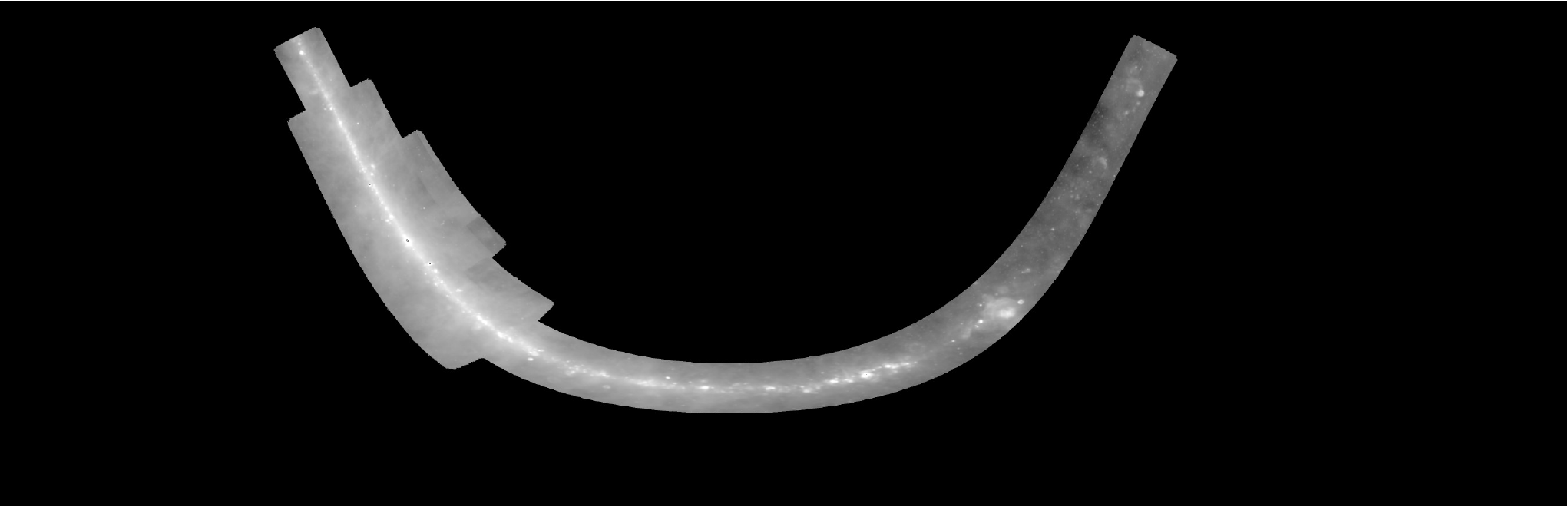}}
    \noalign{\vspace{1mm}}
    \noalign{\includegraphics[width=\textwidth]{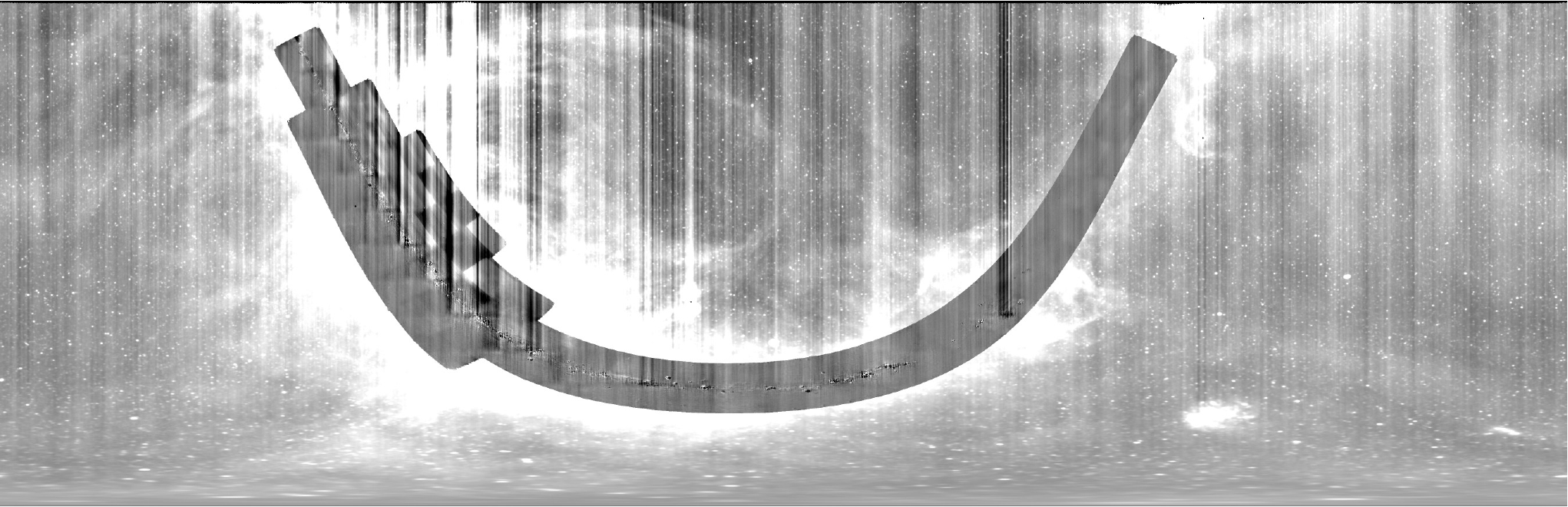}}
  \end{tabular}
  \vskip -450pt
  \begin{picture}(460,450)
    \put(3,334){\figlab{a: HIPASS}}
    \put(3,170){\figlab{b: ZOA}}
    \put(3,  7){\figlab{c: HIPASS $-$ ZOA}}
  \end{picture}
  \caption[]{Graphical depiction of the first iteration of HIPASS zone level
    correction based on ZOA.  In each case the data are gridded onto a plate
    carr\'{e}e projection but this time in equatorial coordinates, $24\hr \ge
    \alpha \ge 0\hr$, $-90\degr \le \delta \le +26\degr$.  The inputs are (a)
    HIPASS zipped using levels deduced from the declination zone overlaps, as
    per Figure \ref{fig:CZOAZIP}a, and (b) ZOA with levels deduced as per
    Figure \ref{fig:CZOAZIP}d.  The difference between these, (c), is used to
    derive level corrections for HIPASS.  In practice this is done separately
    for each HIPASS zone, with the correction derived by averaging in
    declination.  Logarithmic greyscale as for previous figures except for (c)
    which is a linear scale from -1 to +1\,Jy/beam.}
  \label{fig:CHIZIP}
\end{figure*}

The zone levels obtained are shown in Figure \ref{fig:chipassLevs} with
obvious streamers extending northwards from the Galactic plane, particularly
in the vicinity of the Galactic centre.  The result of applying them when
gridding may be seen in Figure \ref{fig:CHIZIP}a.  This is essentially what
would be obtained by adding Figure \ref{fig:chipassLevs} to Figure
\ref{fig:ElCorr} (on the same projection), the zone level correction being
additive.  Streamers are clearly evident.  The next step is to refine these
levels by crossing HIPASS with ZOA.

%----------------------------------------------------------------- floating --

% /DATA/GRUS_1/CHIPASS/median8/-
%   CHIZIP1: CHIPASS_CARZ/pass-1/CHIPASS_CARZ_WGTMED.continuum.fits
%   CHIZIP2: CZOA_HIPASS/pass-1/CZOA_HIPASS_WGTMED.continuum.fits
%   CHIZIP3: generated by kvis
%
% kvis levels 0 to 50, logarithmic 3-cycle, Greyscale1
% except for CHIZIP3, -1 to +1, linear.

%----------------------------------------------------------------- floating --

% /DATA/GRUS_1/CHIPASS/median8/-
%   H-Z?: CHIPASS_ZOA/pass-?/CHIPASS_ZOA_WGTMED.continuum.fits
%         - CZOA_CARZ/pass-?/CZOA_CARZ_WGTMED.continuum.fits
%         (generated by kvis)
%
% kvis levels 0 to 50, logarithmic 3-cycle, Greyscale1.
\begin{figure*}[t]
  \centering
  \begin{tabular}{l}
    \noalign{\includegraphics[width=\textwidth]{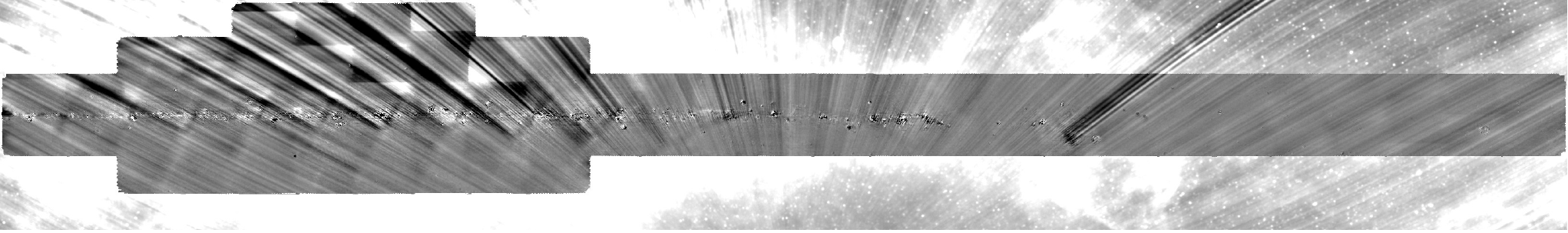}}
    \noalign{\vspace{1mm}}
    \noalign{\includegraphics[width=\textwidth]{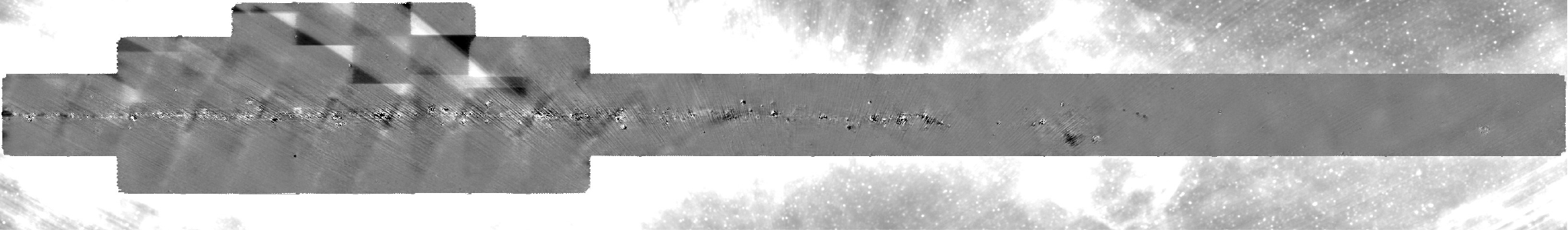}}
    \noalign{\vspace{1mm}}
    \noalign{\includegraphics[width=\textwidth]{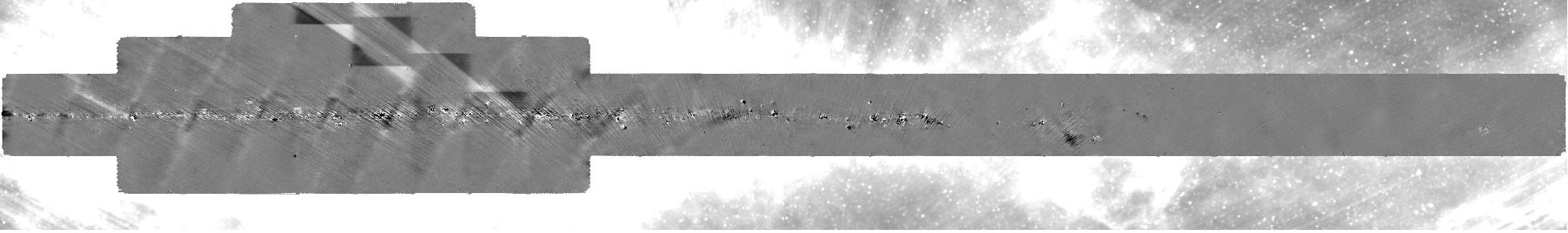}}
    \noalign{\vspace{1mm}}
    \noalign{\includegraphics[width=\textwidth]{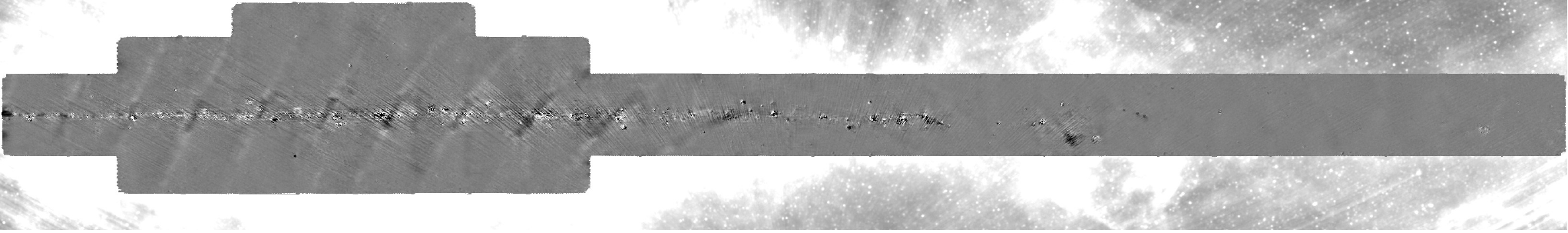}}
    \noalign{\vspace{1mm}}
    \noalign{\includegraphics[width=\textwidth]{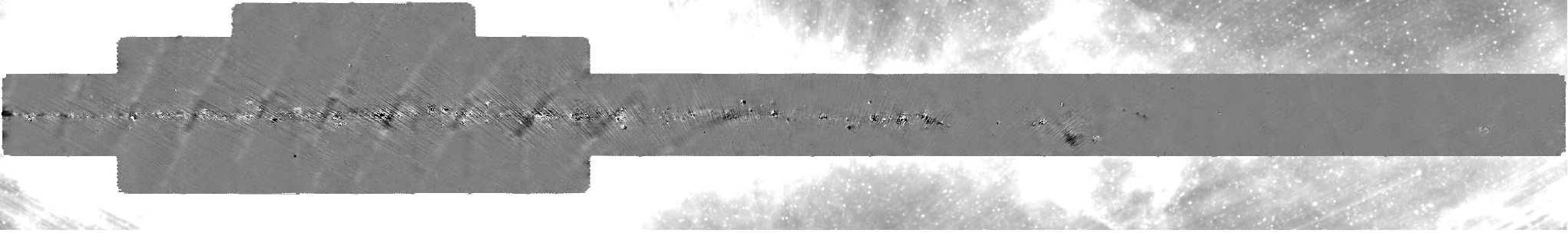}}
  \end{tabular}
  \vskip -350pt
  \begin{picture}(460,350)
    % \put(0,0){\line(1,0){460}}
    % \put(0,0){\line(0,1){350}}
    \put(3,367){\figlab{a}}
    \put(3,292){\figlab{b}}
    \put(3,216){\figlab{c}}
    \put(3,141){\figlab{d}}
    \put(3, 65){\figlab{e}}
  \end{picture}
  \caption[]{Progress of the iteration depicted graphically -- HIPASS minus
    ZOA after (a) zero, (b) one, (c) two, (d) four, and (e) ten passes of
    refining the HIPASS levels.  (Figure (a) is equivalent to Figure
    \ref{fig:CHIZIP}c.)  Most of the narrow streamers have disappeared after a
    single pass leaving only intransigents that emanate from the Galactic
    centre region.  These are slowly eaten away in subsequent passes.  The
    subtle saw-tooth residual, which appears below the Galactic centre region,
    becomes more noticable as the streamers are corrected, but is gone in the
    final pass with special handling.  Same projection as Figure
    \ref{fig:CZOAZIP}.  Linear greyscale from -1 to +1\,Jy/beam.}
  \label{fig:iters}
\end{figure*}

%==================================================== Computing the ZOA levels

\subsubsection{Crossing HIPASS and ZOA: (a) computing the ZOA levels}
\label{sec:ZOA_levels}

After bandpass calibration, ZOA scans have an arbitrary offset just as do the
HIPASS scans.  The first step in improving the HIPASS zone levels is to derive
these ZOA zoan levels.  Figure \ref{fig:CZOAZIP} illustrates that process
graphically.  Of course, the zoan levels are also needed so that the large ZOA
data set can be included in the final map, substantially reducing the noise in
the interesting Galactic plane region.

For each ZOA zoan, zone-adjusted HIPASS data were gridded onto a plate
carr\'{e}e projection in Galactic coordinates, dimensioned so that it just
covered the zoan.  ZOA data was also gridded onto an identical projection.
The image width, which is the critical dimension, was 129 pixels spaced by
$4\arcmin$.  The difference between these two maps, HIPASS minus ZOA, should
be constant in Galactic longitude, though varying in Galactic latitude.  This
is the zoan level, i.e.\ the offset that was lost during bandpass calibration.

As in Section \ref{sec:zoneOverlaps}, iterative gridding (Section
\ref{sec:iterative}) was not used in order to avoid edge effects.  However,
the whole procedure was iterated as described in Section \ref{sec:iteration}.

In practice, because of errors in the HIPASS zone levels, the difference maps
are not quite constant in Galactic longitude, particularly in being affected
by residual streamers, as seen in Figure \ref{fig:CZOAZIP}c.  However, as the
streamers cross at an angle to the parallels of Galactic latitude, mostly they
only affect a part of the scan in Galactic longitude from which the zoan level
is determined.

Unlike the HIPASS levels, which must be derived by summing offsets, the ZOA
levels are obtained directly.  As for the HIPASS zone offsets, they are
computed as the median value across the scan in the difference map (Figure
\ref{fig:CZOAZIP}c).   In contrast to HIPASS, there is a full $8\fdg5$ of
overlap to work with, rather than $45\arcmin$, and smoothing of the levels was
not warranted.  However, censoring of outliers was performed as before.  That
is, the zoan levels were subtracted from the difference map to produce a
residual map.  Pixels outside $3 \times S_n$ in this map were deemed to be
outliers, where $S_n$ was computed from the map as a whole, and the zoan
levels then recomputed.  Visual inspection showed that this was particularly
effective at removing outliers associated with areas of strong emission that
did not cancel due to small relative errors between the HIPASS and ZOA maps.

Figure \ref{fig:CZOAZIP}d was produced from zoan-adjusted ZOA data, a
remarkable development of the uncorrected map seen above it.  It bears a close
resemblance to the first-pass HIPASS map of the Galactic plane in Figure
\ref{fig:CZOAZIP}a, but much less affected by streamers.  This suggests that
it may be used for improving the HIPASS zone levels.

%================================================== Refining the HIPASS levels

\subsubsection{Crossing HIPASS and ZOA: (b) refining the HIPASS levels}
\label{sec:HIPASS_levelz}

Differencing Figures \ref{fig:CZOAZIP}a and \ref{fig:CZOAZIP}d produces a map,
seen in Figure \ref{fig:iters}a, that consists mostly of residual artefacts --
streamers -- in the HIPASS map, together with strong sources that don't quite
cancel.  However, Galactic coordinates are not the best choice because the
streamers are oriented at different angles depending on Galactic longitude.
In equatorial coordinates on a cylindrical projection the residual streamers
all run north-south, as seen in Figure \ref{fig:CHIZIP}a.  When the
zoan-corrected ZOA data is gridded onto an identical projection, and the
difference taken, the residual streamers manifest themselves as vertical
stripes as seen in Figure \ref{fig:CHIZIP}c, and as such are amenable to
measurement and subsequent correction.

The procedure was similar to the forgoing.  Maps on a plate carr\'{e}e
projection, this time in equatorial coordinates, were produced separately from
HIPASS and ZOA data for each HIPASS zone that intersects the ZOA, namely
$-66\degr$ and northwards.  These are long thin strips, $5401 \times 161$
pixels, of which the ZOA data only fills a portion.  A correction to the
levels was obtained as the median value across the scan in declination, with
censoring of outliers applied as before.  Again, visual inspection confirmed
the validity of the censoring.  A correction was computed if, after censoring,
at least five pixels remained in the declination strip.

At this point we have a set of corrections to the zone levels.  In fact, what
is required is a correction to the {\em core set} of offsets that were derived
in Section \ref{sec:zoneOverlaps}.  While there was only one place to apply
the 133\,mJy/beam offset introduced to ensure non-negativity in zone
$-82\degr$, namely in the zone $-87\degr$ offsets (Section
\ref{sec:HIPASS_levels}), for more northerly zones the residual could be
applied to any of the zones further south.  In practice, levels were adjusted
by distributing the residual amongst the offsets in zones south of the level
being corrected on a pro-rata basis according to the variance ($S_n^2$),
excluding zones $-87\degr$ and $-82\degr$.  This tended to place almost all of
the correction in the Galactic plane.

This process provided corrections to 13.8\% of the HIPASS levels.  Wherever no
correction was obtained from ZOA, the offsets were replaced with the
corresponding offset from the core set.  In particular this included all
offsets to the east, north, and west of the ZOA survey area.

%=================================================== Iteration and convergence

\subsubsection{Crossing HIPASS and ZOA: (c) iteration and convergence}
\label{sec:iteration}

After refining the HIPASS levels it seems obvious that they could be used to
produce a better estimate of the ZOA levels.  Accordingly, the procedure
described in Section \ref{sec:ZOA_levels} was repeated.  In turn, the new
estimate of the ZOA levels could then be used to improve the HIPASS levels.
In effect such an iterative process exploits the fact that HIPASS scans may
form a bridge between adjacent ZOA zoans from which the relative zoan levels
can be determined.  Likewise, the ZOA scans may form a bridge between adjacent
HIPASS scans.  However, unlike the HIPASS zone overlaps where the overlapping
scans are parallel to each other, these bridges cross at an angle.  Thus,
instead of providing the offset between a {\em pair of scans} on opposite
sides of a discontinuity, they provide the offset between {\em sets of scans}
on either side.  Complicated as the geometry may be, it seems reasonable to
expect that convergence should be obtainable.  That is, we expect that there
is a set of HIPASS and ZOA levels for which the difference between the HIPASS
and ZOA maps is zero plus noise, and we anticipate that iteration will find
them.  However, there are complications.

In successive iterations, as the HIPASS levels are expected to be better
determined, the streamers should reduce in size and generally have less of an
effect on the determination of the ZOA levels.  Indeed, very few streamers
survived even a single iteration.  However, it was found that if the streamers
were wide enough on the first pass, and the angle acute enough, then they
could cause the iteration to fail to converge over part of the map.  The
problem is evident in sections of the ZOA north of the Galactic centre region
in Figure \ref{fig:iters}.

The characteristic blocky structure seen in Figure \ref{fig:iters}b indicates
that wide streamers have corrupted the zoan levels over a range of Galactic
latitude in adjacent zoans.  In fact, this can be seen directly in Figure
\ref{fig:CZOAZIP}d, while close inspection of Figure \ref{fig:iters}b reveals
that all zoans between $344\degr$ and $48\degr$ were affected to some degree.
These {\em intransigents} persist through successive passes, indicating that
they form part of a stable, yet clearly incorrect solution of the iteration.
The cure, applied in the second and third passes, was simply to identify and
excise the affect ranges of Galactic latitude in each zoan of the ZOA maps
from the next pass of the iteration.  The bad latitudes are those associated
with the dark blocks in Figure \ref{fig:iters}b.  The bright corners are so
because they correctly identify the positive excess due to the positive
streamer (in this case).  Effectively these bright corners eat into the
streamer on the next pass, eliminating the smaller intransigents, and reducing
the width of the larger ones in Figure \ref{fig:iters}c.  After four passes
the intransigents have been completely eaten away in Figure \ref{fig:iters}d.

The HIPASS maps produced in each pass always applied the best estimate of the
levels available, namely those obtained from the previous pass.  Thus
subsequent iterations only provided {\em corrections} to the levels from the
previous pass.  As mentioned in Section \ref{sec:zoneOverlaps}, iterating
corrects for the subtle effect that the levels computed from the gridded maps
are necessarily a smoothed version of the levels that must be applied to the
ungridded data.  However, until the intransigent streamers had been completely
removed, the ZOA maps used to compute the zoan levels (Section
\ref{sec:ZOA_levels}) were all produced without levels.  That is, the
uncorrected map shown in Figure \ref{fig:CZOAZIP}b was used each time.  This
provided a fixed reference point from which the convergence behaviour could be
assessed.  The remaining iterations were then done with levels applied.

As the difference between the HIPASS and ZOA maps should consist solely of
noise together with a few outliers associated with strong sources, Figure
\ref{fig:iters} shows that, when present, the effect of streamers or other
defects is readily apparent on visual inspection.  Thus the convergence
behaviour could be monitored qualitatively at a glance.  After dealing with
the intransigent streamers, a new type of artefact assumed prominence albeit
at a lower level.  It appears in Figure \ref{fig:iters}d below the Galactic
centre region as a saw-tooth pattern with an amplitude of about 100\,mJy/beam.
This pattern arises from a different mechanism to the intransigent streamers
and disappears slowly in successive iterations.  However, after eight passes
it was apparent that the convergence was slow enough to warrant investigation
of a more direct solution.

Thus far we have not considered the overlap between adjacent ZOA zoans.
Although they offer an independent mechanism for tying together the ZOA zoans,
and hence the HIPASS zones, these overlaps are problematic for two reasons.
Firstly, unlike HIPASS where the scans all converge on the SCP, ZOA scans when
tied together have no common point so there is no way to relate scans in
neighbouring Galactic latitudes.  Secondly, strong sources within a degree or
two of the Galactic plane render the determination of offsets very uncertain,
with the formation of prominent streamers in Galactic longitude akin to those
in declination in the HIPASS levels.  However, on further investigation it was
apparent that the saw-tooth pattern was generated by steps in the levels
between adjacent zoans.  Only one value is needed per zoan to eliminate the
pattern, and there is sufficient information in the ZOA overlaps to measure
these steps directly.

Although a faint pattern is also visible in the Galactic anti-centre region,
it seemed best to allow iteration to deal with that naturally.  Thus we
focus on the Galactic bulge region, measuring the offsets for each latitude
only in the extensions where $|b| > 6\degr$, well away from bright sources in
the plane.  These measurements were averaged into one single offset for each
pair of zoans from $336\degr$ to $32\degr$.  The offsets were then converted
to zoan level corrections as in Section \ref{sec:HIPASS_levels}, and
renormalised to zero for zoan $336\degr$ and the zoans preceding it.
Corrections for the $48\degr$ and $40\degr$ zoans were set equal to that for
the $32\degr$ zoan.  Thus only seven independent corrections for the ZOA zoan
levels were computed, amounting to 300\,mJy/beam from zoan $336\degr$ to zoan
$32\degr$.  Once applied, the saw-tooth disappeared quickly; on the next
iteration the difference in the levels had dropped by a factor of $\times 15$,
with a further factor of $\times 3$ in the iteration after that, reducing it
to a negligible value.

Iteration continued until there was no substantive change in the difference
maps shown in Figure \ref{fig:iters}.  Convergence was considered to have been
achieved after ten passes of refining the HIPASS levels and one further round
of computing the ZOA levels from them.  Bright source ghosts form the most
prominent residual artefacts visible in Figure \ref{fig:iters}e.  These are
expected as previously explained.  Fine-scale HIPASS streamer noise, the
leftovers of imperfect streamer removal, can be detected over most of the
difference map.  Away from the Galactic centre it has a peak height of about
50\,mJy/beam, but may become more prominent where it crosses bright extended
sources.  There is no evidence of any residual from the ZOA levels.

HIPASS zone boundaries are also visible in the difference map in the Galactic
centre region, mostly a positive excess of up to 100\,mJy/beam but
systematically going negative where the boundary crosses the plane.  This
behaviour seems to be confined to regions of very strong extended emission.
It was not removed by iterative gridding and currently its cause is not
understood.  Again, following treatment of the saw-tooth residual, there is
almost no evidence of the ZOA zoan boundaries.  In areas well away from
residual artefacts the rms of the difference map is 20\,mJy/beam.  Assigning
the relative contribution to this from HIPASS and ZOA in the ratio
$\sqrt{5}:1$, the rms of the component maps would be 18\,mJy/beam and
8\,mJy/beam as best case noise figures.

As previously explained, the error in the levels determined from the HIPASS
zone overlaps in Section \ref{sec:HIPASS_levels} is expected to be highest in
the $+22\degr$ zone, and this was confirmed, with corrections in the range
$\pm 1.5$\,Jy/beam, an rms of 320\,mJy/beam, and a mean value consistent with
zero.

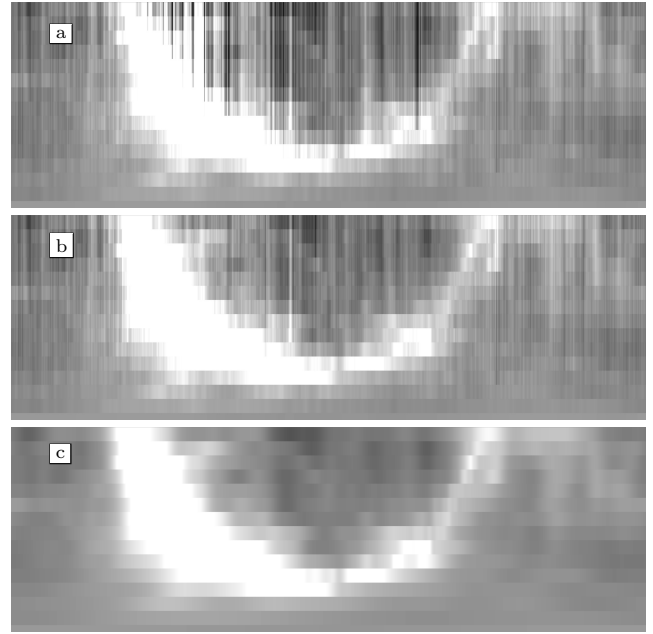
\begin{figure}[t]
  \centering
  \begin{tabular}{l}
    \noalign{\includegraphics[width=\columnwidth]{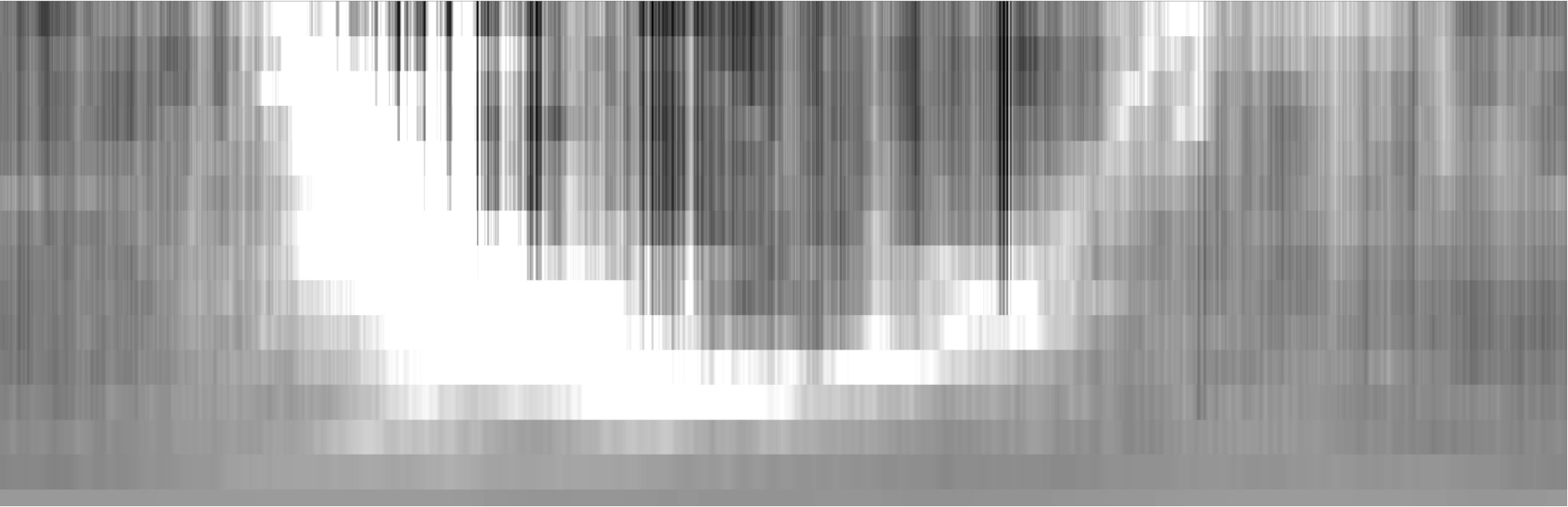}}
    \noalign{\vspace{1mm}}
    \noalign{\includegraphics[width=\columnwidth]{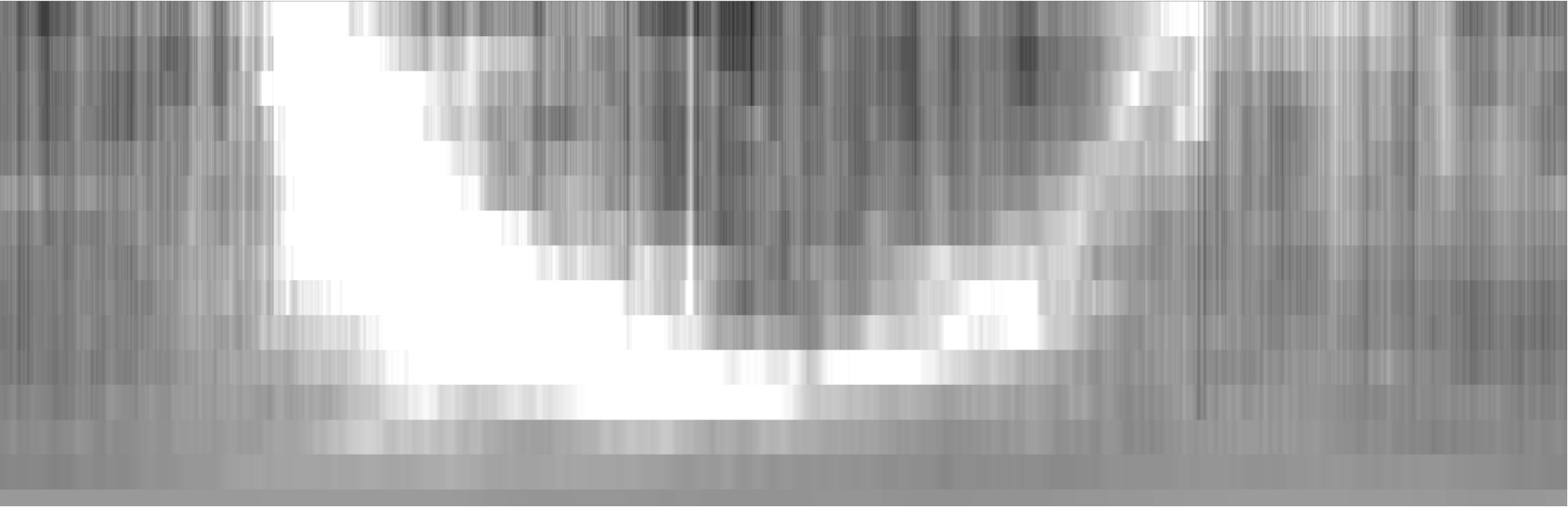}}
    \noalign{\vspace{1mm}}
    \noalign{\includegraphics[width=\columnwidth]{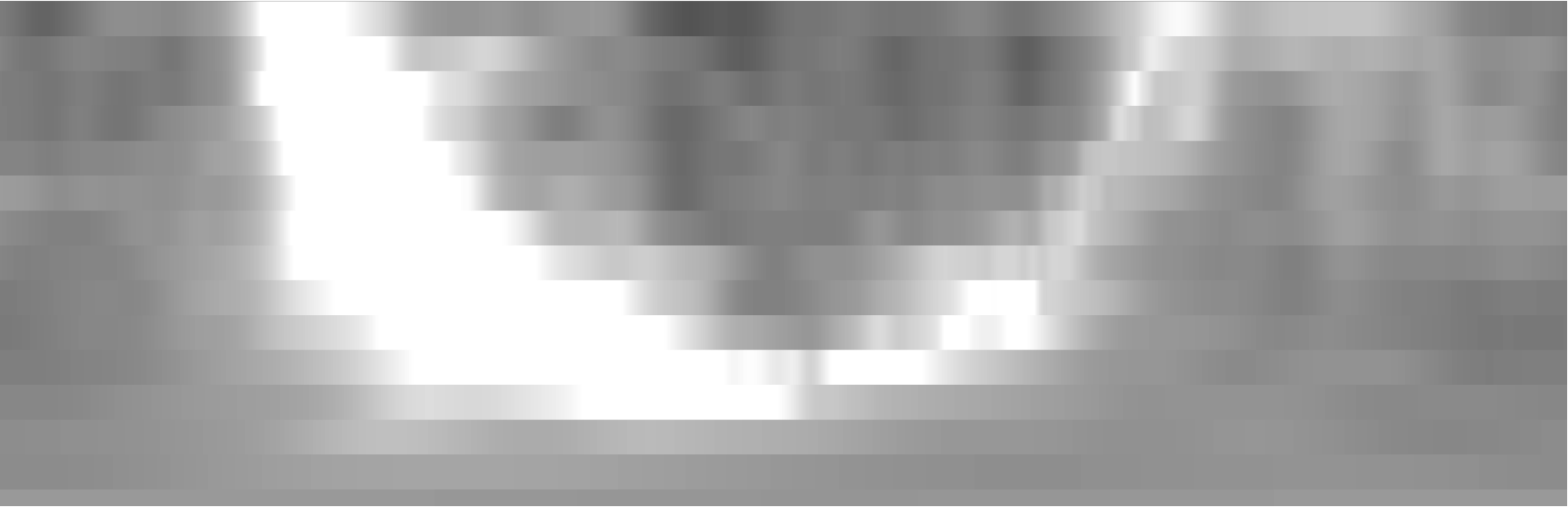}}
  \end{tabular}
  \vskip -217pt
  \begin{picture}(217,217)
    % \put(0,0){\line(1,0){217}}
    % \put(0,0){\line(0,1){217}}
    \put(3,226){\figlab{a}}
    \put(3,145){\figlab{b}}
    \put(3, 67){\figlab{c}}
  \end{picture}
  \caption[]{HIPASS levels (a) before correction (same as Figure
    \ref{fig:chipassLevs}), (b) after correction by crossing with ZOA, and (c)
    the final result after conditioning.  Strong Galactic streamers have all
    disappeared in (b) leaving the prominent Cen\,A streamer visible just left
    of centre, together with low-level streamer noise.  Same projection and
    linear greyscale as Figure \ref{fig:chipassLevs}.}
  \label{fig:chipassLevz}
\end{figure}

Figure \ref{fig:chipassLevz} presents the uncorrected HIPASS levels obtained
from the declination zone overlaps together with those resulting from ten
passes of crossing HIPASS and ZOA.  The worst of the streamers originating in
the Galactic plane have disappeared from Figure \ref{fig:chipassLevz}b, but
some prominent streamers remain, clearly originating from strong
extra-Galactic continuum sources, notably Cen\,A.  The final step is to fix
these individually and then smooth the levels.

%----------------------------------------------------------------- floating --

% /DATA/GRUS_1/CHIPASS/median8/CHIPASS_Overlaps/pass-6/-
%   chipass_levels.fits
% /DATA/GRUS_1/CHIPASS/median8/-
%   CZOAxHIPASS_EQU/pass-10/chipass_levelz.fits
%   CZOAxHIPASS_EQU/pass-10/chipass_smooth.fits
%
% kvis levels -1 to 1, linear, Greyscale1

%============================================== Conditioning the HIPASS levels

\subsubsection{Conditioning the HIPASS levels}
\label{sec:conditioning}

Figure \ref{fig:chipassLevz}b shows the HIPASS levels obtained after
correction by crossing HIPASS and ZOA.  Residual streamer noise is visible
with peaks to about 100\,mJy/beam, though with several stronger.  The source of
five of these was readily identified, including Cen\,A, and it was verified
that the offset computed from the relevant HIPASS zone overlaps in Section
\ref{sec:HIPASS_levels} had a high value of $S_n$ due to these sources.  As
they lie outside the ZOA survey, none of them were amenable to correction by
the previous methods.  However, as these streamers have an obvious source, it
was straightforward to apply an offset that would make them disappear into the
background streamer noise.  For Cen\,A this required an adjustment of
-550\,mJy/beam (Gaussian peak) over 19 levels (FWHM) spaced by 16s in RA; for
3C273, +300\,mJy/beam over 9 levels; for 30\,Doradus with two separate
streamers, +200\,mJy/beam and +140\,mJy/beam both over 10 levels; for PKS
J1935-4620, -300\,mJy/beam over 12 levels; and for PKS J0303-6211,
+130\,mJy/beam over 10 levels.  The correction was applied entirely to the
offset between the zone containing the source and its neighbour to the north.

After dealing with these outliers, the remaining streamers appear to have a
noiselike distribution, with as many positive as negative and no obvious
correlation between them.  As explained previously, the process of computing
HIPASS levels amounts to a simple one-dimensional random walk for which the
variance at any point is equal to the sum of the variances of all steps
leading to it.  Hence the streamer noise is expected to be lower in the south
and higher in the north, as is observed qualitatively.  Indeed, all of the
processing to this point has carefully treated positive and negative streamers
equally so as not to introduce biasses that might disturb this random
distribution.

The final step then was to smooth the levels in right ascension.  Previous
processing only used very light smoothing, but considering that zone levels
arise from emission that is significantly extended with respect to the
$8\fdg5$ scan length, much heavier smoothing in RA is justified on the basis
that there are no long, narrow sources of continuum emission that are oriented
north-south on the sky.  Bear in mind that only the levels are smoothed, not
the final map.  The question then arises of the maximum valid smoothing
length, and what length is sufficient to remove the streamers.

The ZOA levels were used for guidance in assessing a legitimate smoothing
length.  The strong peak that occurs right on the Galactic plane was found to
be the only narrow feature.  The zoan levels on either side of it change
gradually, with only a few, relatively weak features with a scale length in
the range from $1\degr$ to $2\degr$.  It was also apparent that the ZOA levels
are well determined and do not require smoothing.  Considering that the HIPASS
sampling rate is overwhelmed by a factor of $\times 5$ in the region of the
ZOA survey, the choice of smoothing length for the HIPASS levels is not a
significant concern in this region.  Moreover, in light of the results of
Section \ref{sec:zonelevs} in using ZOA to eliminate HIPASS streamers,
smoothing of the HIPASS levels within the ZOA survey area should be kept to a
minimum.  In practice it was only done to prevent discontinuities on the
boundary of the survey.

Several smoothing strategies were investigated, the simplest being to smooth
the levels in right ascension.  In fact, this is the least effective method
because it does nothing to stop streamer formation.  By smoothing the offsets,
streamers are attenuated close to their source, before they can propagate
northwards.  However, unless the ZOA region is smoothed identically, offsets
do not balance between zones northwards of the Galactic plane, leading to the
formation of streamers.  Thus the most effective smoothing method was found to
be a combination of the two approaches.  Starting from the south, the levels
for one zone are smoothed, then the levels for the next zone (outside the ZOA)
are recomputed from these smoothed levels and the original offsets.  The next
zone is then smoothed in turn.  This strategy attacks streamers at their
source, and also has the benefit of smoothing a quantity that will appear
directly as emission on the final map.

In practice an empirical approach was used to determine the smoothing length.
Essentially the levels were put through a high-pass Gaussian filter and the
smoothing length adjusted to best effect, giving little weight to the ZOA
survey area.  Starting with a short smoothing length, only streamer-like noise
made it through the filter -- long filaments oriented north-south with no
obvious structure that could be related to any celestial emission.  As the
smoothing length was increased, shorter and wider features appeared that could
be related to genuine cosmic emission.  A length of $4\degr$ FWHM was found to
be the best compromise for streamer removal and retention of genuine emission
in regions outside the ZOA.  The rms of the high frequency streamer components
was about 100\,mJy/beam in the $+22\degr$ zone being the worst case.  Having
determined the appropriate filter length, the levels were then smoothed using
a low-pass Gaussian filter of the same filter length.  Within the ZOA, a
length of $1\degr$ was sufficient to avoid discontinuities at the edges of the
survey area.

As can be appreciated from Figure \ref{fig:CHIZIP}, the ZOA overlaps the
HIPASS scans to a variable extent.  Some HIPASS scans and their neighbours
have the benefit of a complete intersection, while others intersect only
partially.  In particular, only a small corner of the $200\degr$ and
$48\degr$ ZOA zoans intersect the $+22\degr$ HIPASS zone, with the $48\degr$
corner proving troublesome.  As stated in Section \ref{sec:HIPASS_levelz}, a
correction was computed if, after censoring, at least five pixels remained in
the overlap.  However, compared with peak values of around 130 pixels,
corrections based on such a small number of pixels cannot be said to be fully
within the ZOA survey area, nor be treated with the same degree of confidence.
Thus, in order to prevent discontinuities on the boundary of the ZOA survey
and to clean up nuisance streamers generated in the boundary regions, it was
found necessary to use a graduated definition of inclusion.  A level was
considered to be fully inside the ZOA if the correction was computed from a
certain minimum number of pixels, $n_0$, and fully outside if there was no
correction (in practice, less than 5).  Progressive smoothing operated within
the margins, with the FWHM of the smoothing length varied pro rata between
$1\degr$ (inside) and $4\degr$ (outside) according to the number of pixels
used to compute the correction.  Noting the gradual effect that changing $n_0$
had on the results, a value of $n_0 = 50$ pixels was determined empirically.

The final HIPASS levels are shown in Figure \ref{fig:chipassLevz}c.  At this
stage the levels ranged from -0.34 to +9.1 Jy/beam.

%================================================================= Final steps

\subsection{Final steps}
\label{sec:final}

The final map was produced by iterative gridding using a single iteration and
a loop gain of 1.2, as discussed in Section \ref{sec:iterative}.  As mentioned
in Section \ref{sec:validation}, some sources were strong enough to saturate
the receiver power detection system, including several bright HII regions in
the Galactic plane.  This was more apparent in the relative sensitivity map
(Figure \ref{fig:BeamRSS}) than in the brightness map itself.  The pixel value
in the sensitivity map is the square root of the sum of squares of the beam
weights used in producing the brightness map.  A value of $n$ is essentially
equal to $n^2$ independent boresight observations.  Pixels at, or in the
neighbourhood of saturated peaks appeared as shadows in the sensitivity map.
The brightness computed for these pixels would be affected by the absence, or
asymmetric distribution of data, though with regard to the latter it should be
noted that iterative gridding does tend to correct for this naturally.  Steps
were taken to repair these areas of the map as far as possible, and to blank
any remaining unreliable pixels.

%----------------------------------------------------------------- floating --

\begin{table}
  \caption[]{A total of 300 pixels were blanked from the following 16
    saturated sources.}
  \begin{center}
  \protect\begin{tabular}{rrrl}
    \hline\hline
    \noalign{\smallskip}
      \multicolumn{1}{c}{$\ell$} &
      \multicolumn{1}{c}{$b$}   &
      \multicolumn{1}{c}{$n$}   &
      \multicolumn{1}{c}{Source} \\
    \hline
    \noalign{\smallskip}
       -92.05 &  -1.07 &   4 & RCW38  \\
       -75.73 &  -0.35 &  11 & RCW49  \\
       -72.53 &  -0.64 &  28 & RCW53  \\
       -68.40 &  -0.53 &   9 & RCW57  \\
       -26.67 &  -0.33 &   1 & RCW104 \\
       -23.13 &  +0.03 &   2 & RCW108 \\
       -21.93 &  +0.00 &   1 & Ke44   \\
        -8.70 &  +0.70 &  16 & RCW127 \\
        -6.85 &  +0.74 &  39 & W22    \\
         0.06 &  -0.06 & 122 & Sgr\,A \\
        15.07 &  -0.73 &  26 & W38    \\
        30.77 &  -0.04 &   1 & W43    \\
    \noalign{\smallskip}
      -175.40 &  -5.80 &  20 & Tau\,A \\
      -151.00 & -19.40 &  14 & Ori\,A \\
       -50.48 & +19.42 &   4 & Cen\,A \\
       -76.20 & +74.50 &   2 & Vir\,A \\
    \noalign{\smallskip}
    \hline\hline
  \end{tabular}
  \end{center}
  \label{ta:saturated}
\end{table}

%================================================== Blanking unreliable pixels

\subsubsection{Blanking unreliable pixels}
\label{sec:blanking}

As can be seen in Figure \ref{fig:tracks}, the reliability of pixels north of
$\delta = +25\degr$ reduces as the density of the data gradually diminishes.
These pixels were blanked by imposing a cut-off in the value of the relative
sensitivity over the whole of the brightness map.  A value of 4.0 units was
found to be an effective level.  On this basis the northern boundary of the
map has been allowed to remain ragged.  This filter also blanked eight pixels
at the SCP, and 60 pixels known to be associated with saturated sources.

As explained in Section \ref{sec:validation}, data flagged by the multibeam
correlator were properly handled in this analysis.  Pixels affected by
flagging were found by producing a map representing the proportion of flagged
data that would have (but didn't) contribute to each pixel in the brightness
map, this type of map being more effective at highlighting the influence of
flagged data than the sensitivity map.  Apart from source saturation, data
might also be flagged because of strong RFI, and its effect was certainly
evident in this map.  However, the criterion set to exclude obviously
saturated pixels also reliably ignored all RFI-affected pixels.  Thus, 256
pixels (0.004\% of non-blank pixels) were found where more than 15\% of the
data was flagged.  These pixels convincingly matched all of the bright sources
suspected of being saturated.

Next, new maps were produced as before, but this time with a $5\arcmin$
cut-off radius for the gridding kernel instead of $6\arcmin$.  The density of
data in the Galactic plane (ZOA) region may be high enough even with a
$2\arcmin$ radius to be above the relative sensitivity cut-off of 4.0 units.
Reducing the cut-off radius should exclude flagged data for pixels on the
periphery of strong sources, and indeed 6 of the 256 pixels were reclaimed.
Matching pixels in the sensitivity map were updated with the reduced value
from the smaller gridding kernel.  The radius was further reduced to
$4\arcmin$, $3\arcmin$, and then $2\arcmin$ with the reclamation of a modest
2, 3, and 5 pixels respectively.  Inevitably, a remainder of 240 pixels
(additional to the original 60) were too close to strong peaks to be
recoverable.  These were blanked and the matching pixels in the sensitivity
map were zeroed.  Saturated sources and the associated number of blanked
pixels are identified in Table \ref{ta:saturated}.

%====================================================== Brightness calibration

\subsubsection{Brightness calibration}
\label{sec:fluxCal}

The raw HIPASS data is calibrated reasonably accurately for flux density, all
brightness values quoted so far are with respect to this nominal scale.  The
calibration was checked via Gaussian fits to the peak height of two sources
in the compact source map (Figure \ref{fig:Compact}): 1934-638 (14.9\,Jy), and
Hydra\,A (40.6\,Jy).  The mean of the calibration factors (1.11, 1.07) was
1.09.

The final steps required to obtain a fully calibrated flux density image are:
(a) to determine and apply a zero level; and (b) to apply the overall flux
density scale.  Non-negativity required a global offset of +310 mJy/beam, with
the minimum occuring on the northern edge of the north polar spur, close to
the NGP.  The above scale factor of $\times 1.09$ was then applied.

The final map prior to scaling is presented in Figure \ref{fig:Final} for
comparison with the previous maps.  It should be considered in conjunction
with the relative sensitivity map of Figure \ref{fig:BeamRSS}.

Conversion to absolute brightness temperature depends on the unknown zero
level and the Jy to K scale factor.  The former can be measured using
calibration horns; the latter ratio can be calculated if the beam has been
accurately measured out to distant sidelobes.  Fortunately, we can obtain both
values by comparison with the lower resolution all-sky 1420\,MHz
map\footnote{Patricia Reich (2013), private comm.} derived from the surveys of
Reich \shortcite{Reich1}, Reich \& Reich \shortcite{Reich2}, and Reich,
Testori, \& Reich \shortcite{Reich3}.  This plate carr\'{e}e map was regridded
onto a Hammer-Aitoff projection, it being necessary to use an equiareal
projection for the purpose.  The present map was first regridded onto an
equatorial plate carr\'{e}e projection; then blurred to $35\farcm4$
resolution, first in RA with a $32\farcm3$ FWHM Gaussian kernel (the number of
pixels therefore varying with declination) then in declination; and then
regridded to match the other map.  As the least-squares analysis of offset and
scale was distorted by the preponderence of data below 5\,K, the data was
sorted into 10mK bins.  A factor of 0.44\,K/(Jy/beam) with offset 3.30\,K was
obtained and used to convert the final map to units of absolute brightness
temperature.

\begin{figure}
  \centering
  \includegraphics[width=\columnwidth]{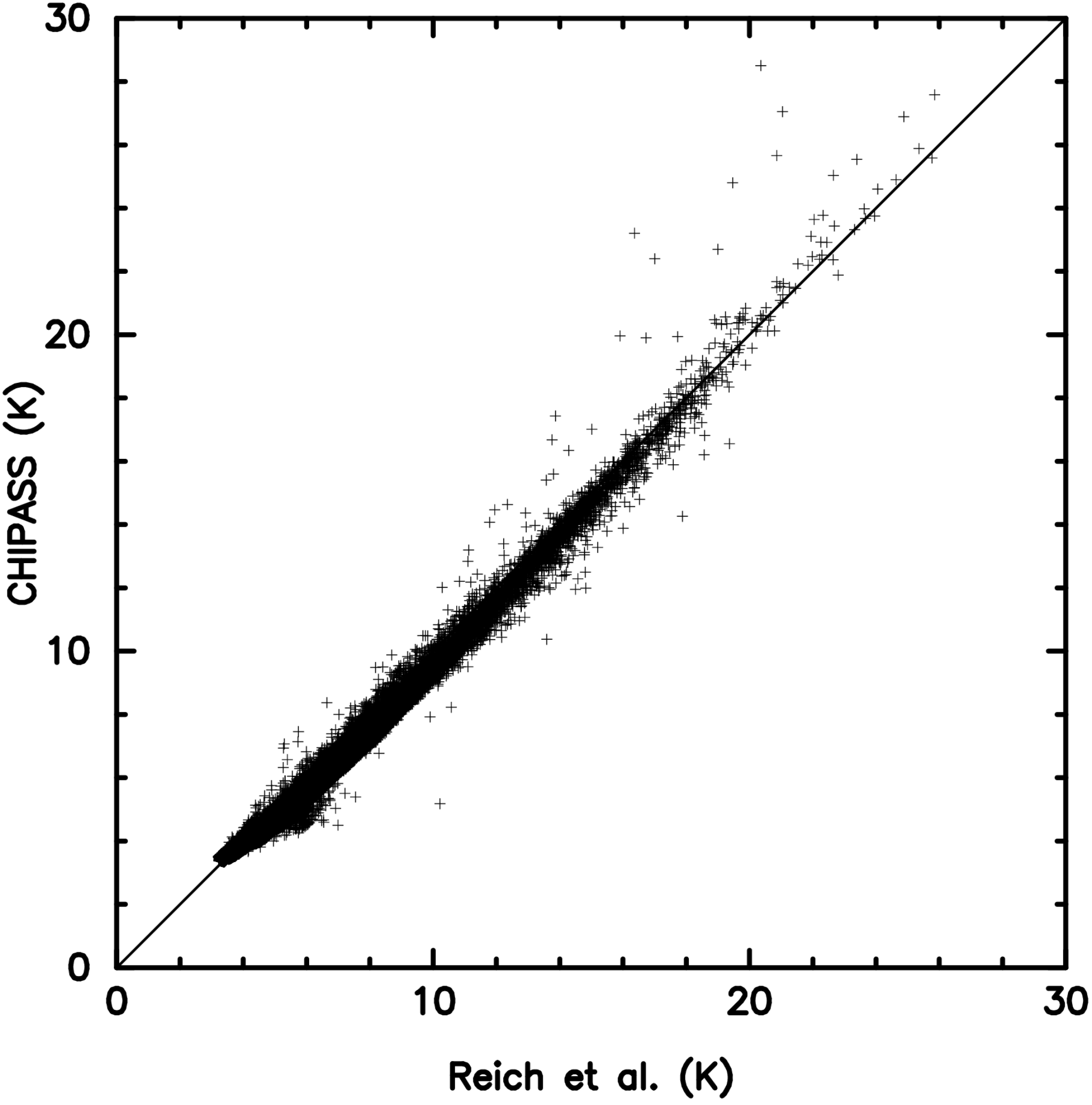}
  \caption[]{Cross-plot of the present HIPASS/ZOA map blurred to $35\farcm4$
    resolution versus the 1420\,MHz all-sky map of Reich et al.}
  \label{fig:xplot}
\end{figure}

\begin{sidewaysfigure*}
  \vspace{180mm}
  \includegraphics[width=665pt]{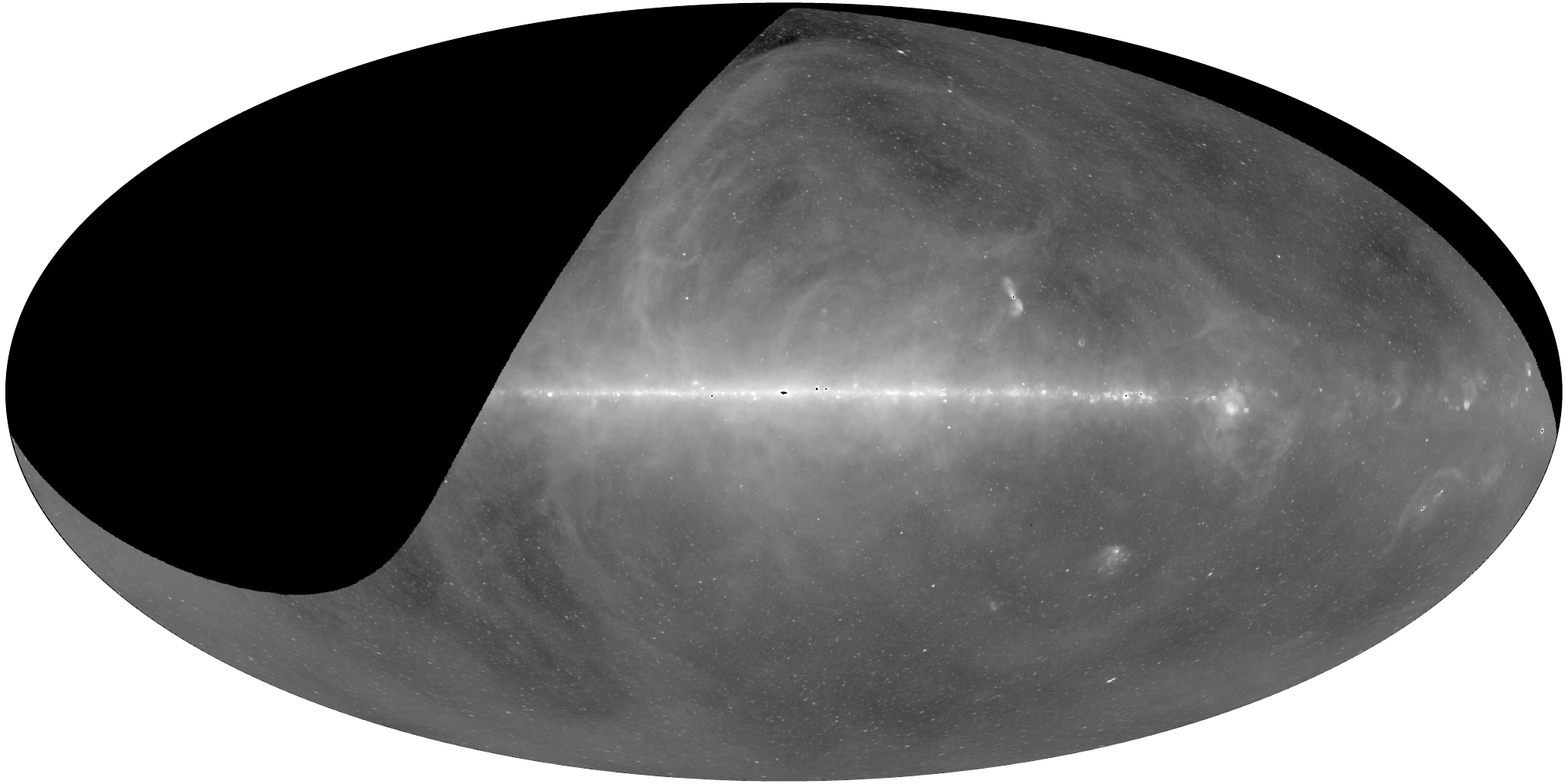}
  \caption[]{HIPASS and ZOA data combined to produce the final version of the
    1.4\,GHz ``CHIPASS'' continuum map.  This map has been rezeroed but not
    rescaled so as to be comparable with Figure \ref{fig:ElCorr}, using the
    same logarithmic greyscale.}
  \label{fig:Final}
\end{sidewaysfigure*}

\begin{figure*}
  \centering
  \includegraphics[width=460pt]{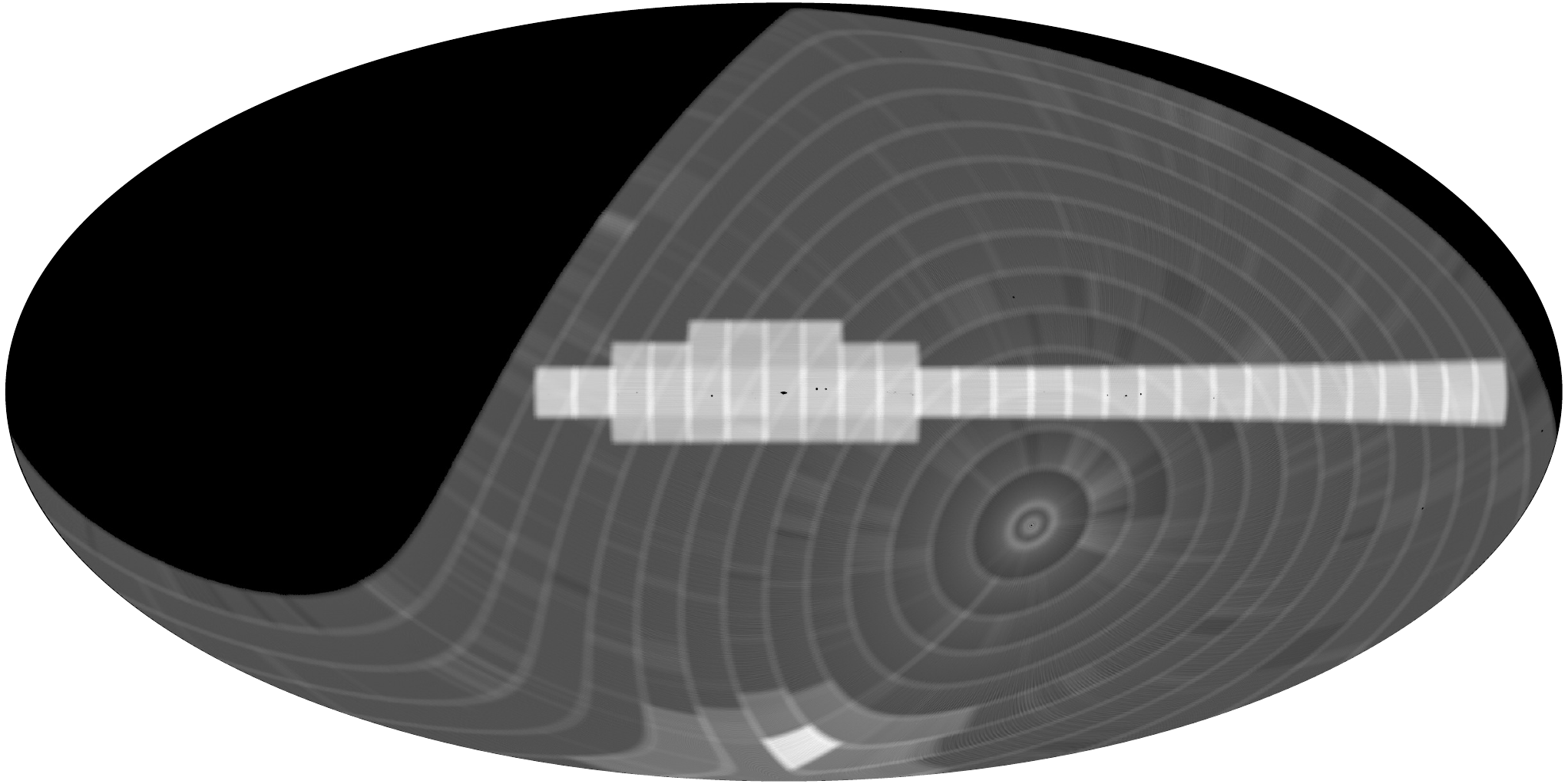}
  \caption[]{Sum in quadrature of the beam weights (BeamRSS) used in producing
    the final map, a measure of sensitivity.  A value of $n$ is essentially
    equal to $n^2$ independent boresight observations.  A typical value for
    HIPASS outside the ZOA survey area is around 7, increasing to 9 in the
    zone overlaps and the RA~= $0^\mathrm{h} / 24^\mathrm{h}$ seam.  Near the
    SCP and in the zone $-87\degr$ / $-82\degr$ overlap it increases to 11,
    but note the pinhole right at the SCP itself.  Deeper mapping was done in
    the SGP region, with values in the range 10 to 12, and up to 21 in the
    Sculptor deep field.  Values range between 18 and 23 within the ZOA.
    Saturated sources may be identified as local depressions (when viewed at
    full resolution).  Linear greyscale ranging from 0 to 23.}
  \label{fig:BeamRSS}
\end{figure*}

The resulting cross-plot of some 470,000 pixels is shown in Figure
\ref{fig:xplot} and shows excellent agreement.  Like Haslam et al.\
\shortcite{Haslam}, the Reich et al.\ maps use a {\em full-beam} brightness
temperature computed for a very extended beam of $7\degr$. The brightness
temperature scale for the Parkes map is therefore consistent with this.  A
conversion value of $T_B/S = 0.44$\,K/(Jy/beam) also allows compact source
fluxes to be measured accurately (though it is recommended to use the compact
source map in Figure \ref{fig:Compact} for this purpose).  Nevertheless, the
implied main beam efficiency is only 51\% with respect to a Gaussian beam with
a HPBW of $14\farcm4$.  This is even lower than the quoted Stockert beam
efficiency of 55\% \cite{Reich1}.  Values of $T_B/S = 0.80$\,K/(Jy/beam) are
typically used in HI observations of extended objects at Parkes, e.g.\
Staveley-Smith et al.\ \shortcite{Staveley-Smith2}.  Some of the lower
efficiency is due to the use, in the present work, of all the off-axis beams,
the outer of which is 26\% less efficient than the central beam.  As discussed
in Barnes et al.\ \shortcite{Barnes}, there are similar factors which arise
from the nature of the gridding function and the subsequent variation of beam
size with the angular scale of the source, some of which was absorbed in the
previously discussed flux calibration factor of 1.09.  Nevertheless, a
conversion value closer to $T_B/S \approx 0.57$\,K/(Jy/beam) might have been
expected.  It therefore remains possible that some of the $\sim 30$\% residual
difference is due to sidelobe power on intermediate scales ($<7\degr$).
Structures on this scale may be slightly less prominent than they would be
with a telescope with a pure Gaussian beam.  However, the excellent match with
the lower resolution data (Figure \ref{fig:xplot}) after application of a
single overall scale factor, implies high consistency with previous work in
the field.

%----------------------------------------------------------------- floating --

% /DATA/GRUS_1/CHIPASS/median8/CHIPASS+Scl+CZOA_GALZ_spots/-
%   CHIPASS_reich.fits
%   Reich_AIT.fits
%
% Cross-plot created by chipass_xplot and saved as PostScript.  Converted to
% jpeg using gimp which reduced the size substantially.

%----------------------------------------------------------------- floating --

% /DATA/GRUS_1/CHIPASS/median8/CHIPASS+Scl+CZOA_GALZ/-
%   CHIPASS+Scl+CZOA_GALZ_WGTMED1.continuum.fits
%
% kvis levels 0 to 54.5 (1.09*50), logarithmic 3-cycle, Greyscale1.

%----------------------------------------------------------------- floating --

% /DATA/GRUS_1/CHIPASS/median8/CHIPASS+Scl+CZOA_GALZ/-
%   CHIPASS+Scl+CZOA_GALZ_BEAMRSS.continuum.fits
%
% kvis levels 0 to 23, linear, Greyscale1.

%----------------------------------------------------------------- floating --

% /DATA/GRUS_1/CHIPASS/median8/CHIPASS+Scl+CZOA_GALZ/-
%   CHIPASS_Reich.fits
%   - Reich_AIT.fits
%   (generated by kvis)
%
% kvis levels -500 to 500, linear, Greyscale1.
\begin{figure*}
  \centering
  \includegraphics[width=460pt]{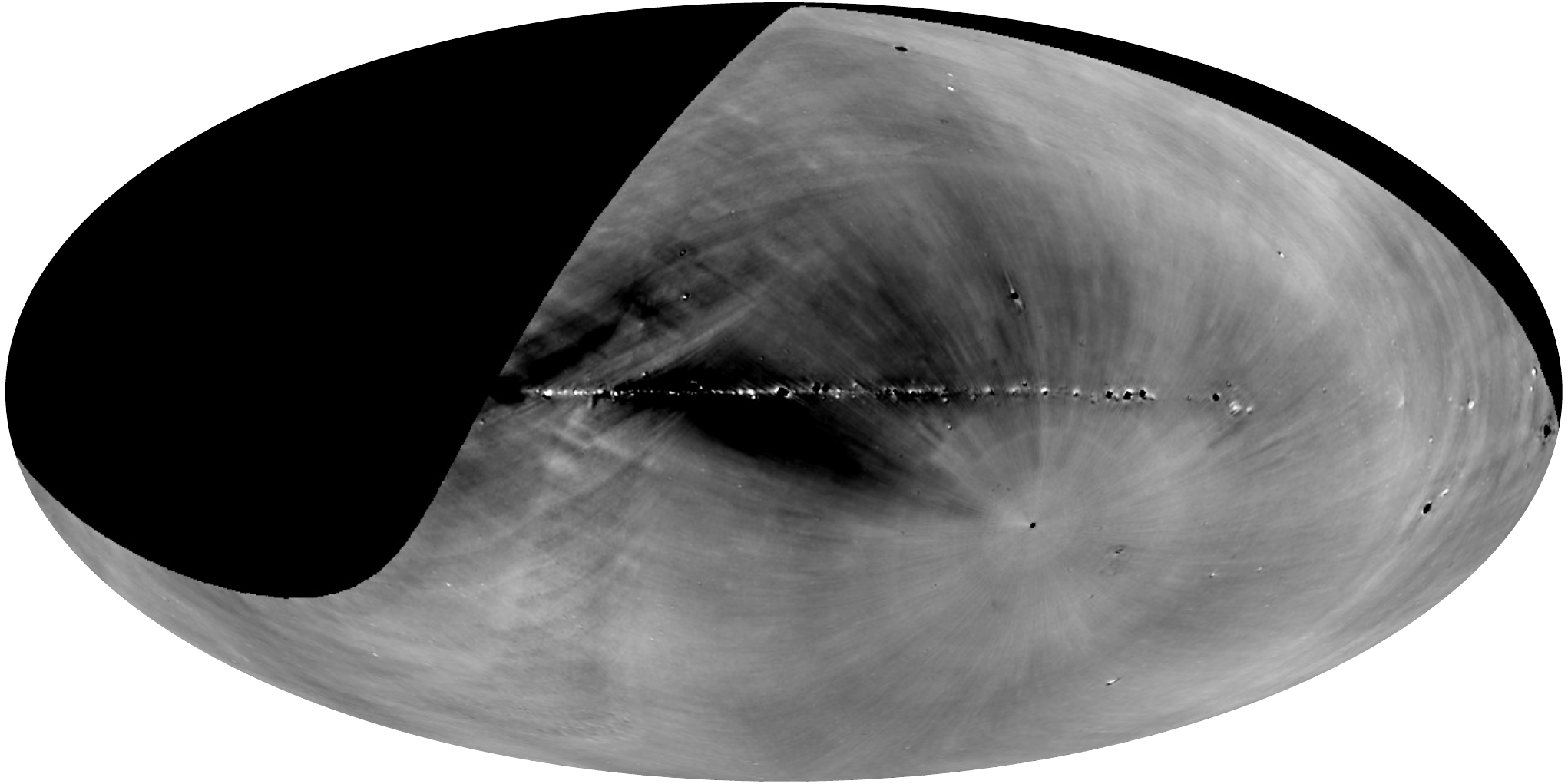}
  \caption[]{The difference map, CHIPASS (blurred to $35\farcm4$) minus Reich
    et al., highlights residual artefacts in both maps.  Inevitably, strong
    point sources fail to cancel exactly, but there is almost no sign of
    medium-scale cosmic emission.  Note that blanked areas around saturated
    sources result from the smoothing operation.  Linear greyscale from -0.5
    to 0.5\,K.}
  \label{fig:C-R}
\end{figure*}

%================================================================== Discussion

% For A (J0322-3714, 240.1-56.7)
% Tau A (J0534+2200, 184.6-05.8)
% Ori A (J0535-0523, 209.0-19.4)
% Vir A (J1230+1223, 283.8+74.5)
% Pic A (J0519-4546, 251.6-34.6)
% Hya A (J0918-1205, 242.9+25.1)
% 3C273 (J1229+0203, 290.0+64.4)
% Her A (J1651+0459,  23.0+28.9)
% 3C353 (J1720-0058,  21.2+19.6)

\section{DISCUSSION}
\label{sec:discussion}

In this section, we discuss the reliability and calibration of the final map
at various angular scales.  In particular, during the course of data
reduction, we have applied several important additive corrections.  What is
the cumulative effect of these?

Fortunately, the main effects we have corrected for manifest themselves
prominently as image artefacts with a definite signature.  The data can be
re-ordered in various ways to accentuate these effects.  Thus, the $\Tsys$
elevation dependence, which has an obvious signature in Figure
\ref{fig:Extended}, is
distilled for measurement in Figure \ref{fig:ZPAEl}.  Judging from the scale
of Figure \ref{fig:TsysEl}, the maximum error across a scan is unlikely to
exceed 100\,mJy/beam in the worst case which is at the SCP and the northern
elevation limit.  In fact, over most of the sky the effect tends to cancel due
to the random orientations of the scans.  Subtracting Figure \ref{fig:ElCorr}
from Figure \ref{fig:Extended} revealed a correction of up to 1\,Jy/beam along
the northern perimeter and 650\,mJy/beam in the vicinity of the SCP.  At
mid-latitudes there are isolated patches where the correction exceeds
100\,mJy, but mostly it is below this.  These are the absolute corrections,
the error would be much less.  If it was as much as 10\% then the residuals
would range from less than 10\,mJy/beam (mostly) up to 100\,mJy/beam depending
on location (5\,mK to 48\,mK).

Similarly, errors in the determination of the HIPASS and ZOA levels manifest
themselves as obvious artefacts -- streamers -- as seen in Figure
\ref{fig:chipassLevz}.  Most of the streamers were eliminated by smoothing the
levels, the only obvious residuals being a short pair associated with
Fornax\,A.  Again the error is highly dependent on location in the map, with
northern declinations the most affected as previously explained.

The most obvious artefacts remaining in the final image of Figure
\ref{fig:Final} resemble fine brushstrokes consistent with the pattern seen in
Figure \ref{fig:tracks}.  Again, these are highly dependent on location,
mostly occurring in isolated patches outside the ZOA at levels of up to
50\,mJy/beam (22mK) at the worst, too low to warrant {\it de jure} smoothing
of the whole map.  However, sections of the map would certainly benefit from
light smoothing at the discretion of the end user.  There is no clear evidence
in the final map of the HIPASS zones or ZOA zoans or the seams between them.
However, these may become apparent on differencing with other surveys.

The cross-plot in Figure \ref{fig:xplot} of CHIPASS versus the 1420\,MHz
all-sky map derived from Reich \shortcite{Reich1}, Reich \& Reich
\shortcite{Reich2}, and Reich, Testori, \& Reich \shortcite{Reich3}
demonstrates excellent agreement overall.  Curiously, most of the worst
outliers at high values, both above and below the line, arise from W51 and
W44.  As shown by the difference map in Figure \ref{fig:C-R}, there is
excellent agreement between the two maps for compact sources and structures up
to about $60\degr$ in angular size, including the North Polar spur, the Gum
nebula, Barnard's Loop, and the many wispy tendrils that emanate from the
plane of the Milky Way.  Reassuringly, some structures at northerly
declinations that looked at first sight as though they might be residuals from
the $\Tsys$ elevation correction and/or HIPASS zone level correction do have
counterparts in the Reich et al.\ map.  The larger depression south of the
plane indicates a discrepancy of up to 0.8\,K from the mean.  Likewise the
more compact 1.2\,K depression at the end of the plane and the dark streak
leading to it.  This area produces the small bulge below the line at low
values in the cross-plot of Figure \ref{fig:xplot}.  The depression is visible
in the 408 -- 1420\,MHz spectral index map shown in Reich, Reich, \& Testori
\shortcite{Reich4} and was previously suspected\footnote{Patricia Reich
(2013), private comm.} to be an artefact caused by the large number of scans
terminating at the Galactic plane, though with relatively small effect on the
spectral index in this bright region of the sky.

CHIPASS produces barely discernible seams, about 20\,mK in height, that
delineate HIPASS zones south of $\delta = -58\degr$.  The CHIPASS zero level
has been set close to that of Reich et al., with the mean difference in more
settled areas away from the Galactic plane at about 0\,K.  The rms computed
over sample areas of about 400 square degrees in settled regions varies from
15\,mK to 60\,mK depending strongly on location, with 40\,mK being typical.
This being a difference map, if shared equally, the error attributable to each
map would be less by $\sqrt{2}$.  On this basis a typical rms error of 30\,mK
may be assigned to the CHIPASS map, excluding the error in the zero level.
However, Figure \ref{fig:C-R} highlights areas where the rms may be much
higher.

The rms of the CHIPASS minus Reich et al.\ difference map is more in the
nature of an absolute error (on all spatial scales) rather than just noise.
It is closely related to the scatter in the cross-plot of Figure
\ref{fig:xplot}, the main contribution to it undoubtedly being localised
baseline error.  It is not simply related to what would normally be quoted as
the {\em rms noise} or {\em sensitivity} for surveys of this type.

Typically the intrinsic rms noise would be computed from the scatter in
source-free regions of a map.  That is impossible here because of the high
source density, but a number can be obtained from the difference between the
HIPASS and ZOA maps (as in Section \ref{sec:iteration}).  It is clear in
Figure \ref{fig:iters}e that strong sources on the Galactic plane do not
cancel exactly in the difference map, and that is readily understood to arise
from small errors in the flux density calibration between the two, as well as
effects that may arise in gridding irregularly sampled data.  Looking in the
very quietest regions of Figure \ref{fig:iters}e and choosing sample areas so
as to exclude source residuals as much as possible, the smallest rms that can
be measured over a few hundred pixels is 20\,mJy/beam, most of which must
arise from the less-well sampled HIPASS map.  Applying a correction of
$\sqrt{5/6}$ for the noise contributed to the difference by the ZOA map brings
this figure down to 18\,mJy/beam.  That gives an rms of 8\,mK for the
HIPASS-only regions outside the ZOA and Sculptor deep field (Figure
\ref{fig:BeamRSS}), and on this basis we have adopted a nominal figure of
40\,mK, or 90\,mJy/beam, (being 5 $\times$ rms) for the sensitivity.

Carretti et al.\ \shortcite{SPASS} kindly provided a pre\-/publication
continuum map derived from S-PASS for comparison with CHIPASS.  This southern
hemisphere ($\delta < 0\degr$) polarisation survey at 2.3\,GHz was made with a
single beam system on the Parkes radio telescope using a
long\-/azimuth\-/scan, basketweaving observing strategy tailored for the
purpose \cite{Carretti}.  The S-PASS continuum map with HPBW $10\farcm75$ was
blurred to CHIPASS resolution.  With a factor of $\times 1.64$ difference in
frequency, the effect of the varying spectral index for thermal and
non-thermal sources makes the comparison less direct than for Reich et al.\
and indeed the cross-plot revealed two distinct populations.  Nevertheless,
excellent agreement on scales up to $60\degr$ was again confirmed.  In fact,
there were only a few minor differences in the complex morphology of the wispy
material extending from the Galactic plane.  However, in places the difference
map showed evidence of the scan lines (brushstrokes) from ZOA as well as
HIPASS.  The depressed area seen in Figure \ref{fig:C-R} also appeared in the
difference map S-PASS minus Reich et al., thus confirming its origin in the
latter.

As an indirect measure of image fidelity, up to four orders of diffraction
rings can be seen around Tau\,A, Ori\,A, and Virgo\,A.  At least the first
order diffraction ring can also be detected around Pictor\,A, Hydra\,A, 3C273,
Hercules\,A, and 3C353.  Three orders were also visible around W38 in the
S-PASS difference map.

The following CHIPASS data products are available from
\url{www.atnf.csiro.au/research/CHIPASS/}:
\begin{itemize}
\item
  The extended source map in Galactic coordinates on a Hammer-Aitoff
  projection as seen in Figure \ref{fig:Final}.  This map, calibrated in mK
  {\em full-beam} brightness temperature, is the main product of this work.

\item
  A subset of the above, the Galactic plane map, in Galactic coordinates but
  on a plate carr\'{e}e projection spanning $+68\degr \ge \ell \ge -180\degr$
  with $|b| < 10\degr$.  This map is provided specifically for studies of the
  many complex sources in the Galactic plane.

\item
  The extended source map in J2000 equatorial coordinates on a plate
  carr\'{e}e projection.  This is the most suitable map to use for blurring
  the HPBW, or regridding onto other coordinate systems and map projections.

\item
  The compact source map in Galactic coordinates on a Hammer-Aitoff
  projection as seen in Figure \ref{fig:Compact}.  This map, calibrated in
  Jy/beam, is best suited for measuring point source positions and flux
  densities.  It is independent of the calibrations needed for the extended
  source maps.
\end{itemize}
\noindent
Each map was produced using all relevant HIPASS, ZOA, and Sculptor deep-field
data.  The compact source map also includes the Centaurus deep-field survey
data.  Each map has an associated relative sensitivity map akin to Figure
\ref{fig:BeamRSS}.  Table \ref{ta:mapparm} summarises important map
parameters.

%----------------------------------------------------------------- floating --

\begin{table}
  \caption[]{Summary of parameters relating to the full-sky, extended-source
    Hammer-Aitoff equiareal map.}
  \begin{center}
  \protect\begin{tabular}{lc}
    \hline\hline
    \noalign{\smallskip}
    \noalign{\smallskip}
    Input spectra (either pol.)        & 131,973,107 \\
    Map dimensions                     & $4901 \times 2451$ \\
    Pixel spacing                      & $4\arcmin$ (at reference point) \\
    Northern boundary: \\
      \hspace{15pt}cut-off declination & $+26\degr00\arcmin$ \\
      \hspace{15pt}averaged over RA    & $+25\degr33\arcmin$ \\
      \hspace{15pt}limit of full coverage
                                       & $+25\degr04\arcmin$ \\
    Blank pixels ($\delta < 25\degr$)  & 308 (incl. SCP)\\
    Gridding kernel radius             & $6\arcmin$ \\
    Gridding kernel area               & \hphantom{1}7.1 pixels \\
    Gridded beam FWHM                  & $14\farcm4$ \\
    Beam effective area$^{\dag\scriptscriptstyle a}$
                                       & 14.7 pixels \\
    Non-blank pixels                   & 6,645,290 \\
    Effective beam areas               & \hphantom{6,}450,000 \\
    Coverage                           & \parbox{65pt}{29,540\,deg$^2$} \\
                                       & \parbox{65pt}{= 8.997\,sr} \\
                                       & \parbox{65pt}{= 0.716\,sphere} \\
    $T_b$ scale (full-beam)            & 0.44\,K/(Jy/beam) \\
    Sensitivity ($5\sigma$)            & 40\,mK (90\,mJy/beam) \\
    Absolute error (rms)               & 30\,mK (70\,mJy/beam) \\
    $T_b$ map statistics: \\
      \hspace{15pt} min : max$^{\dag\scriptscriptstyle b}$
                                       & \parbox{65pt}{3.3\,K : 67.9\,K} \\
      \hspace{15pt} mean $\pm$ rms     & \parbox{65pt}{3.9\,K $\pm$ 1.1\,K} \\
      \hspace{15pt} median $\pm$ $S_n$$^{\dag\scriptscriptstyle ^c}$
                                       & \parbox{65pt}{3.6\,K $\pm$ 0.16\,K} \\
    Relative sensitivity map: \\
      \hspace{15pt} min$^{\dag\scriptscriptstyle d}$ : max
                                       & \parbox{45pt}{4.0 : 23.2} \\
      \hspace{15pt} mean $\pm$ rms     & \parbox{45pt}{8.9 $\pm$ 3.3} \\
      \hspace{15pt} median $\pm$ $S_n$ & \parbox{45pt}{7.7 $\pm$ 0.8} \\
    Spectra counts$^{\dag\scriptscriptstyle e}$: \\
      \hspace{15pt} min : max          & \parbox{50pt}{\noindent\hphantom{1}20
                                                       : 828} \\
      \hspace{15pt} mean $\pm$ rms     & \parbox{50pt}{140 $\pm$ 133} \\
      \hspace{15pt} median $\pm$ $S_n$ & \parbox{50pt}{\noindent\hphantom{1}94
                                                      $\pm$ \hphantom{1}21} \\
    \noalign{\smallskip}
    \hline
    \noalign{\smallskip}
    \multicolumn{2}{l}{\parbox{230pt}{\hangindent=11pt\footnotesize{
      $^{\dag\scriptscriptstyle a}$\hspace{3pt}Area
      of a top-hat function with the same peak height and volume as a Gaussian
      of specified FWHM ($= \pi/(4\ln2)$ FWHM$^2$).}}} \\
    \multicolumn{2}{l}{\parbox{230pt}{\footnotesize{
      $^{\dag\scriptscriptstyle b}$\hspace{3pt}Excluding the brightest 16
      sources which are saturated.}}} \\
    \multicolumn{2}{l}{\parbox{230pt}{\footnotesize{
      $^{\dag\scriptscriptstyle c}$\hspace{3pt}The robust measure of
      dispersion discussed in Section \ref{sec:Sn}.}}} \\
    \multicolumn{2}{l}{\parbox{230pt}{\footnotesize{
      $^{\dag\scriptscriptstyle d}$\hspace{3pt}The imposed cut-off.}}} \\
    \multicolumn{2}{l}{\parbox{230pt}{\hangindent=11pt\footnotesize{
      $^{\dag\scriptscriptstyle e}$\hspace{3pt}Number of input spectra (of
      either polarisation) used to compute the pixel value.}}} \\
    \noalign{\smallskip}
    \hline\hline
  \end{tabular}
  \end{center}
  \label{ta:mapparm}
\end{table}

%============================================================ Acknowledgements

\begin{acknowledgements}
The HIPASS and ZOA surveys were the result of the tireless work, which
extended over many years, of many collaborators and observers, including the
ATNF electronics group who built the beautiful multibeam receiver and
correlator, the Parkes Observatory support staff, and the original authors of
the {\em livedata} and {\em gridzilla} software suites.  For this work Stacy
Mader's assistance with the data archive is gratefully acknowledged.

This work is based on software packages provided by GNU/Linux, specifically
the Debian distribution, and by the Comprehensive \TeX\ Archive Network.

The Parkes radio telescope is part of the Australia Telescope National
Facility which is funded by the Commonwealth of Australia for operation as a
National Facility managed by CSIRO.
\end{acknowledgements}

%================================================================ Bibliography

\end{document}